\documentclass[a4paper,11pt]{book}

\usepackage{amsmath,amsfonts,amsbsy}
\usepackage{multibox}
\usepackage{latexsym}
\usepackage{appendix}
\usepackage{amssymb,latexsym}
\usepackage[mathscr]{eucal}
\usepackage{youngtab}
\usepackage{a4}
\usepackage{amsfonts}
\usepackage{amssymb}
\usepackage{lscape}
\usepackage{cite}
\usepackage{epsfig}
\usepackage{multirow}
\usepackage[all]{xy}
\usepackage{amsmath}
\usepackage{amssymb}
\let\a=\alpha   \let\b=\beta

\def\slash#1{\mbox{$\not \!\! #1$}}

\newcommand{\<}{\langle}
\renewcommand{\>}{\rangle}
\newcommand{\beq}{\begin{equation}}
\newcommand{\eeq}{\end{equation}}
\newcommand{\beqn}{\begin{eqnarray}}
\newcommand{\eeqn}{\end{eqnarray}}

\newcommand{\nn}{\nonumber}

\newcommand{\Tr}{{\rm Tr}}

\newcommand{\mP}{\mathbb{P}}
\newcommand{\mZ}{\mathbb{Z}}
\newcommand{\mC}{\mathbb{C}}
\newcommand{\mR}{\mathbb{R}}
\newcommand{\eps}{\epsilon}

\newcommand{\be}{\begin{equation}}
\newcommand{\ee}{\end{equation}}
\newcommand{\ba}{\begin{eqnarray}}
\newcommand{\ea}{\end{eqnarray}}
\newcommand{\bdm}{\begin{displaymath}}
\newcommand{\edm}{\end{displaymath}}

\def\b{\beta}

\def\a{\alpha}

\def\adot{{\dot\alpha}}

\newcommand{\im}{{\rm Im\,}}

\newcommand{\ie}{{\it i.e.\ }}
\newcommand{\eg}{{\it e.g.\ }}

\newcommand{\calR}{{\mathcal R}}

\DeclareMathAlphabet{\mathpzc}{OT1}{pzc}{m}{it}
\def\bea{\begin{eqnarray}}
\def\eea{\end{eqnarray}}
\def\beas{\begin{eqnarray*}}
\def\eeas{\end{eqnarray*}}
\def\sla{\raise.15ex\hbox{$/$}\kern-.57em}

\def\bea{\begin{eqnarray}}
\def\eea{\end{eqnarray}}
\def\de{\partial}

\def\sla{\raise.15ex\hbox{$/$}\kern-.57em}
\def\ie{{\it i.e.}~}
\def\eg{{\it e.g.}~}
\def\ap{{\alpha^\prime}}

\def\a{\alpha}
\def\b{\beta}

\def\vf{\varphi}

\def\cA{{\cal A}}
\def\cB{{\cal B}}
\def\cC{{\cal C}}
\def\cD{{\cal D}}
\def\cE{{\cal E}}
\def\cF{{\cal F}}
\def\cG{{\cal G}}
\def\cH{{\cal H}}
\def\cI{{\cal I}}
\def\cJ{{\cal J}}

\def\cL{{\cal L}}
\def\cM{{\cal M}}
\def\cN{{\cal N}}
\def\cO{{\cal O}}
\def\cP{{\cal P}}

\def\cR{{\cal R}}
\def\cS{{\cal S}}
\def\cT{{\cal T}}
\def\cU{{\cal U}}
\def\cV{{\cal V}}

\def\cX{{\cal X}}

\def\cZ{{\cal Z}}

\begin{document}
\oddsidemargin 6pt\evensidemargin 6pt\marginparwidth
48pt\marginparsep 10pt
\thispagestyle{empty}

\vspace*{-1.5truecm}

\begin{center}
 \vspace{.7cm}
\makebox[\textwidth]{\raisebox{0.68cm}{
\begin{minipage}[h]{13truecm}
        \hspace*{0.2cm}
        {\sc \LARGE UNIVERSIT\`A DEGLI STUDI DI ROMA\\
        \hspace*{3.37cm}
        ``TOR \rule{0pt}{24pt}VERGATA''}\\
    \end{minipage}}}
\end{center}


\begin{center}
    \includegraphics[height=4cm]{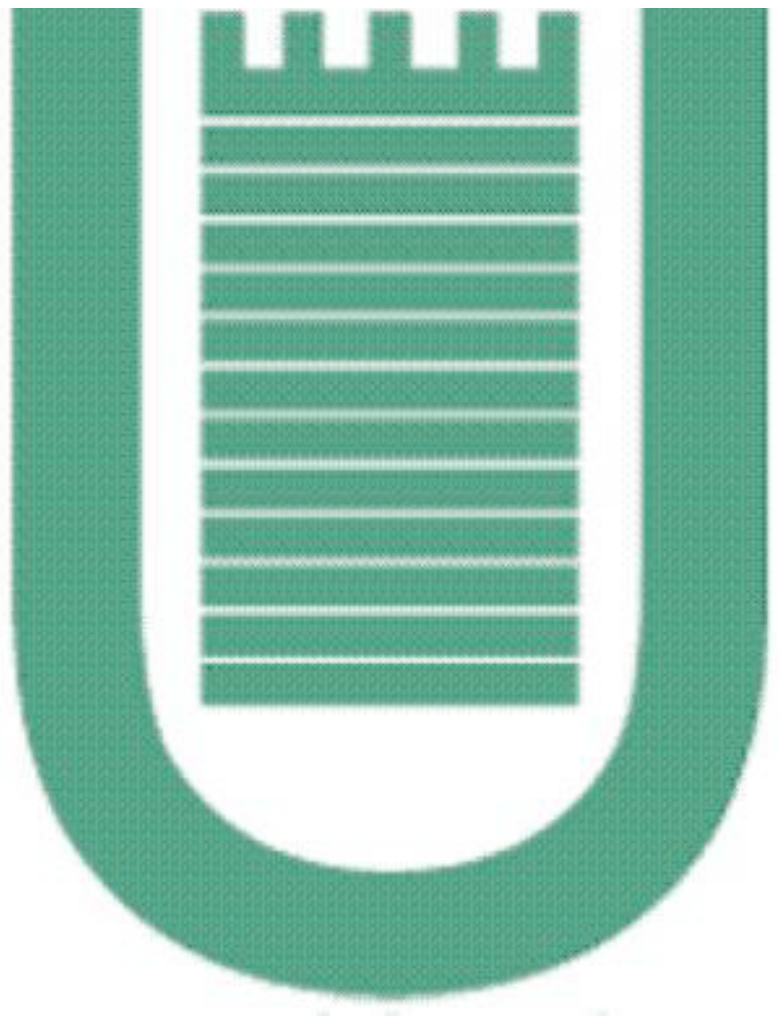}
\end{center}

\begin{center}
FACOLT\`A DI SCIENZE MATEMATICHE, FISICHE E NATURALI \\
Dipartimento di Fisica \\
\end{center}

\vspace{1.5cm}

\begin{center}
{\Large {\bf Non-perturbative aspects of gauge and string theories and their holographic relations}}
\end{center}

\vspace{1.5cm}

\begin{center}
{\large Tesi di Dottorato di Ricerca in Fisica} \\\vspace{.3cm}
\rule{0pt}{22pt}\it {Marine Samsonyan} \rule{0pt}{22pt}
\end{center}

\vspace{1.5cm}  \noindent
{\large
{\centering
  \mbox{Relatore
  \phantom{xxxxxxxxxxxxxxxxxxxxxxxxxxxxxxxxx}
  Coordinatore del dottorato}\\
 \mbox{Prof. {\it Massimo Bianchi} \phantom{xxxxxxxxxxxxxxxxxxxxxxx}
 Prof. {\it Piergiorgio Picozza}}\\}
\quad\\

\vspace{1.0cm}
\begin{center}
\underline{Ciclo XXIII}\\
 Anno Accademico 2010/2011
\end{center}

\newpage
\vspace*{5truecm}
\begin{center}
{\LARGE To my parents}\\
\end{center}
\newpage\thispagestyle{empty}

\begin{center}
{\large \bf Abstract}
\end{center}
In this thesis we discuss non-perturbative phenomena emerging in gauge 
as well as in string/supergravity  theories. We discuss instantons in supersymmetric Yang-Mills theories. 
An interesting class of theories are obtained by adding adjoint hypermultiplets to pure $\cN=2$ theories. 
These theories, called $\cN=2^*$, are massive deformations of $\cN=4$ super Yang-Mills (SYM) and  can be 
thought of as a minimal supersymmetric five dimensional theory compactified on a circle.
We compute the partition function of 5D minimal supersymmetric $U(1)$ gauge theory with extra adjoint matter
in general $\Omega$-background. It is well known that such partition functions encode very rich topological 
information. We show in particular that unlike the case with no extra matter, the partition function with 
extra adjoint at some special values of the parameters directly reproduces the generating function for the
Poincar\`e polynomial of the moduli space of instantons.

Instantons play also a very crucial role in string theory, specifically in the context of string dualities. 
They have also interesting phenomenological implications. We discuss the basic aspects of worldsheet and
penta-brane instantons as well as (unoriented) D-brane instantons and threshold corrections to
BPS-saturated couplings in superstring theories. Then we consider
non-perturbative superpotentials generated by `gauge' and `exotic'
instantons living on D3-branes at orientifold singularities. We also discuss the interplay between 
worldsheet and D-string instantons on $T^4/\mZ_2$. We focus on a 4-fermi amplitude,
give Heterotic and perturbative Type I descriptions, and offer a
multi D-string instanton interpretation.

Furthermore, instantons give a non-trivial check of the $AdS/CFT$ correspondence. $AdS/CFT$ dualities relate 
Type IIB superstring theory or M-theory compactified on an anti-de Sitter space-time times a compact space 
to conformally invariant field theories. In particular, in $AdS_5/CFT_4$, Type IIB D-instantons correspond 
to usual gauge instantons in dual $\cN=4$ SYM theory.
Another interesting application of the holography principle is $AdS_4/CFT_3$ correspondence.
This allows to investigate the worldvolume theory of M2-branes, the basic objects of M-theory. In this 
context we consider $\cN=8$ supergravtiy on $AdS_4\times S^7$, which is the low energy limit of M-theory 
compactified on $S^7$.We revisit Kaluza-Klein compactification of 11-d supergravity on
$S^7/\mZ _k$ using group theory techniques that may find application in other flux vacua with internal coset 
spaces. Among the $SO(2)$ neutral states, we identify marginal deformations and fields that couple to the 
recently discussed world-sheet instanton of Type IIA on $\mC \mP^3$. We also discuss charged states, dual to 
monopole operators, and the $\mZ _k$ projection of the $Osp(8|4)$ singleton and its tensor products. In 
particular, we show that the doubleton spectrum may account for $\cN =6$ higher spin symmetry enhancement
in the limit of vanishing 't Hooft coupling in the boundary Chern-Simons theory.

\vfill

\pagebreak

 \chapter*{Acknowledgments}
This PhD thesis has been done at the University of Rome ``Tor Vergata". I would like to thank the Department 
of Physics and the Coordinator of the PhD program Professor Piergiorgio Picozza for partially supporting my 
work and RTN grant MRTN-CT-2004-503369. 

It is a great pleasure to thank my supervisor Prof. Massimo Bianchi for creating and supporting very 
active research atmosphere, for helping and leading through the fascinating research field, for being ready 
to discuss at any time. Special acknowledgments to Rubik Poghossian for encouraging my work and 
being very nice collaborator. I would like to thank for very 
interesting courses I have attended at the Department of Physics of the University of Rome ``Tor Vergata". 
Many thanks to Francesco Fucito, Jose Francisco Morales Morera, Gianfranco Pradisi, Giancarlo Rossi, Yassen 
Stanev, Gor Sarkissian and Robert Richter.
 
My deepest gratitude goes to my parents Manvel and Anahit, my sister Nune, my brother Hrant, my 
grandfather Hrant and grandmother Nazik. Particularly, I would like to thank Mher for constant support and 
patient.

Thanks to my friends Satenik, Hermine, Paytsar, Arpine, Karapet, 
Ara, Mikayel, Hrach, Vahagn, Manushak for their presence. Thanks to nice people I met in Italy for making my 
stay enjoyable, Gor and his wife, Alessio, Marina, Antonio, Francesco, Maurizio, Simona, Giulia, Flavio, 
Gianluca and Maria, Enrico, Beatrice, Daniel, Andrea Sc., Andrea M., Elena and her mother, signora Maria 
and signor Pompeo, Barbara, Laura, Stefania, Yingying, Carla, Guido and PhD students of {\it {Ciclo XXIII}}.

\pagenumbering{arabic} \setcounter{page}{1}
\tableofcontents
\listoftables

\chapter*{Introduction}

We are introducing the interested reader to the
fascinating subject of non-perturbative effects generated by
unoriented D-brane instantons.
After a very short reminder of Yang-Mills (YM)
instantons and the ADHM construction, we discuss basic aspects of
worldsheet and D-brane instantons and their original applications
to threshold corrections.

Then we discuss non-perturbative superpotentials
generated by `gauge' and `exotic' instantons living on D3-branes
at orientifold singularities.
We consider the interplay between worldsheet
and D-brane instantons on $T^4/\mZ_2$. We focus on a specific
4-hyperini amplitude, give Heterotic and perturbative Type I
descriptions, and offer a multi D-string instanton interpretation.

Several good reviews are available on the subject
\cite{BCKWreview}, that is also covered in textbooks
\cite{Kirbook}. 

We investigate the partition function of a five dimensional supersymmetric $U(1)$ gauge theory with an extra 
adjoint hypermultiplet. Such partition functions encode very rich topological information. As a
manifestation we argue that unlike the case with no extra matter,at some special values of the parameters 
this partition function directly reproduces the generating function of the Poincar\`e polynomial for the 
moduli space of instantons. We check this conclusion explicitly computing the partition function in the case
of gauge group $U(1)$.

We revisit Kaluza-Klein (KK) compactification of 11-d supergravity on $S^7$.
The spectrum of KK excitations in flux vacua plays
an important role both in attempts to embed the Standard Model in
String Theory and in the holographic correspondence. In the spirit
of holography, the seminal observation of Schwarz's
\cite{Schwarz:2004yj} and the subsequent work of Bagger and
Lambert \cite{Bagger:2006sk, Bagger:2007jr, Bagger:2007vi,
Bagger:2008se } and, independently, of Gustavsson
\cite{Gustavsson:2007vu }, motivated Aharony, Bergman, Jafferis
and Maldacena (ABJM) \cite{Aharony:2008ug,Klebanov:2009sg} to propose a duality between
superconformal Chern-Simons (CS) theories in $d=3$ dimensions and String /
M- theory on $AdS_4$.

The duality has been thoroughly tested and extended to cases with
lower supersymmetry \cite{Aganagic:2009zk, Jafferis:2009th, Gaiotto:2009tk, Jafferis:2008em, 
Jafferis:2008qz, Forcella:2009jj}. In particular the
superconformal index has been matched both in the regime $k>>1$
($SO(2)$ singlets) \cite{Bhattacharya:2008bja,
Bhattacharya:2008zy} and at finite $k$ \cite{Kim:2009wb,Kim:2010vw
}. A detailed analysis of the (BPS) spectrum and the
supermultiplet structure is however still incomplete.
We fill in this gap and perform precision spectroscopy of
11-d supergravity on $AdS_4\times S^7/\mZ_k$ or, equivalently, Type
IIA on $AdS_4\times \mC\mP^3$. We will also discuss higher spin
symmetry enhancement in the limit of vanishing 't Hooft coupling
in the boundary $\cN =6$ Chern-Simons theory.

After reviewing the ABJM
model, presenting both bulk and boundary vantage points, we 
revisit KK reduction of 11-d supergravity on $S^7$
\cite{Castellani:1984vv} and then perform the decomposition of
$SO(8)$ into $SO(6)\times SO(2)$ so as to derive the KK
excitations of $\cN=6$ gauged supergravity \cite{Salam:1989fm},
including states charged under $SO(2)$ that are expected to be
dual to `monopole' operators on the boundary
\cite{Aharony:2008ug,Klebanov:2009sg}. Since we rely on group theory
techniques which are not easily found in the available literature,
we try to make this part of the presentation as pedagogical as
possible, also in view of applications to other flux vacua with
internal coset manifolds $G/H$. We then compare the resulting bulk
spectrum with the spectrum of gauge-invariant operators on the
boundary. Finally we compute the partition function of the
boundary theory performing an orbifold projection on the parent
theory ($k=1,2$ cases) and examine the higher spin content of the
theory. Various appendices summarize useful $SO(8)$ and $SO(6)$
group theory formulae.

The plan of the thesis is the following.

In Chapter \ref{chapter1}
we give a review of instantons in gauge theories, discuss ADHM construction. Then we discuss 
instantons in string theories, their possible generated effects in diverse string compactifications. We 
discuss a little bit thresholds in toroidal compactifications, emphasize some phenomenological aspects.

In Chapter \ref{chapter2} we compute the partition function of five dimensional supersymmetric $U(1)$ gauge 
theory with adjoint matter ($\cN=2^*$) in general $\Omega$-background. Such partition functions encode very 
rich topological information. We show that the partition function with extra adjoint matter at some special 
values of the parameters directly reproduces the generating function for the Poincare polynomial of the 
moduli space of instantons.

In Chapter \ref{chapter3} we discuss instantons in string theories. In particular, we explain how the rather 
non trivial ADHM construction arises very intuitively and naturally in string theory in the context of Dp-
branes inside D(p+4) ones. Particularly we discuss D3/D(-1) system. Then we mention the vertex operators for 
`gauge' and `stringy' instantons. D-branes at orbifolds, unoriented projection, in particular $R^6/\mZ_3$ 
projection, non-perturbative superpotential for $Sp(6)\times U(2)$ and for $U(4)$ are discussed as well. 
Then we discuss exotic/stringy instantons. One section is devoted to the effect of fluxes.

Chapter \ref{chapter4} is devoted to the relation of worldsheet and D-brane instantons in different string 
theories. We discuss how perturbatively different string theories may be shown to be equivalent once
non-perturbative effects are taken into account. In particular, Heterotic-Type I duality requires that the 
Heterotic fundamental string and the Type I D-string to be identified. Then we consider compactification on 
$T^4/\mZ_2$ to $D=6$ of Type I and Heterotic theories. We discuss duality and dynamics in $D=6$, where
$\phi _H$ and $\phi _I$ are independent. To further test the correspondence and gain new insights into multi 
D-brane instantons, we consider a four-hyperini Fermi type interaction that is generated by instantons and 
corresponds to a `chiral' (1/2 BPS) coupling in the $\cN=(1,0)$ low energy effective action. 

In Chapter \ref{chapter5} we discuss the role of instantons in AdS/CFT correspondence. First we give 
introductory section 
on gauge theory/string theory dualities, then we discuss the general aspects of $AdS_5\times S^5$ and 
instantons which play an important role in proving the conjectural duality between Type IIB superstring 
theory compactified on $AdS_5\times S^5$ and $\cN=4$ supersymmetric Yang-Mills theory with gauge group 
$SU(N)$. Then we consider other gauge theory/string theory dualities, in particular, we concentrate on 
$AdS_4\times S^7$ case. M-theory on $AdS_4\times S^7$ is dual to three dimensional superconformal field 
theory. This three dimensional theory was identified by Aharony, Bergman, Jafferis and Maldacena and goes 
under the name ABJM model. We give a brief description of the ABJM model, bulk and boundary theories.
Then, by giving group theoretical methods, we show how one can break the supersymmetry from $\cN=8$ of M-
theory on $AdS_4\times S^7$ to $\cN=6$ of Type IIA theory compactified on $AdS_4\times \mC\mP ^3$. This is 
not a spontaneous supersymmetry breaking. We give all KK towers keeping track of $SO(2)$ charge. Among the 
neutral states we identify marginal deformations and fields that couple to world-sheet instanton of Type IIA 
on $\mC\mP^3$, which we discuss later in Chapter 8. Charged states are dual to monopole operators.

In Chapter \ref{chapter6} we discuss the higher spin (HS) extension of $\cN=6$ gauged supergravity in 
$AdS_4$. We perform $\mZ_k$ projection of the $Osp(8|4)$ singleton and its tensor products. We show that the 
spectrum arising from symmetric doubleton is precisely the spectrum of `massless' states of $\cN=8$ gauged 
supergravity on $AdS_4$.

In Chapter \ref{chapter7} we discuss instantons in $\mC\mP^3$.

Appendices include different formulae.

This thesis is based on the papers \cite{Bianchi:2009ij}, \cite{Poghossian:2008ge}, \cite{Bianchi:2010mg}.

\chapter{Instantons from Fields to Strings}
\label{chapter1}

\section{Yang-Mills Instantons: a reminder}

Instantons (anti-instantons) are self-dual (anti-self-dual)
classical solutions of  the equations of motions of pure
Yang-Mills theory in Euclidean space-time. \bea F_{\mu\nu}=\pm
\tilde{F}_{\mu\nu} \label{dantd} \eea with
$\tilde{F}_{\mu\nu}=\frac{1}{2}\epsilon_{\mu\nu\rho\sigma}F_{\rho\sigma}
$. In quantum theory they can be thought of as gauge
configurations bridging quantum tunnelling among topologically
distinct vacua. It is remarkable that self-dual (anti-self-dual)
gauge fields automatically satisfy YM equations {\it in vacuo} as
a result of the Bianchi identities. These solutions are classified
by a topological charge: \bea K =\frac{g^2}{32
\pi^2}\int{d^4xF^a_{\mu\nu}\tilde{F}^a_{\mu\nu}}
\label{Pontryagin} \eea an integer, which computes how
many times an $SU(2)$ subgroup of the gauge group is wrapped by
the classical solution while its space-time location spans the
$S_3$-sphere at infinity. The action of a self-dual (or
anti-self-dual) instanton configuration turns out to be
\be S_I=\frac{8 \pi ^2}{g^2}|K|.\ee
Yang-Mills instantons are interesting from both physical and mathematical point of view. They give
non-perturbative contributions to the functional integral in the semi-classical approximation. Instantons 
are very interesting both in phenomenological models of QCD and for describing exact non-perturbative 
phenomena in supersymmetric gauge theories. In mathematics instanotns play a central role in Donaldson's 
construction of topological invariants of four-manifolds.

In $\cN = 1$ theories, instantons may generate superpotentials,
chiral condensates and lead to the dynamical supersymmetry breaking.

In $\cN = 2$ theories, instantons correct the analytic prepotential, give exact spectrum of $1\over 2$ BPS 
states.
In $\cN=2$ SYM theory the instanton calculus produces the correct coefficients of $\cN=2$ prepotential 
derived in the Seiberg-Witten construction.

In $\cN = 4$ theories, instantons interfere with perturbation
theory (no $R$-symmetry anomaly), give non-perturbative corrections to
correlation functions and anomalous dimensions ($S$-duality).

\subsubsection{ADHM construction}

An elegant algebro-geometric construction of YM instantons was
elaborated by Atiyah, Drinfeld, Hitchin and Manin and goes under
the name of ADHM construction \cite{ADHM}.

For $SU(N)$ groups, the ADHM ansatz for a self-dual gauge field
with topological charge $K$, written as a traceless hermitean
$N\times N$ matrix, reads
\bea (A_\mu)_{uv}(x)=g^{-1}\bar{U}_u^\lambda \partial_\mu U_{\lambda v} \quad ,\eea
where $U_{\lambda u}(x)$ with $u=1,...,N $ and
$\lambda=1,...,N+2K$ are $(N+2K)\times N$ complex `matrices' whose
columns are the basis ortho-normal vectors for the $N$ dimensional
null-space of a complex $2K\times (N+2K)$ `matrix'
$\bar{\Delta}(x)$, \ie satisfy
\be \bar{\Delta}_i^{\dot{\alpha}\lambda}U_{\lambda
u}=0=\bar{U}_u^\lambda \Delta_{\lambda i \dot{\alpha}}\ee for
$i=1,...,K$, $\alpha, \dot{\alpha}=1,2$. Remarkably,
$\Delta_{\lambda i\dot{\alpha}}(x)$ turns out to be at most linear
in $x$. In quaternionic notation\footnote{Any real 4-vector
$V_\mu$ can be written as a `real' quaternion $V_{\a\dot\a} =
V_\mu \sigma^\mu_{\a\dot\a}$ with $\sigma^\mu = \{1,-i\sigma^a\}$.}
for $x$,
\bea \Delta_{\lambda i\dot{\alpha}}(x)=
a_{\lambda i\dot{\alpha}}+b^{\alpha}_{\lambda
i}x_{\alpha\dot{\alpha}}
 \quad , \quad \bar{\Delta}_i^{\dot{\alpha}\lambda}
(x)=\bar{a}_i^{\dot{\alpha}\lambda}+
\bar{x}^{\dot{\alpha}\alpha}\bar{b}_{i\alpha}^{\lambda}\equiv(\Delta_{\lambda
i\dot{\alpha}})^* .\eea The complex constant `matrices' $a$ and $b$
form a redundant set of collective coordinates that include the
moduli space ${\cal{M}}_{K}$. Decomposing the index
$\lambda$ as $\lambda=u+i\alpha$, with no loss of
generality, one can choose a simple canonical form for $b$
\bea b^\beta_{\lambda j}=b^\beta_{(u+i\alpha)j}=\left(\begin{array}{c}
0\\
\delta_{\alpha}^{\beta} \delta_{ij}
\end{array}
\right), \quad \bar{b}_{\beta j}^\lambda=\bar{b}^{(u+i\alpha)}
_{\beta j}=\left(\begin{array}{c} 0 \quad \delta _\alpha ^\beta
\delta _{ij}
\end{array}
\right)\eea One  can also split $a$ in a similar way as:
\bea a_{\lambda j \dot{\alpha}}=a_{(u+i\alpha)j\dot{\alpha}}
=\left(\begin{array}{cc} w_{uj \dot{\alpha}}\\
(X_{\alpha \dot{\alpha}})_{ij}
\end{array}
\right), \quad \bar{a}_j^{\dot{\alpha}\lambda}=
\bar{a}^{\dot{\alpha}(u+i\alpha)}_j= \left(\begin{array}{c}
\bar{w}_{j}^{\dot{\alpha}u} \quad (\bar{X}^{\dot{\alpha}\alpha
})_{ji}
\end{array}
\right)\nn\\
\eea In order to ensure self-duality of the connection, the
`ADHM data' $\{ w, \bar{w}, X, \bar{X}\}$ with
$X_\mu^{\dagger}=X_\mu$ must satisfy algebraic constraints, known
as the ADHM equations, that can be written in the form
\be w_{ui \dot{\alpha}} (\sigma^a)^{\dot\alpha}{}_{\dot\beta}
\bar{w}_{j}^{\dot{\beta}u} + \eta^a_{\mu\nu} [X^\mu,X^\nu]_{ij} =
0\ee for later comparison with the D-brane construction. Note the $U(K)$ invariance of the above $3K\times 
K$ equations. For a recent review of supersymmetric instanton calculus see \cite{MKR}.

The ADHM construction for unitary groups can be generalized to
orthogonal and symplectic groups. It is quite remarkable how the
rather abstract ADHM construction can be made very intuitive using
D-branes and $\Omega$-planes \cite{Douglas'95} as we will see
later on.
\chapter{The $U(1)$ theory with adjoint matter}
\label{chapter2}
Recent progress in
understanding non-perturbative phenomena in supersymmetric
Yang-Mills (SYM) theories due to direct multi-instanton calculations is
quite impressive. Two main ideas played essential role in all this
developments. First was the realization that the SYM action induced to the moduli space of instantons can be
represented in terms of closed, equivariant with respect to the
diagonal part of the gauge group, forms \cite{FPS}. This
observation leads to a crucial simplifications reducing SYM path
integral to an integral over the stable, with respect to the action
of the diagonal part of the gauge group, subset of the moduli space
of instantons. The next brilliant idea, which is the corner stone
for all further developments, was suggested by Nekrasov in
\cite{Nek}. The idea is to generalize the theory involving into
the game in equal setting besides the already mentioned global
diagonal gauge transformations also the diagonal part of the
(Euclidean) space-time rotations. Why this is so crucial because
the subset of the instanton moduli space invariant under this
combined group action appears to consist only of finite number of
points.

In the case of the gauge group $U(N)$ this fixed point set is in
one to one correspondence with the set of array of Young diagrams
$\vec{Y}=(Y_1,...,Y_N)$ with the total number of boxes $|\vec{Y}|$
being equal to the instanton charge $k$. Thus, to calculate path
integral for the various `protected' by supersymmetry physical
quantities one needs to know only the pattern how the combined
group acts in the neighborhoods of the fixed point. All this
information can be encoded in the character of the group action in
the tangent space at given fixed points. An elegant formula for
this character which played a significant role in both physical
and mathematical applications was proposed in \cite{FP} (see eq.
(\ref{char})). Let us note at once, that combining space time
rotations with gauge transformations, besides giving huge
computational advantage due to the finiteness of the fixed point set,
has also a major physical significance generalizing the theory to
the case with certain nontrivial graviphoton backgrounds
\cite{Nek}. In order to recover the standard flat space quantities
(say the Seiberg-Witten prepotential of $\cN=2$
SYM theory) one should take the limit when the space
time rotation angles vanish. It is shown by Nekrasov and Okounkov
\cite{Nek_Okoun} that in this limit the sum over the arrays of
Young diagrams is dominated by a single array with specific
`limiting shape'. This enables one to handle in this limit the
entire instanton sum exressing all relevant quantities in terms of
emerging Seiberg-Witten curve \cite{Sei_Witt}. This is essential,
since only the entire sum and not its truncated part exhibits
remarkable modular properties, which allows one to investigate rich
phase structure of SYM theories. This is why all the attempts to
investigate the instanton sums also in general case seems quite
natural. Unfortunately, there was a little progress till now in this
direction besides the simplest case of the gauge group $U(1)$. Though
the $U(1)$ 4D theory in flat background is trivial, the general 5D
$U(1)$ theory compactified on a circle\footnote{Roughly speaking
the main technical difference between 4D and 5D cases is that in
the former case the above mentioned combined group enters into the
game in the infinitesimal level while in the latter case the main
role is played by finite group elements.} being rather nontrivial
nevertheless in many cases admits full solution.
We investigate the partition function of 5D gauge theory with an
extra adjoint hypermultiplet. It is not surprising that such
partition functions encode very rich topological information. As a
manifestation we argue that unlike the case with no extra matter,
at some special values of the parameters this partition function
directly reproduces the generating function of the Poincare
polynomial for the moduli space of instantons. We check this
conclusion explicitly computing the partition function in the case
of gauge group $U(1)$. We compare our result with that of recently
obtained by Iqbal et. al. \cite{Iqbal} who used the refined
topological vertex method \cite{Iqbal_Vafa} to find the same
partition function and present our comments on discrepancies we
found.

\setcounter{equation}{0} The weight decomposition of the torus
action on the tangent space at the fixed point
$\vec{Y}=(Y_1,\dots,Y_N)$ is given by \cite{FP} \bea \chi=
\sum_{\alpha ,\beta =1}^N e_\beta e_\alpha^{-1} \left\{\sum_{s\in
Y_\alpha}\left(T_1^{-l_\beta(s)}T_2^{a_\alpha(s)+1}\right)+\sum_{s\in
Y_\beta}\left(T_1^{l_\alpha(s)+1}T_2^{-a_\beta(s)}\right)\right\},\label{char}
\eea where $e_1,...,e_N$ are elements of (complexified) maximal
torus of the gauge group $U(N)$ and $T_1, T_2$ belong to the
maximal torus of the (Euclidean) space-time rotations, $a_\alpha
(s)$ ($l_\alpha (s)$) measures the distance from the location of
the box $s$ to the edge of the young diagram $Y_\alpha$ in the
vertical (horizontal) direction.

The 5D partition function can be read off from the above character 
\bea 
\cZ=
\sum_{\vec{Y}}\left\{ \frac{\mathbf{q}^{|\vec{Y}|}}{\prod_{\alpha ,\beta =1}^N \prod_{s\in
Y_\alpha}\left(1-e_\beta e_\alpha^{-1}
T_1^{-l_\beta(s)}T_2^{a_\alpha(s)+1}\right)}\times \right. \nn\\
\left. \frac{1}{\prod_{\alpha ,\beta =1}^N\prod_{s\in
Y_\beta}\left(1-e_\beta e_\alpha^{-1}T_1^{l_\alpha(s)+1}T_2^{-a_\beta(s)}\right)}\right\}
\eea 
From the mathematical point of view this quantity could be
regarded as the character of the torus action on the space of
holomorphic functions of the moduli space of instantons. The
Nekrasov's partition function for 4D theory could be obtained
tuning the parameters $\mathbf{q}\rightarrow \beta^{2N}
\mathbf{q}$, $T_1\rightarrow \exp-\beta \epsilon_1 $,
$T_2\rightarrow \exp-\beta \epsilon_2 $, $e_\alpha \rightarrow
-\beta v_\alpha$ and tending $\beta \rightarrow 0$, where
$v_1,...,v_N$ are the expectation values of the chiral superfield
and $\epsilon_1$, $\epsilon_2$ characterize the strength of the
graviphoton background (sometimes called $\Omega$-background).

Fortunately, instanton counting is powerful enough to handle also
the cases when an extra hypermultipet in adjoint or several
fundamental hypermultiplets are present. In the case with adjoint
hypermultiplet instead of (\ref{char}) one starts with the (super)
character \cite{fucito} \bea &&\chi = (1-T_m)\times\nn\\
&& \sum_{\alpha ,\beta
=1}^N e_\beta e_\alpha^{-1} \left\{\sum_{s\in
Y_\alpha}\left(T_1^{-l_\beta(s)}T_2^{a_\alpha(s)+1}\right)+\sum_{s\in
Y_\beta}\left(T_1^{l_\alpha(s)+1}T_2^{-a_\beta(s)}\right)\right\}.\label{char_adj}
\eea One way to interpret this character is to imagine that each
(complex) 1d eigenspace of the torus action is complemented by a
grassmanian eigenspace with exactly the same eigenvalues of the
torus action. In addition an extra $U(1)$ action is introduced so
that $T_m\in U(1)$ acts trivially on bosonic directions while
acting on each grassmanian coordinate in its fundamental
representation. Then (\ref{char_adj}) is the super-trace of the
extended torus action on the super-tangent space at given fixed
point. The corresponding 5D partition function now reads: \bea
\cZ_{adj}= &&\sum_{\vec{Y}}\mathbf{q}^{|Y|}\prod_{\alpha ,\beta =1}^N
\prod_{s\in Y_\alpha}\frac{\left(1-T_m e_\beta e_\alpha^{-1}
T_1^{-l_\beta(s)}T_2^{a_\alpha(s)+1}\right)}{\left(1-e_\beta
e_\alpha^{-1}
T_1^{-l_\beta(s)}T_2^{a_\alpha(s)+1}\right)}\nn\\
&&\prod_{s\in
Y_\beta}\frac{\left(1-T_m e_\beta
e_\alpha^{-1}T_1^{l_\alpha(s)+1}T_2^{-a_\beta(s)}\right)}{\left(1-e_\beta
e_\alpha^{-1}T_1^{l_\alpha(s)+1}T_2^{-a_\beta(s)}\right)}\nonumber
\\ \label{Zadj}\eea
Each term here could be thought of as trace over the space of local
holomorphic forms, with parameter $T_m$ counting the degrees of
forms. Hence the sum over the fixed points is expected to give the
super-trace over the globally defined holomorphic forms. We see
that $\cZ_{adj}$ is an extremely rich quantity from both physical
and mathematical point of view. It is interesting to note, that at the
special values of the parameters, $\cZ_{adj}$ directly reproduces the
generating function for the Poincare polynomial of the moduli
space of $U(N)$ instantons. Indeed, following \cite{Nak_lect} part
(3.3) let us assume that $T_2\gg T_{a_1}>\cdots >T_{a_N}\gg
T_1>0$. It is easy to see, that in the limit when all these
parameters go to zero, each fraction under the products in
(\ref{char_adj}) tends to $T_m$ or $1$ depending whether we have a
negative weight direction or not (see the classification of
negative directions in \cite{Nak_lect}, proof of corollary 3.10).
We will see this explicitly in the simplest case $N=1$ when the
moduli space of instantons coincides with the Hilbert scheme of
points on $\mathbb{C}^2$.

From now on we will restrict ourselves to the simplest case of
$U(1)$ gauge group, when the partition function could be computed
in a closed way. The partition function of the pure ${\cal N}=2$,
$U(1)$ theory has the form \cite{Nak_blowup} \bea
\cZ=\sum_{Y}\frac{\mathbf{q}^{|Y|}}{\prod_{s\in Y}\left(1-
T_1^{-l(s)}T_2^{a(s)+1}\right)\left(1-T_1^{l(s)+1}T_2^{-a(s)}\right)}
\nonumber \\=\exp \left(\sum_{n=1}^\infty \frac{\mathbf{q}^n}{n
(1-T_1^n)(1-T_2^n)}\right). \eea This remarkable combinatorial
identity in the 4D limit and in `self dual' case
$\epsilon_1=-\epsilon_2$ boils down to the Burnside's theorem \bea
\sum_{|\lambda|=n}(dim \,\mathbf{R}_\lambda )^2=n!, \eea where
$\mathbf{R}_\lambda$ is the irreducible representation of the
symmetric group given by the Young diagram $\lambda$.

Now let us turn to the $U(1)$ theory with adjoint matter. Doing
low instanton calculations using (\ref{Zadj}) is straightforward
and gives \bea \log \cZ_{adj}=\frac{\mathbf{q}(1+T_m
\mathbf{q}+T_m^2 \mathbf{q}^2+T_m^3 \mathbf{q}^3)
(1-T_m T_1)(1-T_m T_2)}{(1- T_1)(1-T_2)}+\nonumber \\
\frac{\mathbf{q}^2(1+T_m^2\mathbf{q}^2) (1-T_m^2 T_1^2)(1-T_m^2
T_2^2)}{2(1- T_1^2)(1-T_2^2)}+\frac{\mathbf{q}^3(1-T_m^3
T_1^3)(1-T_m^3 T_2^3)}{3(1- T_1^3)(1-T_2^3)}\nonumber
\\+\frac{\mathbf{q}^4 (1-T_m^4 T_1^4)(1-T_m^4 T_2^4)}{4(1-
T_1^4)(1-T_2^4)}+\textit{O}(\mathbf{q}^4). \eea These drove us to
the conjecture that the exact formula is \bea \log
\cZ_{adj}=\sum_{n=1}^\infty\frac{\mathbf{q}^n (1-T_m^n
T_1^n)(1-T_m^n T_2^n)}{n(1-
T_1^n)(1-T_2^n)(1-T_m^n\mathbf{q}^n)},\label{log_Z_adj} \eea which
is equivalent to the following highly nontrivial combinatorial
identity \bea \cZ_{adj}=\sum_{Y}\mathbf{q}^{|Y|}\prod_{s\in
Y}\frac{\left(1-
T_mT_1^{-l(s)}T_2^{a(s)+1}\right)\left(1-T_mT_1^{l(s)+1}T_2^{-a(s)}\right)}{
\left(1-T_1^{-l(s)}T_2^{a(s)+1}\right)\left(1-T_1^{l(s)+1}T_2^{-a(s)}\right)}
\\
=\exp \left(\sum_{n=1}^\infty
\frac{\mathbf{q}^n(1-(T_mT_1)^n)(1-(T_mT_2)^n)}{n
(1-T_1^n)(1-T_2^n)(1-(T_m\mathbf{q})^n)}\right). \eea

Indeed calculations with Mathematica code up to 10
instantons further convinced us that this formula is indeed
correct. Note that the 4D limit of this identity with a particular
choice of graviphoton background $\epsilon_1=-\epsilon_2$ is
mentioned earlier in \cite{Nek_Okoun} and was used later in
\cite{FMPT} to calculate the expectation value $tr \langle
\phi^2\rangle$.

As a further check let us go to the limit when $T_1\rightarrow 0$,
$T_2\rightarrow 0$. As we have explained above one expects to find
the generating function of Poincare polynomial for Hilbert scheme
of points on $\mathbb{C}^2$. An easy calculation yields: \bea
\cZ_{adj}|_{T_1,T_2=0}=\exp \sum_{n=1}^\infty
\frac{\mathbf{q}^n}{n(1-T^n_m
\mathbf{q}^n)}=\nonumber \\
\exp\sum_{n=1}^\infty \sum_{k=0}^\infty
\frac{(\mathbf{q}^{1+k}T_m^k)^n}{n}=
\prod_{k=0}^\infty\frac{1}{1-T_m^k\mathbf{q}^{k+1}},\label{Poincare}
\eea which indeed after identifying $T_m$ with Poincare parameter
$t^2$ reproduces the well known result (see e.g. \cite{Nak_book}).
Now let us go back to the general case. In various domains of the
variables $T_1$, $T_2$ we can represent (\ref{log_Z_adj}) as
infinite product as in (\ref{Poincare}). Let us consider
separately
the cases: \\
(a) $|T_1|<1$, $|T_2|<1$, $|T_m \mathbf{q}|<1$\\
In this region (\ref{log_Z_adj}) could be rewritten as \bea
\cZ_{adj}=\exp\left\{\sum_{n=1}^\infty \sum_{k,i,j=0}^\infty
\frac{\mathbf{q}^n}{n}T_1^{ni}T_2^{nj}(T_m
\mathbf{q})^{nk}(1-T_m^n T_1^n)(1-T_m^n T_2^n)\right\}. \eea
Performing summation over $n$ we get \bea
\cZ_{adj}=\prod_{i,j,k=0}^\infty
\frac{(1-\mathbf{q}^{k+1}T_m^{k+1}T_1^{i+1}T_2^{j})
(1-\mathbf{q}^{k+1}T_m^{k+1}T_1^{i}T_2^{j+1})}{(1-\mathbf{q}^{k+1}T_m^{k}T_1^{i}T_2^{j})
(1-\mathbf{q}^{k+1}T_m^{k+2}T_1^{i+1}T_2^{j+1})}.
\label{Z_prod_a}\eea
(b) $|T_1|>1$, $|T_2|<1$, $|T_m \mathbf{q}|<1$\\
In this region we expand (\ref{log_Z_adj}) over $1/T_1$: \bea
\cZ_{adj}=\exp\left\{\sum_{n=1}^\infty \sum_{k,i,j=0}^\infty
\frac{-\mathbf{q}^n}{n}T_1^{-ni}T_2^{nj}(T_m
\mathbf{q})^{nk}(1-T_m^n T_1^n)(1-T_m^n T_2^n)T_1^{-n}\right\},\nn\\
\eea which leads to \bea \cZ_{adj}=\prod_{i,j,k=0}^\infty
\frac{(1-\mathbf{q}^{k+1}T_m^{k}T_1^{-i-1}T_2^{j})
(1-\mathbf{q}^{k+1}T_m^{k+2}T_1^{-i}T_2^{j+1})}{(1-\mathbf{q}^{k+1}T_m^{k+1}T_1^{-i}T_2^{j})
(1-\mathbf{q}^{k+1}T_m^{k+1}T_1^{-i-1}T_2^{j+1})}.\label{Z_prod_b}\eea

Recently Iqbal, Kozcaz and Shabir \cite{Iqbal} have computed
the partition function of these $U(1)$ adjoint theory using the
refined topological vertex formalizm \cite{Iqbal_Vafa}. And, since
the formula (\ref{log_Z_adj}) was known to the present authors for
quite a while, we performed a detailed comparison with their
results. To make contact with the formulae of Iqbal et. al. we
need the following dictionary: $T_m=Q_m(t/q)^{1/2}$, $T_1=1/t$,
$T_2=q$, $\mathbf{q}=Q(q/t)^{1/2}$. In terms of these variables
the
equations (\ref{Z_prod_a}) and (\ref{Z_prod_b}) take the form: \\
(a) $|t|>1$, $|q|<1$, $|QQ_m|<1$\\
\bea \cZ_{adj}=\prod_{i,j,k=1}^\infty
\frac{(1-Q^{k}Q_m^{k}q^{i-1}t^{-j})
(1-Q^{k}Q_m^{k}q^{i}t^{1-j})}{(1-Q^{k}Q_m^{k+1}
q^{i-\frac{1}{2}}t^{-j+\frac{1}{2}})(1-Q^{k}Q_m^{k-1}q^{i-\frac{1}{2}}t^{-j+\frac{1}{2}})},
\eea and \\
(b) $|t|<1$, $|q|<1$, $|QQ_m|<1$\\
\bea \cZ_{adj}=\prod_{i,j,k=1}^\infty
\frac{(1-Q^{k}Q_m^{k+1}q^{i-\frac{1}{2}}t^{j-\frac{1}{2}})
(1-Q^{k}Q_m^{k-1}q^{i-\frac{1}{2}}t^{j-\frac{1}{2}})}{(1-Q^{k}Q_m^{k}q^{i-1}t^{j-1})
(1-Q^{k}Q_m^{k}q^{i}t^{j})}.\eea These equations come rather
close, but certainly do not coincide with those given in
\cite{Iqbal} at the end of the part 3.2. The reason for this
discrepancy seems to us as follows. According to \cite{Iqbal} the
refined topological vertex method for the 5D $U(1)$ theory with
adjoint matter leads to (see eq. (4.6) of \cite{Iqbal}; below we
omit the `perturbative part' $\prod_{i',j'=1}^\infty
(1-Q_mq^{-\rho_{i'}}t^{-\rho_{j'}})$ ) \bea \cZ=\prod_{k=1}^\infty
(1-Q^kQ_m^k)^{-1} \prod_{i,j=1}^\infty
(1-Q^kQ_m^{k-1}q^{-\rho_i}t^{-\rho_j})\nonumber \\
(1-Q^kQ_m^{k}q^{\rho_i-1/2}t^{-\rho_j+1/2})
(1-Q^kQ_m^{k}q^{-\rho_i+1/2}t^{\rho_j-1/2})
(1-Q^kQ_m^{k+1}q^{\rho_i}t^{\rho_j}), \nn\\ \label{Z_RTV}\eea where
$\rho_i=-i+1/2$. But four factors under the product over $i,j$
have different, excluding each other regions of convergence. Thus
this infinite product should be treated very carefully.
Unfortunately the authors of \cite{Iqbal} do not tell what
analytic continuation procedure they have adopted to pass from
their eq. (4.6) to those presented at the end of the part 3.2, but
we will demonstrate now that one, perhaps the simplest approach
directly leads to our conjectural formula (\ref{log_Z_adj}). We
simply examine the product over each factor separately within its
region of convergence and only after that continue analytically to
a common region of the parameters. Thus for the first factor in
(\ref{Z_RTV}) we have \bea \prod_{k=1}^\infty
(1-Q^kQ_m^k)^{-1}=\exp \sum_{n,k=1}^\infty
\frac{(QQ_m)^{nk}}{n}=\exp \sum_{n=1}^\infty
\frac{(QQ_m)^{n}}{n(1-(QQ_m)^n)}. \eea For the next factor
(assuming $q<1$, $t<1$) \bea \prod_{k,i,j=1}^\infty
(1-Q^kQ_m^{k-1}q^{i-\frac{1}{2}}t^{j-\frac{1}{2}})=\exp
\sum_{n,k,i,j=1}^\infty
\frac{-Q^{kn}Q_m^{(k-1)n}q^{(i-\frac{1}{2})n}t^{(j-\frac{1}{2})n}}{n}
\nonumber \\
=\exp \sum_{n=1}^\infty
\frac{-Q^nq^{\frac{n}{2}}t^{\frac{n}{2}}}{(1-(QQ_m)^n)
(1-q^n)(1-t^n)}.\eea Similarly for $q>1$, $t<1$ \bea
\prod_{k,i,j=1}^\infty (1-Q^kQ_m^{k}q^{-i}t^{j})=\exp
\sum_{n=1}^\infty \frac{-Q^nQ_m^{n}q^{-n}t^{n}}{n(1-(QQ_m)^n)
(1-q^{-n})(1-t^n)}, \eea for $q<1$, $t>1$ \bea
\prod_{k,i,j=1}^\infty (1-Q^kQ_m^{k}q^{i}t^{-j})=\exp
\sum_{n=1}^\infty \frac{-Q^nQ_m^{n}q^{n}t^{-n}}{n(1-(QQ_m)^n)
(1-q^{n})(1-t^{-n})}, \eea and, finally for $q>1$, $t>1$ \bea
\prod_{k,i,j=1}^\infty
(1-Q^kQ_m^{k+1}q^{-i+\frac{1}{2}}t^{-j+\frac{1}{2}})=\exp
\sum_{n=1}^\infty
\frac{-Q^nQ_m^{2n}q^{-\frac{n}{2}}t^{-\frac{n}{2}}}{n(1-(QQ_m)^n)
(1-q^{-n})(1-t^{-n})}.\nn\\ \eea Note that the r.h.s.'s of above
expressions are defined also outside of their initial convergence
region. Combining all these together we get \bea  \cZ =\exp
\sum_{n=1}^\infty
\frac{(QQ_m)^n(q^{\frac{n}{2}}t^{\frac{n}{2}}-Q_m^n)
(q^{\frac{n}{2}}t^{\frac{n}{2}}-Q_m^{-n})}{n
(1-(QQ_m)^n)(1-q^n)(1-t^n)}, \eea which in terms of the parameters
$\mathbf{q}$, $T_1$, $T_2$ exactly coincides with our conjectural
result (\ref{log_Z_adj}).

\chapter{Instantons in String Theory}
\label{chapter3}
\section{Worldsheet instantons}

World-sheet instantons in Heterotic and Type II theories
correspond to Euclidean fundamental string world-sheets wrapping
topologically non-trivial internal cycles of the compactification
space and produce effects that scale as $e^{-R^2/\ap}$
\cite{Dine}. Depending on the number of supersymmetries (thus on
the number of fermionic zero modes), they can correct the
two-derivative effective action or they can contribute to
threshold corrections to higher derivative (BPS saturated)
couplings \cite{DualitInst}. For Type II compactifications on CY
three-folds, preserving $\cN=2$ supersymmetry in $D=4$,
holomorphic worldsheet instantons ($\bar\partial X = 0$) correct
the special K\"ahler geometry of vector multiplets (Type IIA) or
the dual quaternionic geometry of hypermultiplets (Type IIB). For
heterotic compactifications with standard embedding of the
holonomy group in the gauge group, complex structure deformations
are governed by the same special K\"ahler geometry as in Type IIB
on the same CY three-fold, that is not corrected by worldsheet
instantons. Complexified K\"ahler deformations are governed by the
same special K\"ahler geometry as in Type IIA on the same CY
three-fold, that is corrected by worldsheet instantons, or
equivalently, as a result of mirror symmetry, by the same special
K\"ahler geometry as in Type IIB on the mirror CY three-fold that
is tree level exact. For standard embedding, the K\"ahler metrics
of charged supermultiplets in the $27$ and $27^*$ representations
of the surviving/visible $E_6$ are simply determined by the ones
of the neutral moduli of the same kind by a rescaling \cite{25}.
For non standard embeddings the situation is much subtler.

\subsubsection{Brane instantons} Euclidean NS5-branes (EN5-branes)
wrapping the 6-dimensional compactification manifold produce
non-perturbative effects in $e^{-c/g_s^2}$ (reflecting the
NS5-brane tension) that qualitatively correspond to `standard'
gauge and gravitational instantons. Euclidean Dp-brane wrapping
$(p+1)$-cycles produce instanton effects that scale as
$e^{-c_p/g_s}$  (reflecting the EDp-brane tension)\cite{Becker}.
In Type IIB on CY three-fold, ED(-1), ED1-, ED3- and ED5-brane
instantons, obtained by wrapping holomorphic submanifolds, correct
dual quaternionic geometry in combination with world-sheet (EF1)
and EN5-instantons. In Type IIA on CY three-folds, ED2-instantons
(D-`membrane' instantons) wrapping special Lagrangian
submanifolds, correct the dual quaternionic geometry, in
combination with EN5-instantons. In both cases, the dilaton
belongs to the universal hypermultiplet.

\subsubsection{Unoriented D-brane instantons} In Type I, the
presence of $\Omega$9-planes  severely restricts the possible
homologically non trivial instanton configurations. Only ED1- and
ED5-branes are homologically stable. Other (Euclidean) branes may
be associated to instanton with torsion (K-theory) charges. For
other un-oriented strings the situation is similar and can be
deduced by means of T-duality: e.g. for intersecting D6-branes one
has two different kinds of ED2-branes (ED0- and/or ED4-brane
instanton require $b_{1,5}\neq 0$), for intersecting D3- and
D7-branes one has ED(-1) and ED3-branes. There are two classes of
unoriented D-brane instantons depending on the stack of branes
under consideration.
\begin{itemize}
\item{ `Gauge' instantons correspond to EDp-branes wrapping
the same cycle $\cal C$ as a stack of background D(p+4)-branes. The
prototype is the D3, D(-1) system \cite{Billo}, \cite{Billo.Inst.CalcinRR} that has 4 N-D
directions. The EDp-branes behave as instantons inside D(p+4)'s:
\be F=\tilde{F}\ee
and produce effects whose strength, given by
\be e^{-W_{p+1}(C)/g_s \ell_s^{p+1}}=e^{-1/g_{YM}^2},\ee
is precisely the one expected from `gauge' instantons in the
effective field-theory.}

\item{ `Exotic' instantons arise from EDp'-branes wrap a cycle
${\cal C}'$ which is not wrapped by any stack of background D(p+4)-branes. The prototype is
the D9, ED1 system with 8 N-D directions and only a chiral fermion
at the intersection. In this case
\be F\neq \tilde{F}\ee
and the strength is given by
\be e^{-W_{p'+1}(C')/g_s \ell_s^{p'+1}} \neq e^{-1/g_{YM}^2}\ee
`Exotic' instantons may eventually enjoy a field theory
description in terms of octonionic instantons or hyper-instantons
with $F\wedge F = *_8 F\wedge F$.}
\end{itemize}

\section{Original Applications and Various Comments}

Let us now list possible effects generated by (un)oriented D-brane
instantons in diverse string compactifications.

\begin{itemize}

\item{In $\cN =8$ theories (\eg toroidal compactifications of
oriented Type II A/B) D-brane instantons produce threshold corrections to $R^4$
terms and other 1/2 BPS (higher derivative) terms}.

\item{In $\cN =4$ theories (\eg toroidal compactifications of Type I
/ Heterotic) D-brane instantons produce threshold corrections to $F^4$ terms
and other 1/2 BPS (higher derivative) terms}.

\item{In $\cN =2$ theories (\eg toroidal orbifolds with
$\Gamma\subset SU(2)$) D-brane instantons produce threshold corrections to
$F^2$ terms and other 1/2 BPS terms}.

\item{In $\cN =1$ theories (\eg toroidal orbifolds with
$\Gamma\subset SU(3)$) D-brane instantons produce threshold corrections and
superpotential terms}.

\end{itemize}

\subsubsection{Thresholds in toroidal compactifications}

We have not much to add to the vast literature on threshold
corrections to $R^4$ terms in $\cN =8$ theories which are induced
by oriented D-brane as well as world-sheet instantons\footnote{In
$D=4$ and lower Euclidean NS5-branes can also contribute.}. We
would only like to argue that in unoriented Type I strings and
alike these corrections should survive as functions of the
unprojected closed string moduli despite some of the corresponding
D-brane or worldsheet instantons be not BPS. These and lower
derivative ($R^2$) couplings may receive further perturbative
corrections from surfaces with boundaries and crosscaps. Viz: ${\cal L}_{II}\approx {\calR}^4f_{II}(\phi ,
\chi)\rightarrow {\cal L}_I\approx {\calR}^4[f_{II}(\phi , \chi=0)+f_I(\phi)]$.

The original application of unoriented D-brane instanton was in
the context of threshold corrections to $F^4$ terms in
toroidal compactifications of Type I strings \cite{BKB}. These are closely related to the threshold corrections to $F^4$
terms for heterotic strings on $T^d$. For later use, let us briefly summarize the structure of the latter.
After
\begin{itemize}
\item{Computing the one-loop correlation function of 4 gauge boson vertex operators $V_{(0)}=A_\mu ^a(\partial
X^\mu+ip\psi\psi^\mu)\tilde{J}_ae^{ipx}$}
\item{Taking the limit of zero momentum in the exponential factors \ie neglecting
the factor $\Pi(z_i,p_i) = \prod_{i,j} exp[-\ap p_i\cdot p_j \cG(z_{ij})] \rightarrow 1$}
\end{itemize}
or, equivalently,
\begin{itemize}
\item{computing the character-valued partition function in a
constant field-strength background $\nu$} \item{taking the fourth
derivative wrt $\nu$}
\end{itemize}
one arrives at the integral over the one-loop moduli space that
receives contribution only from BPS states and schematically reads
\bea \cI_{d} [\Phi]= \cV_d \int_{\cF} {d^2\tau \over \tau_2^2} \sum_M
e^{2\pi i \cT(M)} e^{-{\pi Im\cT(M)\over \tau_2 Im\cU(M)}|\tau -
\cU(M)|^2} \Phi (\tau)\eea
where $M=(\vec{n}, \vec{m})$ represent the embedding of the world-sheet torus in the target $T^d$, $\Phi(\tau)$ is some
modular form. The induced Kahler $\cT(M)$ and complex $\cU(M)$ structures are given by
\bea \cT(M) = \cB_{12} + i \sqrt{\det\cG} \quad, \quad \cU(M) = {1\over \cG_{11}} (\cG_{12} + i \sqrt{\det\cG}) \eea
with $ \cG = M^t G M $, $\cB = M^t B M $ induced metric and
$B$-field \cite{Kiritsis}. The integral can be decomposed into
three terms $\cI_{d}[\Phi]= \cI_{d}^{triv}[\Phi] +
\cI_{d}^{deg}[\Phi] + \cI_{d}^{ndeg}[\Phi]$.  The three different
orbits are classified as follows: the orbit of $M=0$ (trivial
orbit), degenerate orbits with $\det(M^{i,j})=0$ and
non-degenerate orbits with some $\det(M^{i,j})\neq 0$. Let us
consider the various contributions.
\begin{itemize}
\item {\bf Trivial orbit}: $M=0$,
\bea \cI_{d,d}^{triv}[\Phi] =
\int_{\cF}{d^2\tau \over \tau_2^2} \Phi(\tau) \rightarrow
\cI_{d,d}^{triv}[1] = {\pi^2 \over 3} \cV_d \eea

\item {\bf Degenerate orbits}: $M\neq 0$, $\det(M^{i,j}) = n^i m^j
- n^j m^i =0$ $\forall i,j$. One can choose $\vec{n}=0$
representative and unfold $\cF$ to the strip $\cS = \{
|\tau_1|<1/2, \tau_2>0\}$, then

\bea \cI_{d,d}^{deg}[\Phi] = \cV_d \int_{\cS}{d^2\tau \over
\tau_2^2}\sum_{\vec{m} \neq \vec{0}} e^{-{\pi\over \tau_2}
\vec{m}^t G \vec{m}} \Phi(\tau) \rightarrow \cI_{d}^{deg}[1]= \cV_d
\cE^{SL(d)}_{\bf d}(G).\nn\\ \eea

\item {\bf Non degenerate orbits}: at least one $\det(M_{ij}) =
n^i m^j - n^j m^i \neq 0$. The representative for these orbits may
be chosen to be $\vec{n}^\alpha=0$ for $\alpha = 1, .., k$,
$m^\alpha \neq 0$, $n^{\bar\alpha}> m^{\bar\alpha}\ge 0$ and
enlarging the region of integration $\cF$ to the full upper half
plane $\cH^+$ one finds:
\bea \cI_{d,d}^{ndeg}[\Phi] = \cV_d
\int_{\cH^+} {d^2\tau \over \tau_2^2}
\sum_{(n^{\bar\a},0;m^{\bar\a},m^{\a})} e^{2\pi i \cT(M)} e^{-{\pi
Im\cT(M)\over \tau_2 Im\cU(M)}|\tau - \cU(M)|^2} \Phi(\tau)\nn\\
\rightarrow \cI_{d,d}^{deg}[1]= \cV_d \cE^{SO(d,d)}_{{\bf
V}, s=1}(G,B) \quad { \rm {(generalized \, Eisenstein \, series)}}.\nn\\ \eea

\end{itemize}

Thanks to Type I / Heterotic duality, heterotic worldsheet
instantons are mapped into ED-string instantons. Since $F^4$ terms
are 1/2 BPS saturated, matching the spectrum of excitations,
including their charges, was believed to be sufficient to match
the threshold corrections even in the presence of (non)commuting
Wilson lines \cite{BKB,MBEGJFMKN} or after T-duality
\cite{Gutperetal, KFSS}. More recently, thanks to powerful localization
techniques, a perfect match between threshold corrections in
Heterotic and Type I (with D7-branes) has been found on $T^2$
for the specific choice of commuting Wilson lines breaking
$SO(32)$ to $SO(8)^4$ \cite{Lerda_etal}. The somewhat
unsatisfactory results of \cite{Fucito:2009rs} for different breaking
patterns with orthogonal or symplectic groups can be either
interpreted as a failure of localization or as the need to include
higher order terms. Notice that only for $SO(8)$, `exotic'
string instantons should admit a field theory interpretation in
terms of `octonionic' instantons. It would be nice to further
explore this issue in this or closely related context of $\cN=1,
2$ theories in D=4 where heterotic worldsheet instantons
correcting the gauge kinetic function should be dual to ED-string
(or other ED-brane) instantons \cite{ABDFPT}. A short review of the strategy to
compute similar threshold corrections will be presented later
on when we discuss Heterotic / Type I duality on $T^4/\mZ_2$.

\subsubsection{Phenomenological considerations}

Despite some success in embedding (MS)SM in vacuum configurations
with open and unoriented strings, there are few hampering
properties at the perturbative level:

\begin{itemize}
\item{Forbidden Yukawas in $U(5)$ (susy) GUT's
$$H^d_{{\bf 5}^*_{-1}}F^c_{{\bf 5}^*_{-1}}A_{{\bf 10}_{+2}} \quad
OK \quad {\bf but} \quad H^u_{{\bf 5}_{+1}} A_{{\bf 10}_{+2}}
A_{{\bf 10}_{+2}}  \quad KO $$ forbidden by (global, anomalous)
$U(1)$ invariance, though compatible with $SU(5)$ (yet no way
$\eps^{abcde}$ from Chan-Paton)}

\item{R-handed (s)neutrino masses $W_M=M_R NN$ forbidden by \eg
$U(1)_{B-L}$ in Pati-Salam like models $SO(6)\times SO(4)
\rightarrow SU(3)\times SU(2)_L\times SU(2)_R \times U(1)_{B-L}$}

\item{$\mu$-term in MSSM $W_{\mu}=\mu H_1 H_2$ typically forbidden
by extra (anomalous) $U(1)$'s}
\end{itemize}

All the above couplings can be generated by `stringy' instantons
after integrating over the `non-dynamical' moduli living on the
world-volume of the EDp'-branes under consideration. These effects
scale as $e^{-T_{EDp'} V_{EDp'}}$ and are non-perturbative in
$g_s$, since $T_{EDp'}\approx 1/g_s (\ap)^{p+1/2}$. Yet {\it a
priori} they depend on different moduli (through the dependence of
$V_{EDp'}$ on variuos $Z$'s) from the ones appearing in the gauge
kinetic function(s) of background Dp'-brane, so they cannot in
general be identified with the standard `gauge' instantons.
Relying on the $g_s$ power counting introduced in \cite{Billo},\cite{Billo.Inst.CalcinRR}
the relevant are disks with insertions of the non-dynamical vertex
operators $V_\Theta$ (connecting EDp'-EDp') and $V_\lambda$
(connecting EDp-Dp') with or without insertions of dynamical
vertex operators $V_A$ etc, which correspond to the massless
excitations of the vacuum configuration of
(intersecting/magnetized) unoriented Dp-branes \cite{Blum}, \cite{Blumik}. Disks
without dynamical insertions yield the `instanton action', with
one dynamical vertex they produce classical profiles for $A$ etc.
Disks with more insertions contribute to higher-order corrections.
One loop diagrams with no insertions produce running couplings and
subtle numerical prefactor that can cancel a given type of
non-perturbative F-terms \cite{Silverstein, BWnieist}.

\subsubsection{Anomalous $U(1)$'s and gauged PQ symmetries}

In general, a `naked' chiral field $Z$  whose pseudoscalar axionic
components $\zeta=\im Z$ shift under some local anomalous $U(1)$
cannot appear in a (super)potential term if not dressed with other
chiral fields charged under $U(1)$. $U(1)$ invariance puts tight
constraints on the form of the possible superpotential terms.
Since the axionic shift is gauged it must be a symmetry of the
kinetic term. This is only possible when no non-perturbative
(world-sheet or D-brane instanton) corrections spoil the tree
level (in fact perturbative) PQ symmetry. This means that the
gauging procedure corresponds to turning on fluxes such that the
potential instanton corrections in $Z$ are in fact disallowed. In
practice, this means the corresponding wrapped brane is either
anomalous (\`a la Freed-Witten) \cite{FW} or destabilized due to
the flux \cite{KPT}.

Moreover, background fluxes (for both open and closed strings) can
lift fermionic zero-modes. Various `perturbative' studies have
been carried out \cite{Billo':2008sp}, \cite{Billo':2008pg}.

\section{ADHM from branes within branes}

As already mentioned, the ADHM construction has a rather intuitive
description in open string theory, whereby the gauge theory is
realized on a stack of Dp-branes. D(p-4)-branes which are
localized within the previous stack of branes behave as instantons
\cite{Douglas'95}.

Indeed, the WZ couplings on the Dp-brane worldvolume schematically
reads \be S_{WZ} = \int C_{p+1} + \int C_{p-1} \wedge Tr(F) + \int
C_{p-3} \wedge Tr(F\wedge F) + ... \ee

In particular a localized source of $C_{p-3}$ within a Dp-brane
behaves like an instanton density $Tr(F\wedge F)$. Moreover, the
ADHM data are nothing but the massless modes of open strings
connecting the D(p-4)-branes with one another or with the
background Dp-branes.

Let us take $p=3$ for definiteness. The low-energy effective
theory on the world-volume of $N$ parallel D3-branes is $\cN=4$
supersymmetric Yang-Mills theory with gauge group $U(N)$.
Instanton moduli are described by the massless modes of open
strings with at least one end on the D(-1)-brane stack.  In this
system of D3 and D(-1) branes there are three sectors of the open
string spectrum to be considered. $U(N)$ gauge fields and their
superpartners are provided by the strings that start and end on
D3-branes. Strings stretching between two D(-1)-branes give rise
to $U(K)$ non-dynamical gauge fields and their superpartners.
These `fields' represent part of the (super) ADHM data. The
remaining (super) ADHM data are provided by the strings with one
end on the D3-branes and the other one on the D(-1)-branes and
{\it vice-versa}.

In the presence of D3-branes, (Euclidean) Lorentz symmetry is
broken $SO(10)\rightarrow SO(4)\times SO(6)$ and it is convenient
to split ten `gauge bosons'', $A_M$, into four gauge bosons
$a_\mu$, and six real `scalars', $\chi_i$. Similarly the $d=10$
gauginos produce four non-dynamical Weyl `gauginos'',
$\Theta_\alpha^A$ as well as their antiparticles
$\bar{\Theta}_A^{\dot{\alpha}}$.

Introducing, for later convenience, three auxiliary fields $D^c$,
the D(-1)-D(-1) $U(K)$ `geometric' supermoduli are given by
$$a_\mu,\quad \chi_i;\quad
\Theta_\alpha^A,\quad \bar{\Theta}_A^{\dot{\alpha}};\quad D^c$$
while the $4KN$ D(-1)-D(3) `gauge' supermoduli are
$$w_{\dot{\alpha}i}^u,\quad \bar{w}_{\dot{\alpha}u}^i;\quad
\nu_{i}^{Au},\quad \bar{\nu}_{u}^{Ai}$$ with $\mu=1,...,4,\quad
\alpha,\dot{\alpha}=1,2$ vector and spinor indices of $SO(4)$,
$i=1,...,6,\quad A=1,...,4$ are vector and spinor indices of
$SO(6)$ respectively and $c=1,2,3$. The matrices $a_\mu,\, \chi_i$
describe the position of the instanton along the longitudinal and
transverse directions to the D3-brane respectively.
$\Theta_\alpha^A$ and $\bar{\Theta}_A^{\dot{\alpha}}$ are their
superpartners. $w_{\dot{\alpha}}$, $\bar{w}_{\dot{\alpha}}$
represent the D3-D(-1) open string in the $NS$ sector, accounting
for instanton sizes and orientations, and $\nu^A$, $\bar{\nu}^A$
are their fermionic superpartners.

\section{The D3-D(-1) action}

By computing scattering amplitudes on the disk, one can determine
the complete action that governs the dynamics of the light modes
(or moduli) of the system of D(-1) branes in the presence of
D3-branes. It schematically reads \cite{Dorey}
\be S_{K,N}={Tr}_{k}\left[{1\over g_0^{2}}S_{G}+S_{K}+S_{D}\right]\ee
with
\be S_{G}=-[\chi_i,\chi_j]^2+ i\bar{\Theta}_{\dot{\alpha}
A}[\chi_{AB}^\dagger,\bar{\Theta}^{\dot{\alpha}}_B]-D^{c}D^{c}\ee
\be S_{K}=-[\chi_i,a_{\mu}]^2+\chi^i\bar{w}^{\adot}
w_{\adot}\chi_i-i\Theta^{\alpha A}[\chi_{AB},
\Theta^{B}_{\alpha}]+2i\chi_{AB}\bar{\nu}^{A}\nu^{B}\ee
\be S_{D}=i\left(-[a_{\alpha\dot{\alpha}},\Theta^{\alpha A}]
+\bar{\nu}^A w_{\dot{\alpha}} +\bar{w}_{\dot{\alpha}}\nu^A \right)
\bar{\Theta}^{\dot{\alpha}}_A+D^{c}\left(\bar{w}\sigma^c w-i
\bar{\eta}_{\mu\nu}^c [a^\mu,a^\nu] \right)\ee where
$\chi_{AB}\equiv{1\over 2}\Sigma^i_{AB}\chi_i$ and
$\Sigma^i_{AB}=(\eta^c_{AB},i\bar{\eta}^c_{AB})$ are given in
terms of t'Hooft symbols and
$g_0^2=4\pi(4\pi^2\alpha^{\prime})^{-2}g_s$. Note that the action
$S_{K,N}$ arises from the dimensional reduction of the D5-D9
action in six dimensions down to zero dimension. If there are
v.e.v. for the six $U(N)$-adjoint scalars $\varphi_a$ belonging to
the D3-D3 open string sector one has to add the term
\be S_\varphi=tr_{k}\left[\bar{w}^{\dot{\alpha}}(\varphi^i \varphi_i+2
\chi^i\varphi_i)w_{\dot{\alpha}}+2i\bar{\nu}^A \varphi_{AB}\nu^B
\right]\ee to the action $S_{K,N}$. In the limit $g_0\sim
(\alpha^{\prime})^{-1}\rightarrow \infty$ ($g_0$ fixed) gravity
decouples from the gauge theory and there are no contributions
coming from $S_G$. $\bar{\Theta}_{\dot{\alpha}A}$ and $D^c$ fields
become Lagrange multipliers for the super ADHM constraints:
\bea D^a:\quad [a_\mu,a_\nu]\eta_a^{\mu\nu}+w\sigma_a\bar{w}=0 \qquad {\rm ADHM \quad Eqs} \eea
\bea \bar{\Theta}_A^{\dot{\alpha}}:\quad
[a_\mu,\Theta^A]\sigma^{\mu}+w\bar{\nu}^A+\nu^A\bar{w}=0\qquad
{\rm super \quad ADHM \quad Eqs}\eea

In this limit the multi-instanton `partition function' becomes
\be \cZ_{k,N}=\int_{\mathfrak{M}}e^{-S_{k,N}-S_\vf}=\frac{1}{{\rm
Vol}\,U(k)}\, \int_{\mathfrak{M}} d\chi\, dD\,da\,
d\theta\,d\bar{\theta} dw\, d\nu\, e^{-S_{k,N}-S_\vf}.\ee

\section{Vertex operators}
Classical actions, (super)instanton profiles and non-perturbative
contributions to scattering amplitudes can be derived by computing
disk amplitudes with insertions of vertex operators for
non-dynamical moduli $V_a$, $V_\chi$, $V_w$, $V_{w^\dagger}$ (ADHM
data) and their superpartners \cite{Billo}, \cite{Billo.Inst.CalcinRR}.

\subsubsection{Vertex operators for `gauge' instantons}

Let us first start considering the vertex operator for a non
dynamical gauge boson $a_\mu$ along the four D-D space-time
directions. The vertex operator reads
\be V_a=a_\mu e^{-\varphi}\psi^\mu T_{K\times K}\ee
where $\vf$ arises from the bosonization of the $\beta, \gamma$
worldsheet super-ghosts, $\psi$ are the worldsheet fermions and
$T_{K\times K}$ are $U(K)$ Chan-Paton matrices. For the non
dynamical transverse scalars $\chi_i$ along the six internal D-D
directions, the vertex operator reads
\be V_\chi=\chi_i e^{-\varphi}\psi^i T_{K\times K}\ee
Similarly
\be V_\Lambda=\Theta^a(p) S_a e^{-\varphi / 2} T_{K \times K}\ee
with $a=1,...,16$, produces four non-dynamical Weyl `gauginos'',
$\Theta^A_\alpha$, and their antiparticles,
$\bar{\Theta}_A^\adot$.

Bosonic vertex operators for low-lying D(p-4)-Dp strings, with
multiplicity $K\times N$ and their conjugates are given by
\be V_w=\sqrt{g_s\over v_{p-3}}w_\alpha e^{-\varphi}\prod_\mu\sigma_{\mu}S^\alpha T_{K\times N}\ee
with $S^\alpha$ an $SO(4)$ spin field of worldsheet scaling
dimension $1/4$. $\sigma_\mu$ are $\mZ_2$ bosonic twist fields along
the 4 relatively transverse N-D directions. $\Pi_\mu \sigma_\mu$
has total dimension $1/4=4/16$. $T_{K,N}$ denote the $K\times N$
Chan-Paton `matrices'. The super-partners of $w_\alpha$ are
represented by vertex operators of the form
\be V_\nu=\sqrt{g_s\over v_{p-3}}\nu_A e^{-\varphi / 2}\prod_\mu\sigma_{\mu}S^A T_{K\times N}\ee
where $S^A$ is an $SO(6)$ spin field of dimension 3/8. Note that
the overall normalization $\sqrt{g_s/ v_{p-3}}$ is crucial for the
correct field theory limit $(\alpha^\prime\rightarrow 0)$.

\subsubsection{Vertex operators for `stringy' instantons}

Let us now consider `stringy' instantons. The prototype is the D9,
D1 system which has 8 N-D directions. The multi-(instanton)
configuration of this system was first analyzed in \cite{BKB}. The
lowest lying modes of an open string stretched between N D9 and K
D1 branes are massless fermions with a given chirality (say Right)
along the two common N-N directions. For Type I strings there are
32 such chiral fermions ($\lambda^A$) that precisely reproduce the
gauge degrees of freedom of the `dual' heterotic string
\cite{WitPol}. In addition, in the $\cN = (8,0)$ theory on the D1
world-sheet with $SO(8)$ R-symmetry group, there are 8 transverse
bosons $X^I$ in the $8_v$ and as many Green-Schwarz type fermions
$S^a$ of opposite chirality (say Left) in the $8_s$ of $SO(8)$.
The 32 massless right-moving $\lambda^A$ are inert under the
left-moving susy $Q_{\dot{a}}$ in the $8_c$.

After compactification to $D=4$ on a manifold with non-trivial
holonomy some of the global supersymmetries are broken and the
corresponding D1 world-sheet theory changes accordingly. In
particular $SO(8)$ breaks to some subgroup.

\section{D-branes at Orbifolds}

A particularly promising class of configurations with nice phenomenological perspectives that also allow 
explicit non-perturbative computations are unoriented D-branes at singularities.
Let us consider a stack of D3-branes at the orbifold singularity
$T^d/\Gamma \approx R^d/\Gamma$ (locally), and let us take $\Gamma
=Z_n$ for simplicity. At the singularity $N$ D3-branes group into
stacks of $N_i$ `fractional' branes, that cannot move away from
the singularity, with $i=0,1,2,...$ labelling the conjugacy
classes of $Z_n$. The gauge group $U(N)$ decomposes as $\Pi_i
U(N_i)$. \bea (Z_1, Z_2, Z_3)\approx(\omega^{k_1}Z_1,\omega^{k_2}Z_2,\omega^{k_3}Z_3)
\label{proj} \eea for simplicity we take $k_1+k_2+k_3=0~(mod~n)$ that generically preserves $\cN=1$ 
supersymmetry.

The action on Chan-Paton factors is given by
\be \rho(Z_n)=\rho_0(1_{N_0}, \omega^{1} 1_{N_1},\omega^{2}
1_{N_2}, ..., \omega^{n-1} 1_{N_{n-1}})\ee For $\ap \approx 0$,
keeping only invariant components under (\ref{proj}), the
resulting theory turns out to be an $\cN=1$ quiver gauge theory,
in which vector multiplets $V$ are in the $N_i \bar{N}_i$
representation while chiral multiplets $\Phi_i$ are in the $N_j
\bar{N}_l$ representation with $k_i + j - l = 0~(mod~n)$
\cite{U(4)}.

Twisted RR tadpole cancellation in sectors with non vanishing
Witten index can be written as $tr\rho(Z_n)=0$ that ensures the
cancellation of chiral non-abelian anomalies \cite{Anom&Tad}.

\subsubsection{Unoriented projection}

Possible unoriented projections depend on the parity of $n$ and
the charge of the $\Omega$-plane. For $n$ odd there is only one
possibility
\bea N_0=\bar{N}_0\quad,\quad N_{i}=\bar{N}_{n-i} \eea
For $n$ even there are two possibilities
\bea N_0=\bar{N}_0\quad , \quad N_{i}=\bar{N}_{n-i}\quad , \quad N_{n/2} = \bar{N}_{n/2}\eea
\bea N_0=\bar{N}_{n/2} \quad , \quad N_{i}=\bar{N}_{n/2-i} \eea
One should also impose the twisted RR tadpole cancellation condition (non vanishing Witten index)
$tr\rho(Z_n)=\pm q_n^\Omega$ which from the field theory point of view is just the chiral anomaly
cancellation \cite{AIMU}.

Let us focus on the very rich and instructive case of $T^6/\mZ_3
\approx R^6/ \mZ_3$.

\section{Unoriented $R^6/\mZ_3$ projection}

In the remaining part of this Section, for illustrative purposes,
we will discuss unoriented D-brane instantons on a stack of
D3-branes located at an unoriented $R^6/\mZ_3$ orbifold singularity.

Since $n=3$ is odd, there is only one possible embedding in the
Chan-Paton group up to the charge of the $\Omega 3^{\pm}$ planes.
Introduction of $\Omega 3^-$-plane combined with local R-R tadpole
cancellation leads to a theory with gauge group $G=SO(N_0)\times
U(N_0+4) \times H_{reg}$, where $H_{reg}$ accounts for the
Chan-Paton group of the `regular' branes that can move into the
bulk. We will henceforth assume that regular branes are far from
the singularity and essentially decoupled from the local quiver
theory. For $N_0 = 0$, we have $U(4)$ gauge group with 3 chirals
in ${\bf 6}_{-2}$. In the presence of $\Omega 3^+$-plane we get a
theory with $G=Sp(2N_0)\times U(2N_0-4) \times H_{reg}$ gauge
group, \eg for $2N_0=6$, we have $Sp(6)\times U(2)$ gauge group
with 3 chirals in $({\bf 6}, {\bf 2}_{+1})+({\bf 1}, {\bf
3}_{-2})$.

In both cases the anomalous $U(1)$ mixes with the twisted RR axion
$\zeta$ in a closed string chiral (linear) multiplet $Z$ (gauging
of axionic shift)
\bea \delta A=d\alpha \quad , \quad \zeta =-M_A \alpha \eea
\bea \cL_{ax}=(d\zeta-M_AA)^2+{1\over f_\zeta}\zeta \: F\wedge F \eea
Anomaly cancellation $\delta_\alpha [\cL_{ax}+\cL_{1-loop}]=0
\leftrightarrow M_A/f_\zeta=t_3=Tr_f Q^3$
\footnote{More complicated cases with several (non)anomalous $U(1)$'s, generalized Chern-Simons
terms $\cL_{GCS}=E_{[ij]k}A^i\wedge A^j \wedge F^k$
are needed in the low-energy effective theory with non-trivial
phenomenological consequences \cite{Lionettoetal}.}.

\subsubsection{Field Theory analysis}

As already mentioned `gauge' instantons are expected to generate VY-ADS-like superpotentials.
Neglecting  U(1)'s for the time being, the two choices of
$\Omega$-planes and, thus, of gauge group lead to superpotentials
of the form
\bea SU(4) \quad : \quad  W={\Lambda^9 \over \det_{I,J}(\eps_{abcd}A^{ab}_I A^{cd}_J)}\eea
\bea Sp(6)\times SU(2) \quad : \quad  W={\Lambda^9 \over \det_{6\times 6}(\Phi^i_{aI})}\eea
In string theory, $\Lambda^\beta=M^\beta_s e^{-{S\over f_a}-{Z\over
f_\zeta}}$ ($\beta=9$ here), shift of $Z$ compensates the $U(1)$
charge of the denominator!
The {\it `thumb rule''} is that in each case there are two
exact/unlifted fermionic zero-modes $n(\lambda_0) - n(\psi_0) = 2
$. The rest is lifted by Yukawa interaction $Y_g=g \phi^\dagger
\psi\lambda$.
We now pass to describe the explicit computations with unoriented
D-instantons
\bea U(4)_{D3}\rightarrow U(K)_{D(-1)}
 \quad , \quad Sp(6)_{D3} \rightarrow O(K)_{D(-1)}\eea
In both cases, there are two exact un-lifted fermionic zero-modes
for $K=1$.

\subsubsection{Non-perturbative superpotential for $Sp(6)\times
U(2)$}

After the projection, in the D(-1)-D(-1) sector one has geometric
supermoduli: $a_\mu$ (instanton position) and $\Theta^0_\alpha$
(Grassman coordinate), which yield the $\cN =1$ superspace
measure. There is no room for $D^c$ and
$\bar\Theta_0^{\dot\alpha}$ in the present case, since the
relevant instanton is an $O(1)$ instanton in the $Sp(6)$ group and
as such there are no super ADHM constraints.

In the D(-1)-D3 sector the gauge super-moduli are
$w^u_{\dot\alpha} \: , \nu^{0u}\: , \nu^{Ia}$ with $u=1,...6$
$Sp(6)$, $a=1,2$ $U(2)$, and $I=1,2,3$ $SU(3)$ `pseudo' flavor
indices\footnote{In string theory, $SU(3)$ is an accidental
symmetry of the two-derivative effective action}. Both
$\bar{\Theta}_{0 \adot}$ and $D^c$ are projected out. Taking into
account the interactions with the D3-D3 excitations
$\Phi^{Iua}=\phi^{Iua}+... $, the instanton action can be reduced
to the form
\be S_{D(-1)-D(3)}=w^u_{\dot\alpha}\bar\phi_{Iua}\phi^{Iva}w_{v\dot\alpha}+
\nu^{0u} \nu^{Ia} \bar\phi_{u Ia} \ee
Integrations over gauge super-moduli are gaussian and the final
result can be written as \be \int d^6w d^3\nu^0 d^3\nu^I
e^{S_{D(-1)-D(3)}}= {\det(\bar\phi_{u,Ia})\over
\det(\bar\phi_{u,Ia}\phi^{v,Ia})}={1 \over \det(\phi^{v,Ia})} \ee
Including D(-1)-D(-1) action and one-loop contribution, up to a
non vanishing numerical constant, we get \be \int d^4a d^2\Theta
{\mu^9 e^{2\pi i \tau(\mu)} \over \det(\Phi^{v,Ia})} = \int d^4x
d^2\theta  {\Lambda^9 \over \det(\Phi^{v,Ia})}\ee to a non-zero
numerical constant.

\subsubsection{Non-perturbative superpotential for $U(4)$}

As explained in \cite{U(4)} for $U(4)$ gauge theory  with three
chiral multiplets in the {\bf 6}, the D(-1)-D(-1)  geometric
`supermoduli' are $a^{(0)}_\mu$ (instanton position),
$\bar\chi_{I(-2)},\chi^I_{(+2)}$ (internal) and $\Theta^0_{\alpha
(0)}, \bar\Theta_{0\dot\alpha (0)}, \bar\Theta_{I\dot\alpha (-1)}$
(Grassman coordinates), which give $\cN = 1$ superspace measure.
D(-1)-D(3) gauge `supermoduli' are $w^{\dot\alpha}_{u(-1)},
\bar{w}^u_{\dot\alpha (+1)} , \nu^{0}_{u(+1)},
\bar\nu^{0u}_{(-1)}, \nu^{Iu}_{(+1)}$ with $u=1,...4$ $U(4)$ and
$I=1,2,3$ $SU(3)$ `pseudo' flavor respectively.  Notice that the subscript in parentheses represents the 
charge under $U(1)_{k_1}$. Taking into account the interactions with D3-D3
excitations $\Phi^{Iuv}=\phi^{Iuv}+ ...$, the fermionic
integration will lead to the determinant
\be \Delta_F = \rho^8 \epsilon^{w_1 w_2 w_3 w_4}\epsilon^{u_1
u_2 u_3 u_4}\epsilon^{v_1 v_2 v_3 v_4} X_{u_1 u_2 v_1 v_2} X_{u_3 u_4 v_3 v_4} Y_{w_1 w_2} Y_{w_3 w_4}\ee
with $X=\eps^{IJK} \bar\chi_I \bar\phi_J \bar\phi_K$, $Y_{uv}=\cU_u^{\adot} \cU_{\adot v}$ and $\rho$, $\cU$ 
are defined by $w_{u\dot{\alpha}}=\rho \, \cU_{u\dot{\alpha}}$, $\bar{w}^{u\dot{\alpha}}=\rho \, 
\bar{\cU}^{u\dot{\alpha}}$, 
$\bar{\cU}^{u\dot{\alpha}}\cU_{u\dot{\beta}}=\delta^{\dot{\alpha}}_{\dot{\beta}}$ .
Integration over bosonic `moduli' is more involved. For arbitrary choices of the v.e.v's $\bar{\phi}^{Iuv}$
and $\phi^{Iuv}$, even along the flat directions, integration over $\cU$ represents a difficult task.
Fortunately for the choice $\phi^{Iuv}=\eta^{Iuv}$, the full $\phi$-dependence can be factorized. And after
rescaling $\rho^2 \rightarrow \rho^2/(\phi\bar\phi)$, $\chi^I \rightarrow \phi \chi^I$, $\bar{\chi}_I
\rightarrow \bar{\phi} \bar{\chi}_I$, $X_{u_1 u_2 v_1 v_2}\rightarrow
\eps^{I_1 I_2 I_3}\bar{\chi}_{I_1}\bar{\eta}_{I_2 u_1 u_2}\bar{\eta}_{I_3 v_1 v_2}$  the $\phi$-independent
integral $I_B$ becomes
\be I_B = \int d\rho \rho^9 d^{12}\cU d^3\chi d^3\bar{\chi} \Delta _F e^{-\tilde{S}_B}\ee
where $\tilde{S}_B = -\rho^2(1 + \eta^{Iuv} Y_{uv} \chi_I + \bar{\eta}_{Iuv} \bar{Y}^{uv}\chi^I +
\bar{\chi}_I\chi^I)$. Restoring the $SU(4)$ gauge and $SU(3)$ 'flavor' invariance the
superpotential follows after promoting $\phi^I \rightarrow \Phi^I $:
\be S_W=c\int d^4a d^2\Theta {\mu^9 e^{2\pi i \tau(\mu)} I_B\over \Phi^6}
= \int d^4x d^2\theta {\Lambda ^9 \over \det_{3\times
3}[\epsilon_{u_1...u_4} \Phi^{I u_1 u_2} \Phi^{J u_3 u_4}] }\ee up
to a non-zero numerical constant.

\section{Exotic/Stringy instantons}

EDp-branes on unoccupied nodes of the quiver produce exotic instanton effects. The gauge theory on EDp' is 
of the same kind as on EDp (like 8 N-D directions, periodic sector).

Grassmann integration over chiral fermions $\nu$'s at
intersections produces positive powers of $\Phi$. The resulting
non perturbative superpotential can grow at large VEV's, which is
incompatible with field theory intuition (asymptotic freedom) for standard `gauge' instantons.
Yet it is compatible with gauge invariance and `exotic' scaling
\be e^{-A(C')/\ell_s^{p'+1}} \neq e^{-1/g_{YM}^2}\ee

For generic $K$, there are many unlifted fermionic zero-modes and one gets
higher derivative F-terms, threshold corrections, ... or
dangerous bosonic zero-modes.
For specific $K$, there are only two unlifted zero-modes ($d^2\theta$) and one gets superpotential terms.
For ED1, the relevant $\nu$'s are in the direction of the worldsheet.

\subsubsection{$U(4)$ model: non-perturbative masses} Let us
consider $U(4)$ model, $\Theta^0_\alpha$, $X_\mu$ plus 4
$\nu^u$ that couple to one complex component $\phi_{uv}$ (related
to $C$) through
\be S_{D3-ED3} = \phi_{uv} \nu^u \nu^v + ... \ee
The superpotential generated by ED-strings wrapping 2-cycles $C$ passing
through the singularity schematically reads
\be W(\Phi) = \sum_{C} M_s e^{-A(C)/g_s\ap} \Phi_{C}^2 \ee
and thus represents a mass terms for $\Phi \approx A$.

Effect of multi-instanton are hard to evaluate. Heterotic/Type I duality may help clarifying the procedure.

\section{Effect of fluxes}
Phenomenological applications of string theory and realistic model building require compactifications. 
Four dimensional compactifications of Type II string theories which preserve $\cN=1$ supersymmetry in the 
presence of intersecting or magnetized D-branes are very interesting. Gauge interactions can be realized 
with space-filling D-branes that partially or totally wrap the internal six dimensional space. Adjoint gauge 
fields are given by the massless excitations of open strings that start and end on the same stack of
D-branes. Open strings stretched between different stacks provide bi-fundamental matter fields. From the 
closed string point of view D-branes are sources for Type II supergravity fields, which have a non-trivial 
profile in the bulk. The effective actions of these models describe interactions of both open string 
(boundary) and closed string (bulk) degrees of freedom and have the generic structure of $\cN=1$ 
supergravity in four dimensions coupled to vector and chiral multiplets. Four dimensional $\cN=1$ 
supergravity theories are specified by the choice of the gauge group $\cG$, by a K\"ahler potential $K$ and 
a superpotential $W$. The K\"ahler potential is real and the superpotential is holomorphic function of some 
chiral superfields $\Phi ^i$. The expectation values of these chiral multiplets, which parametrize 
the supergravity vacuum minimizes the scalar potential
\be
V=e^K(D_i\overline{W}D^iW-3|W|^2)+D^aD_a\,\, ,
\label{sugscpot}
\ee
where $D^iW\equiv \partial _{\Phi ^i}W + (\partial _{\Phi ^i}K)W$ is the K\"ahler covariant derivative of 
the superpotential and the $D^a$ with $a=1,...,\dim(\cG)$ are the D-terms. Supersymmetric vacua correspond 
to the solutions of $\partial _{\Phi ^i} V=0$ equations satisfying D-flatness and F-flatness conditions 
$D^a=D_iW=0$. In the case of Type IIB string theory on a Calabi-Yau three-fold in the presence of D3-branes, 
\cite{Billo':2008pg}, the chiral superfields $\Phi ^i$ consist of the fields $U^r$ and $T^m$, which 
parametrize the deformations of the complex and K\"ahler structures of the three-fold, of the axion-dilaton 
field and also of some other multiplets coming from the open strings stretching between D-branes. The axion-
dilaton field $\tau = C_0+ie^{i\varphi}$ is given in terms of the R-R scalar $C_0$ and the dilaton 
$\varphi$. The corresponding low energy $\cN=1$ supergravity theory has a highly degenerated vacuum. One way 
of lifting this degeneracy, at least partially, is to add the internal 3-form fluxes of the bulk theory 
which generate a superpotential of the form: 
\be
W_{flux}=\int G_3\wedge \Omega
\label{W_flux}
\ee
where $G_3$ is the complex 3-form flux given in terms of the R-R and NS-NS fluxes $F$ and $H$ via 
$G_3=F-\tau H$ and $\Omega$ is a holomorphic (3,0)-form of the Calabi-Yau three-fold. The flux 
superpotential (\ref{W_flux}) depends on $\tau$ and on the complex structure parameters $U^r$ which specify 
$\Omega$. An unbroken $\cN=1$ supersymmetry requries the flux $G_3$ to be an imaginary anti-selfdual 3-form 
of (2,1) type, since the F-terms $D_{U^r}W_{flux}$, $D_\tau W_{flux}$ and $D_{T^m}W_{flux}$ are proportional 
to the (1,2), (3,0) and (0,3) components of the $G$-flux, respectively. 
These F-terms can also be interpreted as the `auxiliary' $\theta ^2$-components of the kinetic functions 
for the gauge theory defined on the space-filling branes, and thus are soft supersymmetry breaking terms for 
the brane-world effective action. Such soft terms in flux compactifications give effects like, for instance, 
induced masses for the gauginos and the gravitino. Non-perturbative contributions to the effective actions 
may also play an important role in the moduli stabilization. They have phenomenologically relevant 
implications for string theory compactifications. Non-perturbative effects, coming from wrapped Euclidean 
branes, may lead to the generation of a non-perturbative superpotential of the following form:
\be
W_{n.p.}=\sum _{\{k_A\}}c_{\{k_A\}}(\Phi ^i)e^{2\pi i\sum _A k_A\tau _A}
\label{Wnp}
\ee
where the index $A$ labels the cycles wrapped by the instantonic branes, $\tau _A$ is the complexified gauge 
coupling of a D-brane wrapping the cycle $A$ and the sum is over the instanton numbers $k_A$. 
$c_{\{k_A\}}(\Phi ^i)$ are holomorphic functions of the chiral superfields. The specific forms of these 
functions depend on the details of the model under consideration. The coupling $\tau _A$ generally depend on 
the axion-dilaton modulus $\tau$ and the K\"ahler parameters $T^m$ that describe the volumes of the cycles 
wrapped by D-branes. When the cycle $A$ is wrapped by some physical D-branes one has gauge instantons. The 
case when the cycle is not wrapped by any D-brane in the background corresponds to stringy instantons. In 
both cases the generated superpotential has the form (\ref{Wnp}). Fluxes and non-perturbative effects 
contribute to the total superpotential $W=W_{flux}+W_{n.p.}$. Thus, new possibilities to get supersymmetric 
vacua arise. Indeed, the derivatives $D_{U^r}W_{flux}$, $D_\tau W_{flux}$ and $D_{T^m}W_{flux}$ can now be 
compensated by $D_{U^r}W_{n.p.}$, $D_\tau W_{n.p.}$ and $D_{T^m}W_{n.p.}$ so that also the (1,2), (3,0) and 
(0,3) components of $G_3$ may become compatible with supersymmetry and help to remove the vacuum degeneracy. 

Besides generating perturbative superpotential $W_{flux}$, fluxes lift some of the zero-modes of the 
instanton background. As a result, new types of non-perturbative couplings arise.
Following \cite{Billo':2008pg}, let us consider the case of D3-branes at a $\mC ^3/(\mZ_2\times\mZ_2)$ 
singularity and engineer an $\cN=1$ $U(N_0)\times U(N_1)$ quiver gauge theory with bi-fundamental matter 
fields. This quiver theory can be thought of as a local description of the Type IIB Calabi-Yau 
compactification on the toroidal orbifold $T^6/(\mZ_2\times \mZ_2)$. In this context the numbers $N_0$ and 
$N_1$ of the D3-branes can be arbitrary and orientifold planes for tadpole cancellation are not needed. 
Gauge and stringy instantons are realized by means of D-instantons. When one introduces background fluxes of 
type $G_{(3,0)}$ and $G_{(0,3)}$ and studies the induced non-perturbative interactions in the presence of 
gauge and stringy instantons, one finds a very rich class of non-perturbative effects from `exotic' 
superpotential terms to non-supersymmetric multi-fermion couplings. 

Stringy instantons in the presence of $G$-fluxes can generate non-perturbative 
interactions even for $U(N)$ gauge theories. In the case without fluxes an orientifold projection is 
required to solve the problem of the neutral fermionic zero-modes. Since $G_{(3,0)}$ and $G_{(0,3)}$ 
components of $G_3$ are related to the gaugino and gravitino masses, the non-perturbative flux-induced 
interactions can be regarded as the analog of the Affleck-Dine-Seiberg (ADS) superpotentials for 
gauge/gravity theories with soft supersymmetry breaking terms. In particular, the presence of the 
$G_{(0,3)}$ flux has no effect on the gauge theory at a perturbative level, but it generates new
instanton-mediated effective interactions.  

Besides this supergravity method there is also world-sheet approach. This approach allows to obtain the 
flux induced couplings in a unified way, it is applicable to more general brane configurations, with or 
without magnetization, with twisted or untwisted boundary conditions. 
In the four dimensional compactifications of Type II string theories preserving $\cN=1$ supersymmetry in the 
presence of intersecting or magnetized D-branes one can add internal antisymmetric fluxes both in the NS-NS 
and in the R-R sector of the bulk theory. These fluxes are important for moduli stabilization, supersymmetry 
breaking and may also generate non-perturbative superpotentials. At a perturbative level internal 3-form 
fluxes are encoded in a bulk superpotential from which F-terms can be obtained using standard supergravity 
methods. These F-terms can be viewed as the $\theta ^2$ auxiliary components of the kinetic functions for 
the gauge theory living on the space-filling branes. Thus, these are soft supersymmetry breaking terms for 
the brane-world effective action. The non-perturbative sector of the effective action coming from string 
theory compactifications is also important to study. The computational tool to study these effects using 
systems of branes with different boundary conditions has been developed in \cite{Billo}, 
\cite{Billo.Inst.CalcinRR}, \cite{Green:2000ke}. These techniques allow to reproduce the known instanton 
calculus of supersymmetric field theories and also can be generalized to more exotic instanton 
configurations for which there are no field theory methods available. The study of these exotic instantons 
has led to interesting results related to moduli stabilization, (partial) supersymmetry breaking, fermion 
masses and Yukawa couplings \cite{Blum}, \cite{Blumenhagen:2007zk}. There are neutral anti-chiral fermionic 
zero modes, which totally decouple from all other exotic instanton moduli. In the case of gauge theory 
instanton this does not happen. In this case neutral anti-chiral fermionic zero modes act as Lagrange 
multipliers for the fermionic ADHM constraints \cite{Billo}. Therefore, to get non-vanishing contributions 
to the effective action from exotic instantons, one needs to remove these anti-chiral zero modes 
\cite{U(4)}, \cite{Argurio:2007vqa} or lift them by some mechanism \cite{Blumenhagen:2007bn}. The presence 
of internal background fluxes may allow for such a lifting and gives an idea of the existence of the 
interplay among soft supersymmetry breaking, moduli stabilization, instantons and more generally
non-perturbative effects in the low energy theory.

In \cite{Billo':2008sp},\cite{Billo':2008pg}, by evaluating disk amplitudes involving two open string vertex 
operators at a generic intersection and one closed string vertex representing the background fluxes, authors 
have given the couplings of NS-NS and R-R fluxes to various types of D-branes including instantonic ones 
using world-sheet approach. This approach, being in full agreement with the derivation of the flux couplings 
in the brane effective actions based on supergravity methods, is applicable also to more general brane 
configurations which involve fields with twisted boundary conditions. It allows to study the modification of 
the action by R-R and NS-NS fluxes which gives the measure of integration on the moduli space of instantons. 
Considering an orbifold compactification of Type IIB string theory with fractional
D-branes preserving $\cN=1$ supersymmetry and studying the flux-induced fermionic mass terms on space-
filling and on instantonic branes, it has been shown that there is a relation between the soft supersymmetry 
breaking and the lifting of some instanton fermionic zero-modes. This may lead to new types of
non-perturbative couplings in brane-world models. 

Let us consider string amplitudes between two massless open string fermions and the background closed string 
flux. This is a mixed open/closed string amplitude on a disk with mixed boundary conditions in general. For 
the two fermionic open string vertices and one closed string R-R vertex one has the following amplitude:
\be
\cA_F=\< V_\Theta (x)V_F(z,\bar{z}) V_{\Theta ^\prime}(y)\> =c_F \Theta _{\cA _1}(F \cR _0)_{\cA _2 \cA 
_3}\Theta ^\prime _{\cA _4}\times A^{\cA _1\cA _2\cA _3\cA _4}
\label{3vertexamplitude}
\ee
where $V_\Theta$ is the vertex operator for the lowest fermionic excitation $\Theta_\cA$ of the open string 
and $V_F$ is the vertex operator for the closed string field strengths of the antisymmetric tensor fields in 
the R-R sector of Type IIB theory in the $(-\frac{1}{2},-\frac{1}{2})$ superghost picture. 
$c_F=\cC_{(p+1)}\cN_\Theta \cN_{\Theta ^\prime}\cN_F$ stands for normalizations of the vertex operators and 
$\cC_{p+1}$ is the topological normalization of any disk amplitude with the boundary conditions of a
Dp-brane. $F_{\cA\cB}$ is the bi-spinor polarization which consist of all R-R field strengths of the Type 
IIB theory via 
\be 
F_{\cA \cB}=\sum_{n=1,3,5}\frac{1}{n!}F_{M_1...M_n}(\Gamma ^{M_1...M_n})_{\cA\cB}.
\ee
Note that we discuss only untwisted closed string vertices. 
Thus, in order to get a non vanishing amplitude, open string vertices must have opposite twists.
In the presence of D-branes the left and right moving components of the vertex operator $V_F$ must be 
identified via the reflection rules. As a consequence $F_{\cA \cB}$ is replaced with $(F\cR_0)_{\cA \cB}$, 
where $\cR_0$ is the reflection matrix in the adjoint representation of the rotation group. The 4-point 
correlator $A^{\cA _1\cA _2\cA _3\cA _4}$ is given by:
\be 
A^{\cA _1\cA _2\cA _3\cA _4}=(\Gamma_M)^{\cA_1\cA_4}(\Gamma^M I_1)^{\cA_2\cA_3}+(\Gamma^M I_2)^{\cA_1\cA_3}
(\Gamma^M)^{\cA_2\cA_4}
\label{4pointamplitude}
\ee
where $I_1$ and $I_2$ are $\vec{\vartheta}$-dependent diagonal matrices with entries:
\bea
(I_1)_{\cA_3}^{\,\,\,\cA_3}=\frac{1}{2}e^{-\frac{i\pi\alpha^\prime s}{2}}\left(e^{-2\pi i(\alpha^\prime 
t-\vec{\vartheta}\cdot \vec{\epsilon}_3)}-1\right) B(\alpha^\prime s; \alpha^\prime t-\vec{\vartheta}\cdot 
\vec{\epsilon}_3)\\
(I_2)_{\cA_3}^{\,\,\,\cA_3}=\frac{1}{2}e^{-\frac{i\pi\alpha^\prime s}{2}}\left(e^{-2\pi i(\alpha^\prime 
t-\vec{\vartheta}\cdot \vec{\epsilon}_3)}-1\right) B(\alpha^\prime s+1; \alpha^\prime t-\vec{\vartheta}\cdot 
\vec{\epsilon}_3)
\eea
where $B(a;b)$ is the Euler $\beta$-function. Plugging 4-point correlator (\ref{4pointamplitude}) into 
(\ref{3vertexamplitude}) one finds: 
\be
\cA _F=-8c_F \Theta ^\prime \Gamma ^M \Theta [F\cR _0(2I_1-I_2)]_M + \frac{4c_F}{3!}\Theta ^\prime \Gamma 
^{MNP}\Theta [F\cR _0 I_2]_{MNP}.\label{Ramplitude}
\ee
This amplitude describes the tree-level bilinear fermionic couplings induced by R-R fluxes on a general 
brane intersection.

Fermionic couplings induced by the NS-NS 3-form flux arise from the following mixed disk amplitude:
\be
\cA _H= \< V_\Theta(x) V_H(z,\bar{z}) V_{\Theta ^\prime}(y)\> =c_H \Theta _\cA (\partial B R_0)_{MNP} \Theta 
^\prime _\cB \times A^{\cA\cB ;MNP}
\ee
where the NS-NS 3-form flux $H$ has the vertex operator $V_H(z,\bar{z})$. $A^{\cA\cB ;MNP}$ is the 4-point 
correlator and $c_H=\cC_{p+1}\cN _\Theta\cN_{\Theta ^\prime}\cN _H$ is the normalization factor. $R_0$ is 
the reflection matrix in the vector representation of the rotation group. The NS-NS 
couterpart of the R-R amplitude (\ref{Ramplitude}) on a generic D-brane intersection is
\be
\cA _H=-4c_H \Theta ^\prime \Gamma ^N \Theta \delta ^{MP} [\partial B R_0 (2I_1-I_2)]_{[MN]P} + 2c_H\Theta 
^\prime \Gamma ^{MNP} \Theta [\partial B R_0 I_2]_{MNP}
\label{Hamplitude}
\ee
and shares with it the same type of fermionic structures. At leading order in $\alpha^\prime$ these 
amplitudes describe fermionic mass terms induced at linear order in the R-R and NS-NS fluxes for open string 
modes. $R_\sigma$ and $\cR_\sigma$ are boundary reflection matrices in the vector and spinor representations 
of the rotation group, respectively. The boundary conditions are encoded in the reflection matrices and in 
the open string twists $\vec{\vartheta}$. We consider constant background fluxes coupled 
to untwisted open strings, \ie strings starting and ending on a single stack of D-branes. This corresponds 
to set $\vec{\vartheta}=0$. Constant background fluxes allow to set the momentum of the closed string 
vertices to zero. As a consequence $2I_1=I_2=-i\pi$. Using the amplitudes (\ref{Ramplitude}) and 
(\ref{Hamplitude}) one can see that the fermionic couplings with a single $\Gamma$ matrix vanish and only 
the terms with three $\Gamma$'s survive. Taking into account the fact that in both amplitudes the untwisted 
fermions $\Theta$ and $\Theta^\prime$ describe the same field and differ only by opposite momentum, the 
total amplitude becomes:
\be
\cA\equiv \cA_ F +\cA_ H=-2\pi i \Theta \Gamma ^{MNP}\Theta \left[\frac{c_F}{3}(F\cR _0)_{MNP}+c_H(\partial 
B R_0)_{MNP}\right]
\label{totamplitude}
\ee
It is clear that once the flux configuration is given, the structure of the fermionic couplings for 
different types of D-branes depends crucially on the boundary reflection matrices $R_0$ and $\cR_0$.
Let us consider only 3-form fluxes. In general the R-R piece of (\ref{totamplitude}) is non-zero for 1-form, 
3-form, and 5-form fluxes. For the 3-form flux the bi-spinor is:
\be
F_{\cA\cB}=\frac{1}{3!}F_{MNP}(\Gamma ^{MNP})_{\cA\cB}
\ee
In this case the normalization factors can be better specified and are related via the string coupling 
constant:
\be
c_F=g_sc_H
\ee
The amplitude can be rewritten as:
\be
\cA =-\frac{2\pi i}{3!}c_F \Theta \Gamma ^{MNP}\Theta T_{MNP}
\ee
where 
\be
T_{MNP}=(F\cR_0)_{MNP} + \frac{3}{g_s}(\partial B R_0)_{[MNP]}
\label{tmnp}
\ee
To study the flux induced couplings for gauge theories and instantons in four dimensions, one has to split 
ten dimensional indices $M,N,...=0,1,...,9$ into four-dimensional space-time indices $\mu,\nu,...=0,1,2,3$ 
and six dimensional indices $m,n,...=4,5,...,9$ for the internal space. Background fluxes, which carry 
space-time indices break the four-dimensional Lorentz invariance and generically give rise to deformed 
gauge theories \cite{Billo.Inst.CalcinRR}, \cite{Billo:2004zq}, \cite{Billo:2005jw}. Let us concentrate only 
on the internal 3-form fluxes, like $F_{mnp}$ or $(\partial B)_{mnp}$. They preserve four dimensional 
Lorentz invariance, and the fermionic amplitudes are of the form:
\be  
\cA=-\frac{2\pi i}{3!}c_F\Theta \Gamma ^{mnp}\Theta T_{mnp}
\label{ampl}
\ee
Let us discuss unmagnetized space-filling D-branes, $\vec{\vartheta}=0$. To do so one has to go to 
Minkowskian signature. In this case $\Theta$ becomes a Majorana-Weyl spinor in ten dimensions. For an 
unmagnetized Dp-brane which fills four dimensional Minkowski space and possibly extends also in some 
internal directions, the reflection matrices are very simple:
\be
R_0=diag(\pm 1, \pm 1,...)
\label{vec_ref_unmagnetized}
\ee
for the reflection matrix in the vector representation. Plus sign of the matrix element refers to the 
longitudinal direction and the minus sign to the transverse one. The reflection matrix in the spinor 
representation has the following form:
\be
\cR _0=\Gamma ^{p+1}\cdots \Gamma ^9
\label{spinor_ref_unmagnetized}
\ee
Using this, one observes from (\ref{tmnp}) that $T_{mnp}$ is a real tensor, so that the total fermionic 
amplitude (\ref{ampl}) is also real in view of equation 
$\Theta \Gamma ^{nmp}\Theta=-(\Theta \Gamma ^{nmp}\Theta)^*$, which Majorana-Weyl spinor satisfies in 
ten dimensions. The explicit expression of $T_{mnp}$ is particularly simple in the case of brane 
configurations which respect the 4+6 structure of the spacetime, \ie D3 and D9-branes. For space-filling D3-
branes all the internal directions are transverse, so that $R_0|_{int}=-1$ and
$\cR_0=\Gamma ^4 \cdots \Gamma ^9$. So that one has:
\be
T_{mnp}=(* _6F)_{mnp}-\frac{1}{g_s}H_{mnp}
\label{D3tmnp}
\ee
where $* _6$ denotes the Poincar\'e dual in the six dimensional internal space and $H_{mnp}=3\partial 
_{[m}B_{np]}=\partial _m B_{np}+\partial _n B_{pm}+\partial _p B_{mn}$. 
For D9-branes all internal indices are longitudinal. The internal longitudinal indices we will denote by 
$\hat{m}, \hat{n},...$. In the D9-brane case $R_0=1$ and $\cR_0=1$, so that
\be
T_{\hat{m}\hat{n}\hat{p}}= F_{\hat{m}\hat{n}\hat{p}}+\frac{1}{g_s}H_{\hat{m}\hat{n}\hat{p}}
\ee
For tadpole cancellation there always should be orientifold 9-planes. The corresponding orientifold 
projection eliminates NS-NS flux $H_{\hat{m}\hat{n}\hat{p}}$. Then the coupling tensor for D9-branes reduces 
to  
\be
T_{\hat{m}\hat{n}\hat{p}}= F_{\hat{m}\hat{n}\hat{p}}
\label{D9tmnp}
\ee

For D7-branes the longitudinal internal indices $\hat{m}, \hat{n},...$ take four values and the trasverse 
indices $p,q,...$ take two values. Non vanishing components of the $T$ tensor for D7-branes are the 
following:
\bea
T_{\hat{m}\hat{n}\hat{p}}=\frac{1}{g_s}H_{\hat{m}\hat{n}\hat{p}}, \quad T_{\hat{m}\hat{n}p}= 
F_{\hat{m}\hat{n}}^{\,\,\,\,\,\,\,q}\epsilon _{qp}+\frac{1}{g_s}H_{\hat{m}\hat{n}p}, \quad 
T_{\hat{m}np}=-\frac{1}{g_s}H_{\hat{m}np}
\eea
O7-planes required for tadpole cancellation remove all $F$ and $H$ components with an even number of 
transverse indices, so that one is left with
\be
T_{\hat{m}\hat{n}p}= F_{\hat{m}\hat{n}}^{\,\,\,\,\,\,\,q}\epsilon _{qp}+\frac{1}{g_s}H_{\hat{m}\hat{n}p}
\label{D7tmnp}
\ee

For D5-branes the longitudinal internal indices run over two values and the trasverse indices over four 
values. Non vanishing components of $T$ are:
\bea
T_{\hat{m}\hat{n}p}=\frac{1}{g_s}H_{\hat{m}\hat{n}p}, \quad T_{\hat{m}np}=-\frac{1}{2}F_{\hat{m}}^{\,\,\, 
qr}\epsilon _{qrnp}-\frac{1}{g_s}H_{\hat{m}np}, \quad T_{mnp}=-\frac{1}{g_s}H_{mnp}\nn\\
\label{D5tmnp}
\eea
O5-planes enforce an orientifold projection $\Omega I_4$ which removes the components of $H$ with even 
number of transverse indices and of $F$ with odd number of transverse indices. Thus $T_{\hat{m}np}$ becomes 
\be 
T_{\hat{m}np}=-\frac{1}{2}F_{\hat{m}}^{\,\,\, qr}\epsilon _{qrnp}
\label{D5tmnp1}
\ee
It is interesting to observe that while for D9 and D5-branes the fermionic couplings depend either on $F$ or 
on $H$, for D3 and D7-branes they depend on a combination of the R-R and NS-NS 3-forms. This follows from 
the fact that O3 and O7-planes act on the same way on R-R and NS-NS fluxes. By introducing the complex
3-form
\be
G=F-\frac{i}{g_s}H
\label{complex3form}
\ee
one can rewrite the D3-brane coupling (\ref{D3tmnp}) as
\be
T_{mnp}=(* _6F)_{mnp}-\frac{1}{g_s}H_{mnp}=Re(*_6G-iG)_{mnp}
\label{D3tG}
\ee
This confirms the fact that an imaginary self-dual (ISD) 3-form flux $G$ does not couple to unmagnetized D3-
branes. The fermionic couplings for D7-branes (\ref{D7tmnp}) can also be rewritten:
\bea
T_{\hat{m}\hat{n}i}=iG_{\hat{m}\hat{n}i} \quad T_{\hat{m}\hat{n}\bar{i}}=-iG^*_{\hat{m}\hat{n}\bar{i}}
\label{D7tG}
\eea
where $i$ and $\bar{i}$ denote the complex directions of the plane transverse to the D7-branes. (\ref{D7tG}) 
is in agreement with the structure of soft fermionic mass terms found in \cite{Camara:2004jj}.

Unmagnetized Euclidean branes that are transverse to the four dimensional space-time and extend partially or 
totally in the internal directions are relevant to discuss non-perturbative instanton effects in the branes 
models. In this case it is necessary to work in a space with Euclidean signature. Then the massless fermions 
$\Theta$ are not Majorana-Weyl spinors anymore. There is no issue about the reality of a fermionic amplitude 
and the coupling tensor $T$ is in general complex. For D-instantons (or D(-1)-branes) all ten directions are 
transverse. In this case the reflection matrices in the vector and spinor representations are:
\bea
R_0=-1, \quad \cR _0=\Gamma ^0\Gamma ^1\cdots \Gamma ^9\equiv i\Gamma ^E_{11}
\label{refmatD(-1)}
\eea
where $\Gamma ^E_{11}$ is the chirality matrix in ten Euclidean dimensions. For D-instantons the $T$ tensor 
is:
\be
T_{mnp}=-iF_{mnp}-\frac{1}{g_s}H_{mnp}=-iG_{mnp}
\label{DinsT}
\ee
In the case of Euclidean instantonic 5-branes, E5-branes, extending in the six internal directions the 
reflection matrices are:
\bea
R_0|_{int}=1, \quad \cR_0=\Gamma ^0\Gamma ^1\Gamma ^2\Gamma ^3=-i\Gamma ^4\cdots \Gamma ^9 \Gamma ^E_{(11)}
\label{refmatE5}
\eea
The fermionic coupling $T$ for unmagnetized E5-branes has the following form:
\be
T_{\hat{m}\hat{n}\hat{p}}=i(*_6F)_{\hat{m}\hat{n}\hat{p}}+\frac{1}{g_s}H_{\hat{m}\hat{n}\hat{p}}
\label{E5T}
\ee
In an orientifold model with O9-planes this coupling reduces to:
\be
T_{\hat{m}\hat{n}\hat{p}}=i(*_6F)_{\hat{m}\hat{n}\hat{p}}
\label{E5+O9T}
\ee
E3-branes extend along four of the six internal directions. Flux induced fermionic couplings $T$ for E3 are 
of the form:
\bea
T_{\hat{m}\hat{n}\hat{p}}=\frac{1}{g_s}H_{\hat{m}\hat{n}\hat{p}}, \quad T_{\hat{m}\hat{n}p}=-\frac{i}
{2}\epsilon _{\hat{m}\hat{n}\hat{r}\hat{s}}F^{\hat{r}\hat{s}}_{\,\,\,\,\,p}+\frac{1}
{g_s}H_{\hat{m}\hat{n}p}, \quad T_{\hat{m}np}=-\frac{1}{g_s}H_{\hat{m}np}\nn\\
\label{E3T}
\eea
The appropriate orientifold projections remove $H_{\hat{m}\hat{n}\hat{p}}$ and $H_{\hat{m}np}$. Thus one has 
only
\be
T_{\hat{m}\hat{n}p}=-\frac{i}{2}\epsilon _{\hat{m}\hat{n}\hat{r}\hat{s}}F^{\hat{r}\hat{s}}_{\,\,\,\,
\,p}+\frac{1}{g_s}H_{\hat{m}\hat{n}p}
\label{E3+OT}
\ee
The fermionic couplings for the E1-branes are:
\bea
T_{\hat{m}\hat{n}p}=\frac{1}{g_s}H_{\hat{m}\hat{n}p}, \quad 
T_{\hat{m}np}=-i\epsilon _{\hat{m}\hat{q}}F^{\hat{q}}_{\,\,\,\,np}-\frac{1}{g_s}H_{\hat{m}np}, \quad 
T_{mnp}=-\frac{1}{g_s}H_{mnp}\nn\\
\label{E1T}
\eea
$H_{\hat{m}np}$ gets removed by the orientifold projection when the E1-branes are considered together with 
D5/D9-branes and the corresponding orientifold planes. One observes that in the presence of E-branes the 
space-time filling Dp-branes live in the Euclidean ten-dimensional space. The couplings of such Dp-branes 
are again given by the same linear combinations of $F$ and $H$ like in the Minkowskian case, since $R_0$ and 
$\cR_0$ are trivial along the would be time direction.

One can generalize the discussion to the branes with a non-trivial magnetization on their worldvolume for 
which the longitudinal coordinates satisfy non-diagonal boundary conditions. In the setup discussed above 
one can introduce a worldvolume gauge field $A$ that couples to the open string end-points and obtain a 
magnetization $\cF_0=\cF_\pi=2\pi \alpha ^\prime (dA)$. Above discussed R-R and NS-NS background fluxes can 
be used and one can study the new couplings induced by the worldvolume magnetization using the 
following reflection matrices:
\bea
&& R_\sigma =(1-\cF _\sigma)^{-1}(1+\cF _\sigma)\\
&& \cR _\sigma =\pm \prod _{I=1}^5 e^{i\pi \theta _\sigma ^I \Gamma ^{I\bar{I}}}=\pm \prod _{I=1}^5 
\frac{(1+if_\sigma ^I \Gamma ^{I\bar{I}})}{\sqrt{1+(f_\sigma ^I)^2}}
\eea

To complete the analysis and to make the structure of the flux-induced fermionic masses more clear let us 
write the fermion bilinear $\Theta \Gamma ^{mnp}\Theta$ using a four dimensional spinor notation. The 
anti-chiral ten dimensional spinor $\Theta _\cA$ has the following 4+6 splitting:
\be
\Theta _\cA\rightarrow (\Theta ^{\alpha A}, \Theta _{\dot{\alpha}A})
\label{Theta4+6}
\ee 
where $\alpha$ ($\dot{\alpha}$) are chiral (anti-chiral) indices in four dimensions and the lower (upper) 
indices $A$ are chiral (anti-chiral) spinor indices of the internal six dimensional space. $\Gamma$ matrices 
decompose as:
\be 
\Gamma ^\mu =\gamma ^\mu\otimes {\bf{1}}, \quad \Gamma ^m=\gamma _{(5)}\otimes \gamma ^m
\label{decompgamma}
\ee
Then the fermion bilinear $\Theta \Gamma ^{mnp}\Theta$ can be written as:
\be
\Theta \Gamma ^{mnp}\Theta =-i \Theta ^{\alpha A}\Theta _\alpha ^{\,\,\,B}(\overline{\Sigma}^{mnp})_{AB}-
i\Theta _{\dot{\alpha}A}\Theta ^{\dot{\alpha}}_{\,\,\,B} (\Sigma ^{mnp})^{AB}
\label{fermbil}
\ee
where $\Sigma ^{mnp}$ and $\overline{\Sigma}^{mnp}$ are the chiral and anti-chiral blocks of $\gamma 
^{mnp}$. $\Sigma ^{mnp}$ couples only to an imaginary self-dual tensor (ISD) tensor and 
$\overline{\Sigma}^{mnp}$ only to imaginary anti-self-dual (IASD) tensor, since
$*_6\Sigma ^{mnp}=-i\Sigma ^{mnp}$ and $*_6\overline{\Sigma}^{mnp}=+i\overline{\Sigma}^{mnp}$. One has:
\bea
\Theta \Gamma ^{mnp}\Theta T_{mnp}&=&-i\Theta ^{\alpha A}\Theta _\alpha ^{\,\,\,B}
(\overline{\Sigma}^{mnp})_{AB} T^{IASD}_{mnp}-i\Theta _{\dot{\alpha}A}\Theta ^{\dot{\alpha}}_{\,\,\,B}
(\Sigma ^{mnp})^{AB}T^{ISD}_{mnp}\nn\\
&=&-i\Theta ^{\alpha A}\Theta _\alpha ^{\,\,\,B}T_{AB}-i\Theta _{\dot{\alpha}A}\Theta ^{\dot{\alpha}}_{\,\,
\,B}T^{AB}
\label{TGTT}
\eea
where in the second line we used a $SU(4)\thicksim SO(6)$ notation and defined
\bea
T_{AB}=(\overline{\Sigma}^{mnp})_{AB}T_{mnp}^{IASD}, \quad T^{AB}=(\Sigma ^{mnp})^{AB}T_{mnp}^{ISD}
\label{TAB}
\eea
where 
\bea
T_{mnp}^{ISD}=\frac{1}{2}(T-i*_6T)_{mnp}, \quad T_{mnp}^{IASD}=\frac{1}{2}(T+i*_6T)_{mnp}
\label{TmnpIASD}
\eea
The upper indices $A,B$ run over {\bf{4}} respresentation of $SU(4)$ and the lower ones over $\bf\bar{{4}}$
representation. Fixing a complex structure , the 3-form tensors $T^{ISD}$, $T^{IASD}$ can be decomposed into 
their (3,0), (2,1), (1,2) and (0,3) parts as:
\bea
T^{ISD}\rightarrow T_{(0,3)}\oplus T_{(1,2)NP}\oplus 
T_{(2,1)P}={\bf\bar{{1}}}\oplus{\bf\bar{{3}}}\oplus{\bf\bar{{6}}}\\
T^{IASD}\rightarrow T_{(3,0)}\oplus T_{(2,1)NP}\oplus 
T_{(1,2)P}={\bf{1}}\oplus{\bf{3}}\oplus{\bf{6}}
\label{decIASD}
\eea
The various compnents transform in irreducible representations of the $SU(3)\in SU(4)$, which are given on 
the right hand side. (2,1) components are six primitive ones (P) and three non-primitive ones (NP). A 
similar decomposition holds for the (1,2) part.

Let us now focus on D3 and D-instantons on flat space. For D3-branes one can use a Minkowski signature and 
the fermionic bilinear decomposition is given in (\ref{fermbil}), where the four dimensional chiral and 
anti-chiral components are related by charge conjugation and assembled into four Majorana spinors. These are 
the four gauginos living on the worldvolume of the D3-brane. Let us denote their chiral and anti-chiral 
parts as $\Lambda ^{\alpha A}$ and $\bar{\Lambda}_{\dot{\alpha}A}$ instead of $\Theta ^{\alpha A}$ and 
$\Theta _{\dot{\alpha}A}$ for future notational convenience. Then plugging (\ref{fermbil}) and (\ref{D3tG}) 
into (\ref{ampl}) one finds the following amplitude for D3-branes in flat space:
\be
\cA _{D3}=\frac{2\pi i}{3!}c_F\Tr[\Lambda ^{\alpha A}\Lambda _\alpha ^{\,\,\,\,B}
(\overline{\Sigma}^{mnp})_{AB}G^{IASD}_{mnp}-\bar{\Lambda}_{\dot{\alpha}A}\bar{\Lambda}^{\dot{\alpha}}_{\,\,
\,\,B}(\Sigma ^{mnp})^{AB}(G_{mnp}^{IASD})^*]
\label{AD3}
\ee
which encodes the structure of soft symmetry breaking terms in $\cN=4$ gauge theory induced by NS-NS and RR 
fluxes. From (\ref{AD3}) we see that an imaginary anti-self dual $G$-flux configuration induces a Majorana 
mass for the gauginos leading to the supersymmetry breaking on the gauge theory. The prefactor is: 
\be
c_F=\frac{4}{g_{YM}^2}(2\pi \alpha ^\prime)^{-\frac{1}{2}}\cN _F
\ee
Notice that the mass term for 
the two different chiralities are complex conjugate of each other: $T^{IASD}=-iG^{IASD}$ and 
$T^{ISD}=i(G^{IASD})^*$, which a consequence of the Majorana condition that the four dimensional spinors 
inherit from the Majorana-Weyl condition of the fermions in the original ten dimensional theory. 
If we decompose $G^{IASD}$ as in (\ref{decIASD}), we see that a $G$-flux of type $(1,2)_P$ gives mass to the 
three gauginos transforming non-trivially under $SU(3)$ but keeps the $SU(3)$-singlet gaugino massless, thus 
preserving $\cN=1$ upersymmetry. A $G$-flux of type (3,0) or $(2,1)_P$ gives mass also to the
$SU(3)$-singlet gaugino.

Using (\ref{DinsT}) one finds the coupling of fluxes to D-instantons:
\be
\cA_{D(-1)}=\frac{2\pi i}{3!}c_F(\Theta)[\Theta ^{\alpha A}\Theta _\alpha ^{\,\,\,\,B}
(\overline{\Sigma}^{mnp})_{AB}G^{IASD}_{mnp} + \bar{\Theta}_{\dot{\alpha}A}\bar{\Theta}^{\dot{\alpha}}_{\,\,
\, B}(\Sigma ^{mnp})^{AB}G^{ISD}_{mnp}] 
\label{AD(-1)}
\ee
where 
\be
c_F(\Theta)=\frac{8\pi ^2}{g_{YM}^2}\cN _\Theta ^2\cN _F
\ee
From (\ref{AD(-1)}) one observes that both the IASD and the ISD components of the $G$-flux couple to the
D-instanton fermions. The couplings are different and independent for the two chiralities since they are not 
related by complex conjugation, as always in Euclidean spaces. Comparing $\cA_{D3}$ and $\cA_{D(-1)}$ shows 
that ISD $G$-flux does not give a mass to any gauginos but instead induces a `mass' term for the
anti-chiral instanton zero-modes which are therefore lifted. This effect may play a crucial role in 
discussing the non-perturbative contributions of the so-called `exotic' D-instantons for which the neutral 
anti-chiral zero modes $\bar{\Theta}_{\dot{\alpha}A}$ must be removed or lifted by some mechanism. 
Introducing ISD $G$-flux is one of such mechanisms.

Twisted fermions stretching between D3-brane and D-instanton represent the charged (or flavored) fermionic 
moduli of the $\cN=4$ ADHM construction of instantons and are usually denoted as $\mu$ and $\bar{\mu}$ 
depending on the orientation. The R-R 3-form flux $F_{mnp}$ couplings with these twisted fermions is given 
by:
\be
\cA _F\sim \bar{\mu}^A \mu ^B (\overline{\Sigma}^{mnp})_{AB} F^{IASD}_{mnp}
\label{afinst}
\ee
The couplings to NS-NS 3-form flux $H_{mnp}$ is:
\be
\cA _H\sim \bar{\mu}^A \mu ^B (\overline{\Sigma}^{mnp})_{AB} H^{IASD}_{mnp}
\label{ahinst}
\ee
Collecting (\ref{afinst}) and (\ref{ahinst}) and reinstating the appropriate normalizations, one finds
\be
\cA_{D3/D(-1)}\equiv \cA _F+\cA _H=\frac{4\pi i}{3!}c_F(\mu)\bar{\mu}^A \mu ^B 
(\overline{\Sigma}^{mnp})_{AB}G^{IASD}_{mnp}.
\label{AD3D(-1)}
\ee
where $c_F(\mu)=\cC_{(0)}\cN_\mu\cN_{\bar{\mu}}\cN_F$ with $\cC_{(0)}=\frac{8\pi ^2}{g_{YM}^2}$. Notice that 
$\mu$ and $\bar{\mu}$ are distinct and independent quantities. This amplitude together with $\cA_{D(-1)}$ 
accounts for the flux induced fermionic couplings on the D-instanton effective action.
(\ref{AD(-1)}) and (\ref{AD3D(-1)}) describe deformations of the instanton moduli space of $\cN=4$ gauge 
theory living on the worldvolume of D3-branes. An extension to less supersymmetric cases can be done 
straightforwardly. For pure $\cN=1$ SYM the flux couplings for gauge and exotic instantons follow from 
(\ref{AD3}), (\ref{AD(-1)}) and (\ref{AD3D(-1)}) by restricting the spinor components to $A=B=0$. 
Contributions to fermionic mass terms come only from the components $G_{3,0}$ and $G_{0,3}$ related to the 
soft symmetry breaking gaugino and gravitino masses. For $\cT _6 / (\mZ_2 \times \mZ_2)$ these masses are:
\bea
&&|m_\Lambda |=|4\frac{e^{\varphi /2}}{\nu}G_{(3,0)}|\\
&&|m_{3/2}|=|4\frac{e^{\varphi /2}}{\nu}G_{(0,3)} |
\eea
where $\varphi$ is the dilaton, $\nu$ is the volume of the $\cT _6 / (\mZ_2 \times \mZ_2)$ orbifold.
The fermionic flux couplings
\bea
&&\cA _{D3}=-\frac{i}{16\pi}m_\Lambda e^{-\varphi}\Tr [\Lambda ^\alpha \Lambda _\alpha]+c.c.\\
&&\cA _{D(-1)}=-i\pi m_\Lambda e^{-\varphi}\theta ^\alpha \theta _\alpha +\frac{i\pi}{8}(2\pi\alpha 
^\prime)^2 m_{3/2}e^{-\varphi}\lambda _{\dot{\alpha}}\lambda ^{\dot{\alpha}}\\
&&\cA _{D3/D(-1)}=-\frac{i}{8}m_\Lambda \bar{\mu}_u\mu ^u
\eea
modify the zero mode structure of the instanton and allow for new low energy coupling in the D3-brane 
action. In the above formulae $\Lambda ^\alpha$ is the gaugino, $\theta ^\alpha$ and $\lambda 
^{\dot{\alpha}}$ are two chiral and two anti-chiral zero modes respectively coming from the R sector of 
D(-1)/D(-1) strings. $\mu ^u$, $\bar{\mu}_u$ scalars come from the R sector of D3/D(-1) and D(-1)/D3 
strings.
One observes that the presence of $\lambda$-fermionic zero modes prevents the generation of
non-perturbative superpotentials via exotic instantons. One can overcome this difficulty by introducing an
O-plane. This leads to O(1)-instantons without $\lambda$-modes. In the case of oriented gauge theories, the 
presence of the $\lambda ^2$-term in amplitudes suggests that R-R and NS-NS fluxes can give an alternative 
mechanism.

\chapter{Worldsheet vs D-brane instantons}
\label{chapter4}

\section{Heterotic-Type I duality in $D\le 10$}

Perturbatively different string theories may be shown to be equivalent once non-perturbative effects are 
taken into account. Heterotic and Type I string theories with gauge group $SO(32)$, were conjectured in 
\cite{WitPol, PWDH}, to be equivalent. In fact, up to field redefinitions, they share the same low-energy
effective field theory. It was shown in \cite{EW} that for the equivalence to work,
the strong coupling limit of one should correspond to the
weak coupling limit of the other\footnote{Similar situations
in which the strong coupling limit of one string theory is the weak
coupling of another `dual' string theory were discussed earlier by
Duff \cite{Duff}.}. In $D = 10$ the strong - weak coupling duality takes
the following form \cite{WitPol, PWDH}
\bea g_s^H = 1/g_s^I,~~~
\alpha^\prime_H = g_s^I
\alpha^\prime_I\eea
where $g_s^H$, $\alpha^\prime_H$ and $g_s^I$, $\alpha^\prime_I$ are the heterotic and Type I coupling 
constants and tensions respectively. The simple strong-weak coupling duality $\phi_I = -\phi_H$ in $D=10$ 
changes significantly in lower dimensions. Indeed, since the dilaton belongs to the universal sector of the 
compactification, the relation between the heterotic and Type I dilatons in $D$ dimensions is determined by 
dimensional reduction to be \cite{ABPSS}, \cite{ABFPT} 
\bea \phi_I^{(D)} = {(6-D) \over 4} \phi_H^{(D)} - {(D-2) \over 16} \log\det{G_H^{(10-D)}}\eea
where $G_H^{(10-D)}$ is the internal metric in the heterotic-string frame, and there is a crucial sign 
change at $D=6$ where $\phi_H$ and $\phi_I$ are independent \cite{X}. It is well known that Type I models 
exist with different number of tensor multiplets in $D=6$ \cite{MSGP, DPGJ}. This does not have an analogue 
in perturbative heterotic compactifications on $K3$. In $D=6$, the Type I dilaton belongs to a 
hypermultiplet to be identified with one of the moduli of the $K3$ compactification on the heterotic side. 
In four dimensional $\cN=1$ models on both sides the dilaton appears in a linear
multiplet, and heterotic-type I duality is related to chiral-linear duality. The presence of anomalous 
$U(1)$'s under which R-R axions shift suggests that the latter correspond to changed scalars on the 
heterotic side.

Heterotic-Type I duality requires that the heterotic fundamental string and the Type I D-string be 
identified. The massless fluctuations of a Type I D-string are eight bosons and eight negative chirality
fermions in the D1-D1 sector together with 32 positive chirality fermions in the D1-D9 sector. Thus, the 
world-sheet of the D-string exactly matches the world-sheet of the Heterotic fundamental string. By the same 
token, the Type I D5-brane should be identified with the heterotic NS5-brane. The latter is a soliton of the 
effective low-energy heterotic action and its microscopic description is not fully understood. The tensions 
agree in the two descriptions since $T_{NS5} = 1/(g_s^H)^2 (\ap_H)^3 \equiv 1/g_s^I (\ap_I)^3 = T_{D5}$.

$SO(32)$ Heterotic / Type I duality has been well tested in $D=10$ and in toroidal compactifications. In 
$D=10$ BPS-saturated terms, like $F^4$, $F^2R^2$ and $R^4$, are anomaly related and match in the two 
theories as a consequence of supersymmetry and absence of anomaly. In toroidal compactifications, the
comparison of BPS-saturated terms becomes more involved. The spectra of BPS states  become richer and differ 
on the two sides at the perturbative level.

Non-perturbative corrections to $F^4$, $F^2R^2$ and $R^4$ terms are due to instantons that preserve half of 
the supersymmetry. In the heterotic string they get perturbative corrections at one loop only and the
NS5-brane is the only relevant non-perturbative configuration in $D\leq 4$. Instanton configurations can be
provided by taking the world-volume of the NS5-brane to be Euclidean and to wrap supersymmetrically around a 
compact manifold, so as to keep finite the classical action. This requires at least six-dimensional compact 
manifold. Therefore, BPS-saturated terms do not receive non-perturbative corrections for toroidal 
compactifications with more than four non-compact directions. Thus, the full heterotic result arises from 
tree level and one loop for $D>4$. In the Type I string both D1- and D5-branes can provide instanton 
configurations after Euclideanization. D5-brane will contribute in four or less noncompact dimensions,
D1-brane can contribute in eight or fewer noncompact dimensions. Thus, in nine dimensions the two theories
can be compared in perturbation theory. In eight dimensions the perturbative heterotic result at one-loop 
corresponds to perturbative as well as nonperturbative Type I contributions coming from the D1-instanton via 
duality. The heterotic results can be expanded and the Type I instanton terms can be identified.
The classical action can be written straightforwardly and it matches with the heterotic result. The 
determinants and multi instanton summation can also be performed in the Type I theory. In
general, world-sheet instantons in heterotic string duals of Type I models help clarifying the rules for 
multi-instanton calculus with unoriented D-branes. Two prototypical examples are the $T^4/\mZ_3$ orbifold to 
$D=4$, that we have already encountered \cite{ABPSS}, and the $T^6/\mZ_2$ orbifold to $D=6$ \cite{MSGP},
that we are going to discuss in the following.

\section{Compactification on $T^4/\mZ_2$ to $D=6$}

\subsubsection{Type I description}

The Type I theory is an un-oriented projection of the Type IIB theory. Upon compactification on $T^4/\mZ_2$ 
to $D=6$, the Type IIB theory has $\cN =(2,0)$ spacetime supersymmetry with 16 supercharges, \ie those 
satisfying $Q = R Q$, where $R$ denotes the inversion of the four coordinates of $T^4$. The $\Omega$
projection preserves only the sum of left- and right-moving supersymmetries $Q_\alpha +\tilde{Q}_\alpha$. 
The $\Omega R$ projection preserves the same linear combination since
$ Q_\alpha+ R \tilde{Q}_\alpha \equiv Q_\alpha +\tilde{Q}_\alpha$. The massless little group in six 
dimensions is $SO(4)=SU(2)\times SU(2)$. The massless bosonic content of the unoriented closed
string spectrum contains in untwisted NS-NS sector ({\bf 3,3})+11({\bf 1,1}), in the untwisted R-R sector 
({\bf 3,1})+({\bf 1,3})+6({\bf 1,1}), in the twisted NS-NS sector there are 48 ({\bf 1,1}) and in the 
twisted R-R sector 16 ({\bf 1,1}). This is exactly the bosonic content of the $D=6$ $\cN=(1,0)$ supergravity
coupled to one tensor and 20 hypermultiplets.

Let us now discuss the unoriented open string spectrum. Tadpole cancellation conditions imply that the total 
Chan-Paton dimensionalities of twisted and untwisted sectors both equal to 32. The $U(16)_9\times U(16)_5$ 
model, which arises at the maximally symmetric point, where all the D5-branes are on top of an $\Omega$5-
plane and no Wilson lines are turned on the D9-branes, is of particular interest. This model was first 
discussed by  Bianchi and Sagnotti and later by Gimon and Polchinski in \cite{MSGP}. The D9-D9 sector 
contributes a vector multiplet in the adjoint of $U(16)_9$ and hypermultiplets in the ${\bf 120}_{+2} +{\bf 
120}^*_{-2}$. The D5-D5 gives a vector multiplet in the adjoint of $U(16)_5$ and the hypermultiplet in the 
${\bf \tilde{120}}_{+2} +{\bf\tilde{120}}^*_{-2}$. In the D5-D9 `twisted'\footnote{In the sense that the 4 
N-D directions have half-integer bosonic modes.} spectrum there are half-hypers in the
$({\bf 16}_{+1}, {\bf \tilde{16}}^*_{-1})+({\bf 16}^*_{-1}, {\bf \tilde{16}}_{+1})$ of
$U(16)_9\times U(16)_5$.

\subsubsection{Compactification on $T^4/\mZ_2$ to $D=6$: Heterotic
description}

The Type I model corresponds to a compactification without vector structure \cite{19}, \cite{Berkooz}:
\bea \tilde\omega_{2,YM}^{SW} \neq 0 \approx B_2^{NS-NS} = 1/2 (mod~1)\eea
where $\tilde\omega_{2,YM}^{SW}$ is modified second Stieffel-Whitney class (obstruction to vector 
structure).

The $\mZ_2$ orbifold (besides its geometrical action) acts on the 32 heterotic fermions as
$\lambda_{ws}^A \rightarrow (i) \lambda_{ws}^u, (-i)\lambda_{ws}^{\bar{u}} $, which breaks the gauge group 
$SO(32)$ to $U(16)$. The resulting massless spectrum is as follows. In the untwisted sector we have four 
neutral hypers, charged hypers in ${\bf 120}_{+2} +{\bf 120}^*_{-2}$, vector in the adjoint, one tensor and 
the $\cN =(1,0)$ supergravity multiplet.

The twisted sector (16 fixed points) does not contain neutral hypermultiplets, it has charged half 
hypermultiplets in the ${\bf 16}_{-3}+{\bf 16}^*_{+3}$.

\subsubsection{Matching the spectrum}

In order to match the massless spectrums of the two descriptions one has to distribute one `fractional' D5-
brane per each fixed point, thus breaking the D5-brane gauge group $U(16)_5 \rightarrow U(1)_5^{16}$ 
\cite{Berkooz}.

In six dimensions the full gauge plus gravitational anomaly can be written as \cite{Anomalies&Tadpoles}

\be \cI_8 = \sum_i \left( X^i_2 \wedge X^i_6 + X^i_4 \wedge \tilde{X}^i_4\right)\ee
The GSS counterterm reads as $L_{GSS}=C^{RR}_2 X_4 + \sum_f C^{RR}_{0,f} X^f_6$, so that Type I photons 
become massive by eating twisted RR axions:
$\partial C_{0,f}^{RR} \rightarrow D C_{0,f}^{RR} = \partial C_{0,f}^{RR} + 4 A^{(9)} + A_f^{(5)}$.
The Type I combination $A^I = A^{(9)} - 4 \sum_f A_f^{(5)}$ decouples from twisted
closed string scalars and matches with the heterotic photon $A^{H}$.
The vector multiplets get massive by eating neutral closed string hypers. Thus we have a supersymmetric 
Higgs-like mechanism: full hypers are eaten\footnote{This is an efficient, not fully exploited mechanism for 
moduli stabilization even in $D=4$. The remnant of the $D=6$ anomaly in $D=4$ is massive `non-anomalous' 
$U(1)$'s \cite{Pascal}.}.

\section{Duality and dynamics in $D=6$}

In order to further test the correspondence and gain new insights
into multi D-brane instantons, we are going to consider a
four-hyperini Fermi type interaction that is generated by
instantons and corresponds to a `chiral' (1/2 BPS) coupling in the
$\cN=(1,0)$ low energy effective action. If the four hyperini are
localized at four different fixed points, this coupling is absent
to any order in perturbation theory. This is so, because twisted
fields at different fixed points do not interact perturbatively.
ED1-brane or worldsheet instantons which connect the four fixed
points can generate such a term. The contributions will be
exponentially suppressed with the area of the cycle wrapped by the
instanton.

Let us mention what kind of corrections one expects in the two descriptions before describing the 
computation. In $D=6$ Heterotic / Type I duality implies
\bea \phi_{_H} = \omega_{_I} \qquad , \qquad \phi_{_I} = \omega_{_H} \eea
where  $\phi$ is the dilaton and  $\omega$ is the volume modulus.
Supersymmetry implies that there are no neutral couplings between vectors and hypers.
The gauge couplings can only depend (linearly) on the scalar
$\phi_{_H} = \omega_{_I}$ in the unique tensor multiplet, while
$\phi_{_I} = \omega_{_H}$ belongs to a neutral hyper. For these reasons in the heterotic description the 
hypermultiplet geometry is tree-level exact, but may get worldsheet instanton corrections
$e^{-h(C)/\alpha^\prime}$, where neutral hypers $h$ determine the size of
2-cycles $C$ in $T^4/\mZ_2$. In the Type I description, hypers receive both perturbative (string
loops) and non-perturbative corrections from BPS Euclidean
D-string instantons wrapping susy 2-cycles $C$ in $T^4/\mZ_2$.
The Type I gauge couplings are completely determined by disk amplitudes.
In the heterotic string, they receive (only) a one-loop correction.

\section{Four-hyperini amplitude}

\subsubsection{Computational strategy}

Let us summarize our strategy:
\begin{itemize}

\item Focus on a specific 4-hyperini amplitude
\bea \cA^{f_1 f_2 f_3 f_4}_{4hyper} = \langle V^{\zeta, f_1}_{\bf 16}V^{\zeta, f_2}_{{\bf 16}^*} V^{\zeta,
f_3}_{\bf 16} V^{\zeta, f_4}_{{\bf 16}^*} \rangle \eea absent at tree level for
particular choices of fixed points

\item Compute $\cA^{f_1 f_2 f_3 f_4}_{4hyper}$ in the limit of
vanishing momenta

\item Start with heterotic string, where it is tree level exact and
extract worldsheet instanton corrections

\item Translate into Type I language and interpret the result in
terms of perturbative and non-perturbative contributions

\item Learn new rules for unoriented multi D-brane instantons

\end{itemize}

\subsubsection{Heterotic description}

To compute the four-hyperini Fermi interaction in the heterotic description we need the hyperini vertex 
operators
\bea V^\zeta_{{\bf 16}/{\bf
16}^*} = \zeta^{u/\bar{u}}_{f,a}(p) S^a e^{-\varphi/2}(z)
\tilde\Sigma^{{u/\bar{u}}}(\bar{z}) \sigma_f e^{ipX}(z,\bar{z})\eea
where $\sigma_f$ is the bosonic $\mZ_2$-twist field ($h= 1/4$),
$\tilde\Sigma^{{u/\bar{u}}} = :e^{\pm i \tilde\phi_u}\prod_v
e^{\mp i \tilde\phi_v/4}:$ are twisted ground-states ($h=3/4$) for
heterotic fermions $\tilde\lambda^{u/\bar{u}}$, $S^a$ are $SO(5,1)$ spin fields, $\varphi$ and 
$\tilde\phi_u$ are the bosonizations of the superghost and $SO(32)$ gauge fermions respectively.
One can use $SL(2,\mC)$ invariance on the sphere to set $z_1\rightarrow \infty$,
$z_2\rightarrow 1$, $z_3\rightarrow z$, $z_4\rightarrow 0$ with
cross ratio $z = z_{12} z_{34} / z_{13} z_{24}$. Then the string amplitude will depend on the $SL(2,\mC)$ 
invariant cross ratio $z$.

The $\mZ_2$-twist field correlator is given by \cite{DFMS}
\bea \langle \prod_{i=1}^4 \, \sigma_{f_i} (z_i,\bar{z}_i)
 \rangle \rightarrow
 |z_\infty|^{-1} \,\Psi_{qu}(z,\bar z)  \,\Lambda_{cl}
 \left[^{\vec{f}_{12}}
 _{\vec{f}_{13}}
 \right](z,\bar z)\label{z4twist} \eea
The quantum part $\Psi_{qu}$ is independent of the twist-fields locations \ie
of the choice of 4 out of 16 fixed points $\vec{f}_i= 1/2(\epsilon^1_i,
\epsilon_i^2, \epsilon_i^3, \epsilon_i^4)$ with $\epsilon_i^a=
0,1$ and in order to get a non-trivial coupling the $\vec{f}_i$ should satisfy $\sum_i \vec{f}_i=\vec{0}$  
mod $\Lambda(T^4)$. $\Lambda_{cl}=\sum e^{-S_{inst}}$ is the classical part accounting for worldsheet
instantons depending on the ralative positions $\vec{f}_{ij}=\vec{f}_i-\vec{f}_j$. The $\mZ_2$-twist field 
correlator can be mapped into the torus doubly covering the sphere with two $\mZ_2$ branch cuts using
the relation between the cross-ratio $z$ and the Teichm\"uller parameter of the torus $\tau (z)$
\be z = {\vartheta_3^4(\tau) /\vartheta_4^4 (\tau) } .\ee
The quantum and classical parts of the 4-twist correlator read
\bea \Psi_{qu} (z,\bar z)= 2^{-{8/3}}\,  |z(1-z)|^{-{1/3}}\, \tau_2^{-2}|\eta(\tau)|^{-8} \eea
\bea \Lambda_{cl} \left[^{\vec{f}_{12}}_{\vec{f}_{13}}\right](z,\bar z)=  \sum_{\vec{m}, \vec{n}} e^{ 
-{\pi\over\tau_2(z)} ( \vec{m}+\vec{n}\tau + \vec{f}_{13}+
\vec{f}_{12}\tau ) \cdot (G+B)\cdot ( \vec{m}+\vec{n}\bar\tau + \vec{f}_{13}+ \vec{f}_{12}\bar\tau )} \eea
where $G_{ij}$ is the metric and $B_{ij}$ is the antisymmetric tensor of $T^4/\mZ_2$
(neutral hypers). Writing the $z$-integral as integral over the torus modulus $\tau$ (for $s,t\to 0$) one 
finds
\bea {\cal A}^{f_1,f_2,f_3,f_4}_{u_1 \bar u_2  u_3 \bar{u}_4}
= \cV(T^4) \int_{\cF_2}  {d^2\tau \over \tau_2^2}
\left( {\bar \vartheta_4^4 \over \bar\vartheta_3^4 } \delta_{u_1 \bar{u}_2} \delta_{u_3 \bar u_4}
- {\bar \vartheta_4^4 \over \bar\vartheta_2^4 }\,
\delta_{u_1 \bar{u}_4} \delta_{u_3 \bar u_2}\right)\,
\Lambda_{cl} \left[^{\vec{f}_{12}}_{\vec{f}_{13}}\right].\nn\\ \eea
The integral goes over the fundamental domain $\cF_2$ of the index 6 subgroup $\Gamma_2$ of
$SL(2,\mZ)$, leaving invariant $\vartheta_{even}$ \cite{KM}. The region $\cF_2$ can be decomposed into 6 
domains each of which is an image of the fundamental domain $\cF$  of $SL(2,\mZ)$ under the action of the 6 
elements of $SL(2,\mZ)/\Gamma_2$
\bea \int_{\cF_2} {d^2\tau\over \tau_2^2} \, \Phi(\tau,\bar{\tau})=
  \int_{\cF} {d^2\tau\over \tau_2^2} \, \sum_{s=1}^6 \, \Phi(\tau_s,\bar{\tau}_s)\label{f2f} \eea
where $\tau_s=\gamma_s(\tau)$, $\gamma_s =\{1,S,T,TS,ST,TST\}$.
For the 4-hyperini amplitude one gets
\bea 
\Phi(\tau,\bar{\tau})=  \left( {\bar \vartheta_4^4 \over \bar\vartheta _3^4 } \delta_{u_1 \bar{u}_2} 
\delta_{u_3 \bar u_4} -{\bar \vartheta_4^4 \over \bar\vartheta_2^4 }\,\delta_{u_1 \bar{u}_4} \delta_{u_2 
\bar u_3}\right)\, \Lambda_{cl} \left[^{\vec{f}_{12}} _{\vec{f}_{13}} \right] .\eea

In the special case when all 4-hyperini are located at the same fixed point 
$\vec{f}_{12}=\vec{f}_{13}=(\vec{0})$, the amplitudes receive contribution only from BPS-like modes as in 
Type I (see later). The instanton sum $\Lambda_{cl}\left[^{\vec{0}}_{\vec{0}}\right]$ is modular invariant. 
Sums over 6 images produce
\bea \sum_{s=1}^6 \, {\bar \vartheta_4^4 \over \bar\vartheta_3^4 } (\bar{\tau}_s)=3 \quad , \quad
\sum_{s=1}^6 {\bar \vartheta_4^4 \over \bar\vartheta_2^4 } (\bar{\tau}_s)= -3 \eea
and the final expression for the amplitude with $\vec{f}_1= \vec{f}_2 = \vec{f}_3 = \vec{f}_4$ is given by
\bea
{\cal A}^{f_1,f_1,f_1,f_1}_{u_1 \bar u_2  u_3 \bar{u}_4} = 3 (\delta_{u_1 \bar{u}_2} \delta_{u_3 \bar u_4}
+\delta_{u_1 \bar{u}_4} \delta_{\bar{u}_2 u_3})\, \cV(T^4) \int_{\cF} {d^2\tau\over \tau_2^2} \,
\Lambda_{cl} \left[^{\vec{0}}_{\vec{0}} \right]. \eea
Next consider the case when hyperini are located in pairs at two different fixed point:
\begin{itemize}
\item $\vec{f}_{12}=\vec{f}_{13} = \vec{f}$ for $ \delta_{u_1 \bar{u}_2}
\delta_{u_3 \bar u_4}$ structure in 4-hyperini amplitude
\bea {\cal A}^{f_1,f_2,f_2,f_1}_{u_1 \bar u_1  u_3 \bar{u}_3}
 =\cV(T^4) \int_{\cF} {d^2\tau\over \tau_2^2} \, \left( \Lambda_{cl}
\left[^{\vec{f}}_{\vec{f}}\right] +\Lambda_{cl}\left[^{\vec{f}}_{\vec{0}}\right]
 +\Lambda_{cl} \left[^{\vec{0}}_{\vec{f}}\right] \right). \eea
\item $\vec{f}_{12}=\vec{0}, \vec{f}_{13}=\vec{h}$ for
$\delta_{u_1 \bar{u}_4} \delta_{u_3 \bar u_2}$ structure in
4-hyperini amplitude
\bea{\cal A}^{f_1,f_1,f_3,f_3}_{u_1 \bar u_2  u_2 \bar{u}_1} = \cV(T^4)
\int_{\cF} {d^2\tau\over \tau_2^2} \, \left( \Lambda_{cl}
\left[^{\vec{0}}_{\vec{h}}\right] +\Lambda_{cl}
\left[^{\vec{h}}_{\vec{0}}\right]
+\Lambda_{cl} \left[^{\vec{h}}_{\vec{h}}\right]\right). \eea
\end{itemize}
These are the same integrals as for BPS saturated thresholds to $F^4$ in $T^4$
compactifications (with shifts) \cite{KM}. Since the pieces proportional to
$\delta_{u_1 \bar{u}_2} \delta_{u_3 \bar u_4}$
and $\delta_{u_1 \bar{u}_4} \delta_{u_3 \bar u_2}$ are related by a simple relabeling of the fixed points 
$f_i$'s, we have restricted our attention onto the amplitude with color structure $\delta_{u_1 \bar{u}_2} 
\delta_{u_3 \bar u_4}$ for the first and the amplitude with color structure $\delta_{u_1 \bar{u}_4} 
\delta_{u_3 \bar u_2}$ for the second case. Performing Poiss\`on resummation over $\vec{m}$ in $\Lambda 
_{cl}$ one finds
\bea
\Lambda_{cl} \left[^{\vec{0}}_{\vec{f}}\right] = {\tau_2^2 \over
\cV(T^4)} \sum_{\vec{k}, \vec{n}} (-)^{2\vec{f}\cdot\vec{k}}
q^{p_L^2/2} \bar{q}^{p_R^2/2}\eea
where $\vec{p}_{L/R} = {1\over \sqrt{2}} (E^{-1}\vec{k} + E^t
\vec{n}) $ , $E E^t = G$ and we have set $B=0$ for simplicity. One can recognize the shifted orbifold 
partition function
\bea \Lambda_{cl} \left[^{\vec{0}}_{\vec{f}}\right] (G) +\Lambda_{cl}\left[^{\vec{f}}_{\vec{0}}\right](G)
+\Lambda_{cl} \left[^{\vec{f}}_{\vec{f}}\right](G)=2 \Lambda_{cl} \left[^{\vec{0}}_{\vec{0}}\right] 
(G_{(\vec{f})}) -\Lambda_{cl} \left[^{\vec{0}}_{\vec{0}}\right] (G).\nn\\ 
\eea
The toroidal metric $G_{(\vec{f})}$ is `halved' along the direction $\vec{v} = 2\vec{f}$ by $SO(d,d)$ 
transformation. This is similar to what one gets for the threshold corrections to $F^4$ terms in toroidal  
compactifications.

\subsubsection{Type I description}

As we already noticed, in the Type I description hypers can receive both perturbative and non-perturbative 
corrections since the dilaton belongs to a hypermultiplet. Some scattering amplitudes may vanish in 
perturbation theory and receive only contributions from non-perturbative effects. The four-hyperini Fermi 
interaction term does not get perturbative contributions for $f_i$ all different from one another or for 
$f_1=f_3$ (same charge). When all fixed points $f_i$ are equal or are equal in pairs $f_1=f_2$ or $f_2=f_3$ 
(opposite charge) there is a perturbative correction which matches with the contribution of the degenarate 
orbit in the heterotic description.

The open string vertex operators are given by
\bea V^\zeta_{\bf 16} = \zeta^f_a(p) S^a e^{-\varphi/2}
\sigma_f e^{ipX} \Lambda^u_f \qquad V^\zeta_{{\bf 16}^*} =
\zeta^f_a(p) S^a e^{-\varphi/2} \sigma_f e^{ipX} \bar\Lambda^f_{u}\eea
They involve Chan-Paton matrices $\Lambda ^u_f$ in the bifundamental of $U(16)\times U(1)_5^{16}$ rather 
than heterotic fermions $\lambda$ yielding
\bea Tr(\Lambda^{u_1}_{f_1} \Lambda^{\bar{u}_2}_{f_2}
\Lambda^{u_3}_{f_3}\Lambda^{\bar{u}_4}_{f_4}) = \delta_{f_1f_2}
\delta_{f_3f_4} \delta^{u_1\bar{u}_4} \delta^{u_3\bar{u}_2} +
\delta_{f_1f_4} \delta_{f_3f_2} \delta^{u_1\bar{u}_2}
\delta^{u_3\bar{u}_4}. \eea

Consider $Z_2$-twist field correlator for open strings with 4 N-D boundary
conditions \cite{CPKW}. The quantum part of the 4-twist correlator, which is independent of location $f_i$ 
of twist fields, is given by
\bea \Psi_{qu} = [x(1-x)]^{-1/3} t(x)^{-2} \eta(it)^{-4} \eea
where $ x ={\vartheta_3^4(it) / \vartheta_4^4 (it) }$ is $SL(2,R)$ invariant ratio with $t$ the modular 
parameter of the annulus doubly covering the disk. The classical part from exchange of (massive) open string 
modes stretched between (different) fixed points is
\bea \Lambda_{cl} [^{f_{12}}_{f_{13}}]= \delta^{f_1}_{f_2} \delta^{f_3}_{f_4}
\sum_{\vec{n}} e^{- {\pi t} (\vec{n}+\vec{f}_{14})^t G
(\vec{n}+\vec{f}_{14})} + \delta^{f_1}_{f_4} \delta^{f_3}_{f_2}
\sum_{\vec{n}} e^{- {\pi t} (\vec{n}+\vec{f}_{12})^t G
(\vec{n}+\vec{f}_{12})}.\nn\\
\eea
Plugging into the open string amplitude and taking the limit $s,t\rightarrow 0$ one finds a perfect 
agreement with heterotic degenerate orbits, which is independent of $B_2^H \approx C_2^{R-R}$ but only on 
$G$ and $\phi$ (recall $\omega_H = \phi_I$, $\phi_H = \omega_I$). Terms involving $B_2^H$  have no disk 
counterpart in the Type I description since the dual $C_2^{RR}$ couples to (E)D-strings.

\subsubsection{ED-string corrections}

We then consider non-perturbative corrections to the four-hyperini coupling in the Type I description.
When the four hyperini are located at the different fixed points $f_i$ we have only non-perturbative 
contribution from (regular) ED-strings wrapping supersymmetric (untwisted) two cycles
$C \approx T^2/\mZ_2 = S^2$, passing through the four fixed points. Fractional ED-strings wrapping the 16 
collapsed `rigid' 2-cycles (since corresponding moduli are eaten by anomalous $U(1)_{5}^{16}$) may 
contribute to amplitudes having also perturbative contributions.

Using by now the well established Heterotic / Type I duality we will deduce
the `exact' 4-hyperini amplitude and determine ED-string corrections. Then we interpret these corrections in 
terms of symmetric orbifold CFT \cite{GMB}.

Let us describe the spectrum of ED-strings. The instanton dynamics is governed by a gauge theory describing 
the excitations of unoriented strings connecting E1, D5 and D9 branes. Three sectors of open string 
excitations are
\begin{itemize}
\item E1-E1 strings (2 N-N, 8 D-D): $X^I$, $S^a$,
$\tilde{S}^{\dot{a}}$ with $I=1,...,8_v$, $a, \dot{a}=1,...,4$
\item E1-D9 strings (2 N-N, 8 N-D):
$\lambda^u, \lambda_{\bar{u}}$ with $u, \bar{u} = 1, ..., 16$
\item E1-D5 strings (2 D-D, 8 N-D): $\mu^f, \mu_{\bar{f}}$ with $f=f_1, f_2, f_3, f_4$
\end{itemize}

Alternatively, after T-duality along the wrapped 2-cycle we will have $E1
\rightarrow E(-1)$, $D9 \rightarrow D7_9$, $D5 \rightarrow D7_5$.
The residual (super)symmetry of the spectrum is
\bea \cN =(8,0) \rightarrow \cN = (4,0).\eea
And the spacetime symmetry breaks according to
\bea 
SO(9,1) \rightarrow SO(5,1)\times SU(2) \times SU(2) \rightarrow SO(5,1) \times SO(2)_E\times SO(2).\nn\\
\eea
$\cN = (4,0)$ gauge theory in IR flows to symmetric product CFT
\bea ({ R}^6\times T^4/\mZ_2)^k/S_k.\eea
ED-string wraps two-cycle $C$ inside $T^4/\mZ_2$ which is
specified by the two vectors $M_k=(\vec{k}_1,\vec{k}_2)$ each made out of four integers with greatest common 
divisor 1. $\vec{k}_{1,2}$ show how many times the two 1-cycles of $C$ wrap around 1-cycle of $T^4/\mZ_2$.

Heterotic vertex operator can be derived from interaction term with hyperino:
\bea \cL_{4F} = (\zeta_a)^u_f \mu^f_i \lambda^i_u S^a = V^H_\zeta \eea
Only $(\ell)^m$-twisted sectors (with $m\ell=k$ and ${Z}_{\ell}^s$
projection) with exactly four fermionic zero modes of $S^a$
contribute. So, we can fold the $k$ copies of fields and form a single field on a worldsheet with the 
following Kahler and
complex structures:
\bea \cT(M)=k \,\cT(M_k) \quad , \quad \cU(M)={m\,\cU(M_k)+s \over \ell}\eea
where $M=M_k \left( \begin{array}{cc}
l & s \\
0 & m  \\
\end{array} \right) $.
This is in perfect agreement with the heterotic result $\cI_{d,d}^{ndeg}$ for the four-hyperino coupling on 
$T^4/\mZ_2$.

This is the generalized Hecke transform as in the Heterotic computation!

\chapter{AdS/CFT correspondence and instantons}
\label{chapter5}
\section{Gauge theory/string theory dualities}
Gauge theory/string theory duality states that string theory or M-theory in the near-horizon geometry of a 
collection of coincident D-branes or M-branes is equivalent to the low-energy world-volume theory of the 
corresponding branes.
AdS/CFT dualities relate Type IIB superstring theory or M-theory compactified on an anti-de Sitter space-
time times a compact space to conformally invariant field theories. Anti-de Sitter space is a maximally 
symmetric space-time with a negative cosmological constant. The space-time manifold of the conformal field 
theory is associated with the conformal boundary of the AdS. This boundary lies at the 
infinity of the AdS space-time. Usually AdS/CFT dualities are such that when one description is weakly 
coupled, the dual descripiton is strongly coupled. Using an 
information in the weakly coupled theory 
allows to learn non trivial facts about strongly coupled dual theory. There are three basic examples of 
AdS/CFT duality. They all have maximal supersymmetry (32 supercharges). Superconformal field theory on the 
world-volume of $N$ parallel D3-branes corresponds to the type IIB theory on $AdS_5\times S^5$. M-theory on 
$AdS_7\times S^4$ is dual to superconformal field theory on $N$ M5-branes. M-theory on $AdS_4\times S^7$ 
corresponds to superconformal field theory which lives on the world-volume of $N$ parallel M2-branes. In 
each case the sphere surrounds the branes. Each of these branes has one unit of the appropriate type of 
charge. Thus, the background has nonvanishing antisymmetric tensor gauge field with $N$ units of flux. 
Gauss's law requires that these fluxes thread the sphere.

\section{General aspects of $AdS_5\times S^5$}
Let us consider how the symmetries of bulk and boundary theories are related in this case.
The isometry group $SO(6)$ of $S^5$ and the isometry group $SO(4,2)$ of $AdS_5$ correspond resepctively to 
R-symmetry group and conformal group of the boundary theory. The radii of $AdS_5$ and $S^5$ are equal and 
are related to the 't Hooft parameter $\lambda =g_{YM}N$ of the gauge theory by $L=\lambda ^{1\over 4}l_s$ 
with $l_s=\sqrt{\alpha ^\prime}$. 
The dual theories have the same symmetry. In 
each case the supergroup is $PSU(2,2|4)$ which has a bosonic subgroup $SU(2,2)\times SU(4)$ and 32 fermionic 
generators transforming as $({\bf 4}, {\bf 4})+({\bf \bar{4}}, {\bf \bar{4}})$ under
$SU(2,2)\times SU(4)$. First let us consider the string theory side. The covering groups of $SO(4,2)$ and of 
$SO(6)$  are $SU(2,2)$ and $SU(4)$ respectively. 
Since there are fermions in the theory belonging to the spinor representations one needs to consider these 
covering groups. Thus, the bosonic subgroup is realized by the geometry. The 32 supersymmetries of type IIB 
superstring theory are realized as vacuum symmetries. The conserved charges transform as 
$({\bf 4}, {\bf 4})+({\bf \bar{4}}, {\bf \bar{4}})$ under $SU(2,2)\times SU(4)$ and together with spacetime 
isometries give $PSU(2,2|4)$. In the dual $\cN=4$ SYM theory $SU(4)$ symmetry is a global R-symmetry, which 
does not commute with supersymmetries. The four fermions of one chirality are in the ${\bf 4}$ 
representation and the others with opposite chirality are in the ${\bf \bar{4}}$ of $SU(4)$. The six scalar 
fields transform as ${\bf 6}$ of $SU(4)$. The 32 supersymmetries are also realized. So one gets $PSU(2,2|4)$ 
superconformal algebra.

As already mentioned for D3-branes AdS/CFT conjecture says that type IIB superstring theory on $AdS_5 
\times S^5$ is dual to $\cN=4$, $d=4$ super Yang-Mills theory with $SU(N)$ gauge group. It is known that in 
type II superstring theories the world-volume theory of $N$ coincident BPS D-branes is a maximally 
supersymmetric $U(N)$ gauge theory at the lowest order in $\alpha ^\prime$ and in the absence of background 
fields. The low energy effective action on the world-volume of $N$ coincident Dp-branes is given by 
dimensional reduction of supersymmetric $U(N)$ gauge theory in ten dimensions down to $p+1$ dimensions. A 
$U(N)$ gauge theory is equivalent to a free $U(1)$ vector multiplet times an $SU(N)$ gauge theory, up to 
some $\mZ_N$ identifications, which affect only global issues. In the dual stirg theory all modes interact 
with gravity, so there are no decoupled modes. Hence, the bulk AdS theory describes the $SU(N)$ part of the 
gauge theory. There are also some zero modes living in the region connecting the near horizon region (the 
`throat') with the bulk. These zero modes correspond to the above mentioned $U(1)$ degrees of freedom. The 
$U(1)$ vector supermultiplet includes six scalars related to the center of mass motion of all the branes. 
These zero modes live at the boundary from the AdS point of view. It seems one might or might not decide to 
include them in the AdS theory. Depending on this choice one could have a correspondence to an $SU(N)$ or 
$U(N)$ theory. The $U(1)$ center of mass degree of freedom is related to the topological theory of 
$B$-fields on AdS. If one imposes local boundary conditions for these $B$-fields at the boundary of AdS one 
finds a $U(1)$ gauge field living at the boundary, as in Chern-Simons theories. These modes living at the 
boundary are sometimes called singletons or doubletons.
There is a distinction between these two theories with $SU(N)$ and $U(N)$, which is actually a 
subleading effect in the large $N$-limit.

Gauge fields are the massless modes of open strings. Hence the super Yang-Mills and the open string coupling 
constants are the same. For Dp-branes with $p\neq 0$ the relation is the following: 
\be
g_{YM}^2=\frac{4\pi g_s}{V_{p-3}/(\alpha ^\prime)^{\frac{p-3}{2}}}
\ee
where $V_{p-3}$ is the volume of the cycle wrapped by the branes and where one has taken into account the 
relation between open and closed string coupling constants.
The 't Hooft coupling constant is: \be \lambda = g_{YM}^2N \ee which 
is held constant in the large $N$ expansion of the gauge theory. The D3-branes are sources for non-vanishing 
Ramond-Ramond five form field strength $F_5=\star F_5$. The five-sphere surrounds D3-branes and according to 
the Gauss's law there is a flux through $S^5$. There is a very nice fact of the duality saying that the rank 
of the gauge group (the numder of branes with unit charge) corresponds to the 
five-form flux through the five-sphere: \be \int _{S^5}F_5=N\ee 
Constant Ramond-Ramond axionic background $\tilde{C}^{(0)}$ is proportional to the Yang-Mills vacuum angle 
via \be 2\pi \tilde{C}^{(0)}=\theta _{YM}\ee Therefore, the complex Yang-Mills coupling is identified with 
the constant boundary value of the complex scalar field of the type IIB superstring 
\be 
\tau \equiv\frac{\theta _{YM}}{2\pi}+\frac{4\pi i}{g_{YM}^2}= \tilde{C}^{(0)}+\frac{i}{g_s}
\ee

Boundary values of the bulk superstring theory fields are in correspondence with guage invariant operators 
of four-dimensional boundary $\cN =4$ supersymmetric Yang-Mills theory. The lowest Kaluza-Klein modes of the 
graviton supermultiplet couple to the superconformal multiplet of Yang-Mills currents. To every Kaluza-Klein 
excitation one can associate a gauge-singlet composite Yang-Mills operator. The type IIB supergravity 
effective action evaluated on a solution of the equations of motion with boundary conditions is equal to the 
generating functional of connected gauge-invariant correlation functions in the Yang-Mills theory. The 
connection between bulk and boundary theory can be schematically given as
\be
\exp (-S_{IIB}[\Phi _m(J)])=\int DA \exp (-S_{YM}[A]+\cO _\Delta [A]J)
\ee
where $S_{IIB}$ is the effective action of the IIB superstring or its low energy supergravity limit,
$\Phi _m(J)$ denote boundary values of the bulk `massless' supergravity fields and their Kaluza-Klein 
descendents, $A$ denotes fluctuating boundary $\cN=4$ supersymmetric Yang-Mills fields and $\cO(A)$ is the 
set of gauge-invariant composite operators to which $J$ couples. Each field propagating on $AdS$ space has 
its correspondent operator in the field theory. Each field propagating on AdS space is in one to one 
correspodence with an operator in the field theory. There is a relation between the mass of the field 
and the conformal dimension of the corresponding operator in the conformal field theory. The mass-dimension 
formula changes for particles with spin. These formulae in $AdS_{d+1}$ with unit radius are the 
following:
\bea 
\begin{array}{ll}
\Delta _\pm =\frac{1}{2}(d\pm \sqrt{d^2+4m^2}) &  \text{scalars}\\
\Delta =\frac{1}{2}(d+2|m|) & \text{spinors} \\
\Delta _\pm =\frac{1}{2}(d\pm \sqrt{(d-2)^2+4m^2}) & \text{vectors} \\
\Delta =\frac{1}{2}(d\pm \sqrt{(d-2p)^2+4m^2}) & \text{p-forms} \\
\Delta =\frac{1}{2}(d+2|m|) & \text{first order d/2 forms (d even)} \\
\Delta =\frac{1}{2}(d+2|m|) & \text{spin-3/2} \\
\Delta = d &  \text{massless spin-2}
\end{array}
\eea
The choice $\Delta =\Delta _+$ is clear from the unitary bound.
Only the 
positive branch $\Delta =\Delta _+$ is relevant for the lowest `mass' supergravity multiplet. To compute 
correlation functions one needs bulk-to-boundary Green functions. These are specific normalized limits of 
bulk-to-bulk Green functions when one point is taken to the $AdS$ boundary. The precise forms of these 
propagators depend on the spin and mass of the field. For scalar field with conformal dimension $\Delta$ the 
normalized bulk-to-boundary Green function has the following form:
\be
G_\Delta(x,\rho ,\omega ;x^\prime , 0, \omega ^\prime)=c_\Delta K_\Delta (x^\mu , \rho ;x^{\prime \mu},0)
\label{green}
\ee
where $\omega$ are the coordinates on $S^5$ and $z^M\equiv (x^\mu, \rho)$  
are the $AdS_5$ coordinates, $M=0,1,2,3,5$ and $\mu =0,1,2,3$, $\rho\equiv z_5$ is the coordinate transverse 
to the boundary. 
\be 
c_\Delta =\frac{\Gamma (\Delta)}{\pi ^2 \Gamma (\Delta -2)}
\ee 
and
\be
K_\Delta (x^\mu , \rho ; x^{\prime \mu}, 0)=\frac{\rho ^\Delta}{(\rho ^2 +(x-x^\prime)^2)^\Delta}
\label{bulk-to-bound-prop}
\ee
Note that due to (\ref{green}) this Green function is independent of $\omega$. The bulk field is given by:
\be
\Phi _m(z;J)=c_\Delta \int d^4x^\prime K_\Delta (x^\mu , \rho ; x^{\prime \mu}, 0) J_\Delta (x^{\prime \mu})
\ee
In the limit $\rho \rightarrow 0$, $\rho ^{\Delta -4}K_\Delta$ reduces to the $\delta$-function on the 
boundary and one has
\be
\Phi _m(x,\rho ;J)\thickapprox \rho ^{4-\Delta}J_\Delta (x)
\ee
In particular, for the massless scalar ($\Delta_+=4$) in the limit $\rho \rightarrow 0$ the 
propagator reduces to $\delta ^{(4)}(x^\mu -x^{\prime \mu})$. It is crucial to observe that in the case 
$\Delta_+=4$ 
the propagator (\ref{bulk-to-bound-prop}) is exactly equal to $\Tr (F_{\mu\nu}^-)^2$, where $F_{\mu\nu}^-$ 
is the field strength corresponding to the Yang-Mills instanton with $\rho$ identified with instanton size.
On the other hand (\ref{bulk-to-bound-prop}) has precisely the same form as the five-dimensional profile of 
a D-instanton centered on the point $z^M$ and evaluated at the boundary point $(x^{\prime \mu},0)$. This 
allows to identify D-instanton effects of the bulk theory with those of YM instanton of the boundary 
theory. The correspondence between YM and D-instantons is nicely clarified in \cite{Bianchi:1998nk}, where 
the classical D-instanton solution of the Type IIB supergravity equations in the $AdS_5\times S^5$ is 
presented. It is shown that though D-instanotn solution does not affect the $AdS_5\times S^5$ geometry but 
it creates wormhole (in string frame) which leads to interesting modification of the geometry in the large 
instanton number limit.

\section{General aspects of $AdS_7\times S^4$}
There is a six dimensional superconformal field theory living on the world-volume of M5-branes and it is 
dual to M-theory compactified on $AdS_7\times S^4$. The isometry group of $AdS_7\times S^4$ metirc is 
$SO(6,2)\times SO(5)\thickapprox Spin(6,2)\times USp(4)$. After including supersymmetries one gets $OSp(6,2|
4)$ as a complete isometry superalgebra which contains 32 fermionic generators transforming as $({\bf 8}, 
{\bf 4})$ under $Spin(6,2)\times USp(4)$. The `mysterious' $\cN=(2,0)$ superconformal theory is related to 
SCFT in $d=4$ \'a la AGT \cite{Alday:2009aq} after compactification on Riemann surfaces. We will not discuss 
it anymore here.

\section{General aspects of $AdS_4\times S^7$}
A stack of M2-branes has $AdS_4\times S^7$ near-horizon geometry. M-theory on $AdS_4\times S^7$ is dual to 
three dimensional superconformal field theory. M-theory does not contain a dilaton field, which means that 
there is no weak-coupling limit. Hence the dual field theory is strongly coupled and as a result does not 
need to have a classical Lagrangian descripiton. One can think about this three dimensional conformal field 
theory in the following way. Remember that the low energy effective theory on the world-volume of $N$ 
coincident D2-branes of type IIA superstring theory is a maximally supersymmetric three dimensional Yang-
Mills theory with gauge group $U(N)$. Yang-Mills coupling in three dimensions is dimensionful and introduces 
a scale. This means this theory is not conformal. From the other side type IIA coupling constant is 
proportional to the radius of a circle on which eleventh dimension is compactified. When this coupling 
constant becomes large the gauge theory coupling constant also increases. This corresponds to going to 
the infrared in the gauge theory. Also the circular eleventh dimension is increasing. In the limit where the 
coupling constant becomes infinite one reaches the conformally invariant fixed-point theory describing a 
stack of coincident M2-branes in 11 dimensions. This theory has $SO(8)$ R-symmetry which corresponds to the 
rotations in the eight dimensions transverse to the M2-branes in 11 dimensions. The isometry group of 
$AdS_4\times S^7$ metric is $SO(3,2)\times SO(8)\thickapprox Sp(4)\times Spin(8)$. $SO(3,2)$ is the symmetry 
of $AdS_4$ and in the dual theory it corresponds to the conformal symmetry group. The isometry group $SO(8)$ 
of $S^7$ corresponds to the R-symmetry of the dual gauge theory. There are 32 conserved supercharges 
(maximal supersymmetry). In the dual gauge theory 16 of these supersymmetries are realized linearly and the 
other 16 are conformal supersymmetries. The isometry superalgebra becomes $OSp(8|4)$ once these 
supersymmetries are included. The 32 fermionic generators transform as $({\bf 8}, {\bf 4})$ under 
$Spin(8)\times Sp(4)$. It was mysterious how to get the dual $d=3$ SCFT.
ABJM have conjectured that 11-d supergravity on $AdS_4\times
S^7/\mZ_k$, corresponding to the near horizon geometry of $N$
M2-branes at a $\mC^4/\mZ_k$ singularity, be dual to $\cN=6$ CS
theory in $d=3$ with gauge group $U(N)_k\times U(N)_{-k}$ and
opposite CS couplings $k_1=k=-k_2$ \cite{Aharony:2008ug, Klebanov:2009sg}.
Remind that the near-horizon geometry of a stack of $N$ M2-branes is
$AdS_4\times S^7$ with $N$ units of $F_4$ flux along $AdS_4$ and
as many units of its dual $F_7$ along $S^7$ \cite{Malda}. The
metric reads \be ds^2_{11} = {1\over 4} L^2 ds^2_{AdS} + L^2
ds^2_{S^7} \ee for later use, note that $L_{AdS} = L/2$ with $L$
the radius of $S^7$ and henceforth the metrics of the subspaces are
for unit curvature radii.

\section{Supergravity description}

The Type IIA solution corresponding to the ABJM model reads \be
ds^2_{IIA} = 4 {\rho^2\over L^2} dx\cdot dx + 4 {L^2\over 4
\rho^2} d\rho^2 + L^2 ds^2_{\mC\mP^3} = {1\over 4} L^2 ds^2_{AdS} +
L^2 ds^2_{\mC\mP^3} \ee where \be L = \left(32\pi^2 N \over k
\right)^{1/4} \ee is the curvature radius in string units. The
string coupling, related to the VEV of the dilaton, is given by
\be g_s = L/k = \left(32\pi^2 N \over k^5 \right)^{1/4} \ee Thus
the perturbative Type IIA description should be valid for $L>>1$
and $g_s<<1$ \ie for $N^{1/5}<<k<<N$ while $\lambda=N/k$ is the 't
Hooft coupling of the boundary CS theory.

In the 11-d uplift, ${\mC\mP^3}$ becomes the base of a Hopf fibration
$S^7 = \mC\mP^3 \ltimes S^1$ whose metric reads \be ds^2_{S^7} =
ds^2_{\mC\mP^3} + (d\tau + \cA)^2 \ee with $d\cA = 2 \cJ_{\mC\mP^3}$, the
K\"ahler form on $\mC\mP^3$ normalized so that $dV(\mC\mP^3)=\cJ\wedge \cJ
\wedge \cJ/6$ and $V(\mC\mP^3) = \pi^3/6$. The solution is supported
by R-R fluxes \be g_s F_2= 2L \cJ \quad , \quad g_s F_4= {6 L^3}
dV(AdS_4) \quad , \quad g_s F_6= 6 L^5 dV(\mC\mP^3)\ee

In the ABJM model, corresponding to $\cN = 6$ $U(N)_{k}
\times U(N)_{-k}$ CS theory on the boundary, $B_2 = 0$. For fractional
M2-branes, one has the ABJ model corresponding to $\cN = 6$ CS
theory with $U(N)_{k} \times U(N+k-l)_{-k}$ \cite{Aharony:2008gk} on
the boundary, $B_2 = \cJ l/k$, with $l=1, ..., k-1$. Boundary CS
theories with $\sum_i k_i \neq 0$ and lower susy should be dual to
turning on a non-zero Romans mass ($F_0\neq 0$) in the bulk Type
IIA description \cite{Gaiotto:2009mv, Petrini:2009ur, Gaiotto:2009yz}.

The 11-d supergravity approximation should be valid in the
double-scaling limit $k\rightarrow \infty$, $N\rightarrow \infty$
with $\lambda = N/k$ fixed and large. The CFT description, to
which we momentarily turn our attention, should instead be valid
when $\lambda<<1 $, \ie $k>>N$. As $\lambda \rightarrow 0$ higher
spin symmetry enhancement takes place as we will eventually see.

\section{Boundary CFT description}

$\cN =6$ CS theories are conveniently constructed from $\cN =3$ CS
theories. The case $\cN =3$ arises in turn from the $\cN =4$ case
obtained after dimensional reduction of $\cN' = 2$ in $d=4$. In
this way, each vector multiplet includes an $\cN = 2$ (\ie
$\cN=1'$ in $d=4$) chiral multiplet in the adjoint $\Phi = \Phi_a
t^a$ and couples to various hypers $Q$ and $\tilde{Q}$ in real
(reducible) representations. Adding to the `standard' $\cN = 4$
superpotential \be W = \tilde{Q} \Phi Q \ee the CS term, giving a
mass $m = g_{YM}^2 {k\over 4\pi}$ to the vectors, and a CS
superpotential \be W = {k\over 8\pi} Tr\Phi^2 \ee breaks $\cN =4$
to $\cN =3$. Integrating out $\Phi$ yields \be W= {4\pi \over k}
(\tilde{Q} t^a Q) (\tilde{Q} t^a Q) \ee The resulting $\cN =3$
theory has no marginal susy preserving deformations
\cite{Gaiotto:2009yz, Petrini:2009ur, Gaiotto:2009mv}. In
the process R-symmetry is reduced to $SO(3)\approx SU(2)$ for $\cN
=3$ from the original $SO(4)$ of $\cN =4$.

The case $\cN = 6$ is special. Starting with the $\cN =3$ theory
with $G=U(N)_k \times U(N)_{-k}$ and two pairs of hypers, $A_r \in
({\bf N}, {\bf N}^*)$ and $B_{\dot{m}} \in ({\bf N}^*, {\bf N})$
and integrating out $\Phi_1$ and $\Phi_2$ one gets \be W = {2\pi
\over k} \epsilon^{rs} \epsilon^{\dot{m} \dot{n}} Tr(A_r
B_{\dot{m}} A_s B_{\dot{n}}) \ee Since the manifest `flavour'
symmetry of $W$ under $SU(2)\times SU(2) \times U(1)_B$ does not
commute with R-symmetry $SO(3)\approx SU(2)$ under which $A$ and
$B$ form doublets, the full theory has a larger $SU(4)\approx
SO(6)$ symmetry which is the R-symmetry of $\cN=6$. To expose the
symmetry it is convenient to define $X^i = (A_1, A_2,
B^*_{\dot{1}}, B^*_{\dot{2}})$ and their conjugate $X_i^*$ that
together transform as ${\bf 4}_{+1} + {\bf 4}^*_{-1}$ of
$SO(6)\times SO(2)$. As we will momentarily see, $SO(2)\sim U(1)$
acts as a baryonic symmetry. Further (super)symmetry enhancement
to $\cN = 8$ with $SO(8)$ R-symmetry takes place for $k=1$ and
$k=2$. The former corresponds to compactification on $S^7$ the
latter to $S^7/\mZ_2$ (only `even' spherical harmonics).

\section{A quick look at the spectrum}

The (ungauged) $\cN = 6$ supergravity multiplet consists of the
graviton $g_{\mu\nu}$, 6 gravitini $\psi_\mu^i$, 16 graviphotons
$A_\mu^{[ij]}$ and $A_\mu^{0}$, 26 dilatini $\lambda^{[ijk]}$ and
$\lambda_i$, and 30 scalars $\phi^{[ijkl]}$ and $\phi_{[ij]}$. The
latter parameterize the moduli space $\cM = SO^*(12)/U(6)$.  After
`gauging' $SO(6)\times SO(2)$ a scalar potential is generated and
the two sets of ${\bf 15}_0$ scalars become `massive' or rather
`tachyonic' \ie $(ML_{AdS})^2 = -2$, safely above the B-F bound
$(ML_{AdS})^2=-9/4$.

Compactification of Type IIA supergravity on $\mC\mP^3$ was studied in
\cite{Nilsson:1984bj}. KK excitations with $Q=0$, \ie neutral wrt
$SO(2)$, were identified there. The non-perturbative spectrum,
contains various wrapped branes, including D0-branes that are
charged wrt $SO(2)$. The latter correspond to 11-d KK modes along
the compact circle that can be obtained by a $\mZ_k$ projection of
the M-theory compactification on $S^7$. The dual to $SO(2)$
charged states are monopole operators on the boundary
\cite{Klebanov:2008vq, Benna:2009xd, Klebanov:2009sg}. Although
the fundamental fields ($A_r, B_{\dot{s}}$) are neutral wrt the
diagonal $U(1)$ that couples to $A_\mu^+ = A_\mu^1 + A_\mu^2$, the
orthogonal combination $A_\mu^- = A_\mu^1 - A_\mu^2$ acts as a
baryonic symmetry. The corresponding current, $J_B = *F^+$, is
conserved thanks to Bianchi identities. Due to the CS coupling $
k\int A^-\wedge F^+$, configurations with $A^+$ magnetic charge
are electrically charged wrt $A^-$. Alternatively one can
introduce a Lagrange multiplier $\tau$ for $dF^+=0$ (on-shell
$kA^-= d\tau$) and form combinations $e^{in\tau}$ that can screen
the baryonic charge of matter field composites. In general one can
consider magnetic monopoles charged under $U(1)^N\subset U(N)$
with $H=(Q_1, ..., Q_N)$. Without loss of generality one can take
$Q_1\ge Q_2 \ge ... \ge Q_N$. Since elementary fields have unit
charges and transform in the fundamental of $SU(N)$, these
monopole operators correspond to Young Tableaux with $kQ_i$ boxes
in the $i^{th}$ row. For $k=1,2$ dressing composite vector
currents in the ${\bf 6}_{\pm 2}$ and scalar operators in the
${\bf 10}_{\pm 2}$ and ${\bf 10}^*_{\mp 2}$ (with
$\Delta_{\pm}=1,2$) with charge 2 monopole operators is crucial to
the enhancement of supersymmetry to $\cN = 8$ with full $SO(8)$
R-symmetry \cite{Klebanov:2009sg}. Monopole and anti-monopole
operators however appear in the spectrum even when $k\ge 3$ and no
(super)symmetry enhancement takes place \cite{Klebanov:2008vq,
Benna:2009xd}.

Before concluding this preliminary look, let us note that out of
the two $U(1)$ in the boundary CS theory only the Baryonic
$U(1)_B=U(1)_-$ is visible as a global symmetry, whose $\mZ_k$
subgroup is gauged, in the bulk description. The fate of the other
$U(1)$ is a sort of Higgs mechanism, under which $A_M \rightarrow
A_\mu$ and $C_{MNP} \rightarrow C_\mu \cJ_{ab}$ mix. Only the
combination $kA_\mu + N C_\mu$ remains massless and couples to
$U(1)_B$ while the orthogonal combination $NA_\mu - k C_\mu$
becomes massive by `eating' the (pseudo)scalar $\beta$ from $B_2 =
\beta \cJ$. A 5-brane instanton is thus expected to mediate
processes in which $k$ D0-branes transform into $N$ D4-branes
wrapped around $\mC\mP^2\subset \mC\mP^3$ \cite{Aharony:2008gk}.

\section{Compactification on $S^7$ revisited}

For the later use let us briefly review the mass spectrum of the
Freund-Rubin solution of $d=11$ supergravity on $S^7$
\cite{Sezgin:1983ik, Castellani:1984vv, VanNieuwenhuizen:1985be}.
The gravitino field  as well as all the fermions are set to zero,
the $AdS_4$ Riemann tensor and the three-form field strength are
given by: \bea
R_{\mu\nu\rho\sigma}=-4(g_{\mu\rho}(x)g_{\nu\sigma}(x)-g_{\mu\sigma}(x)g_{\nu\rho}(x))\\
F_{\mu\nu\rho\sigma}=3\sqrt{2}\sqrt{-\det g_{\mu\nu}(x)}\varepsilon_{\mu\nu\rho\sigma}
\eea
where $\varepsilon_{0123}=-1$.
The metric and the three form field with mixed indices vanish:
\bea
g_{\mu\alpha}=F_{\mu\nu\rho\alpha}=F_{\mu\nu\alpha\beta}=F_{\mu\alpha\beta\gamma}=0
\eea
and also
\bea
F_{\alpha\beta\gamma\delta}(y)=0\\
R_{\alpha \beta}=-6g_{\alpha \beta}(y)
\eea
$\mu, \nu, \rho =0,...,3$ are
$d=4$ indices, $\alpha, \beta, \gamma =1,...,7$ are internal indices.

Let us then consider fluctuations around the Freund-Rubin
solution. The linearized field equations are obtained by replacing
the background fields in the $d=11$ field equations by background
fields plus arbitrary fluctuations. An elegant and quite general
method to determine the complete mass spectrum on any coset
manifold relies on generalized harmonic expansion. In our case,
one expands the fluctuations in a complete set of spherical
harmonics of $S^7 = SO(8)/SO(7)$. The coefficient functions of the
spherical harmonics correspond to the physical fields in $d=4$. In
order to diagonalize the linearized equations it turns out to be
convenient to parameterize the fluctuations as follows: \bea
&& g_{\mu \nu}(x,y)=g_{\mu \nu}(x)+h_{\mu \nu}(x,y)\\
&& h_{\mu \nu}(x,y)=h_{\mu \nu}^\prime(x,y)-\frac{1}{2}g_{\mu \nu}(x)h_\alpha ^{\,\,\, \alpha}
(x,y)\label{Weylresc}\\
&&g_{\alpha \beta}(x,y)=g_{\alpha \beta}(x)+h_{\alpha \beta}(x,y)\\
&& g_{\mu \alpha}(x,y)=h_{\mu \alpha}(x,y)\\
&& A_{\mu \nu \rho}(x,y)=A_{\mu \nu \rho}(x)+a_{\mu \nu \rho}(x,y)
\eea In particular the Weyl rescaled spacetime metric appears in
(\ref{Weylresc}) so as to put the $d=4$ Einstein action  in
canonical form. The spherical harmonic expansions of the
fluctuations of the metric and of the antisymmetric tensor fields
are given by: \bea
&& h_{(\mu \nu)}^\prime(x,y)=\sum H_{\mu \nu}^{N_1}(x)Y^{N_1}(y)\nn\\
&& h_{\mu \alpha}(x,y)=\sum B_\mu ^{N_7}(x)Y_\alpha ^{N_7}(y)+B_\mu ^{N_1}(x)D_\alpha Y^{N_1}(y)\nn\\
&& h_{(\alpha \beta)}(x,y)=\sum \phi ^{N_{27}}(x)Y_{(\alpha \beta)}^{N_{27}}(y)+\phi ^{N_7}
(x)D_{(\alpha}Y_{\beta)}^{N_7}(y)
+\phi ^{N_1}(x)D_{(\alpha}D_{\beta)}Y^{N_{1}}(y)\nn\\
&& h_\alpha ^{\,\,\,\alpha}(x,y)=\sum \pi^{N_1}(x)Y^{N_1}(y)\nn\\
&& A_{\mu \nu \rho}(x,y)=\sum a_{\mu\nu\rho}^{N_1}(x)Y^{N_1}(y)\nn\\
&& A_{\mu \nu \alpha}(x,y)=\sum a_{\mu \nu}^{N_7}(x)Y_\alpha ^{N_7}(y)+a_{\mu \nu}^{N_1}(x)D_\alpha Y^{N_1}
(y)\nn\\
&& A_{\mu \alpha \beta}(x,y)=\sum a_\mu ^{N_{21}}(x)Y_{\alpha \beta}^{N_{21}}(y)+a_\mu ^{N_7}
(x)D_{[\alpha}Y_ {\beta
]}^{N_7}\nn\\
&& A_{\alpha \beta \gamma}(x,y)=\sum a^{N_{35}}(x)Y_{\alpha \beta
\gamma}^{N_{35}}(y)+a^{N_{21}}(x)D_{[\alpha}Y_{\beta
\gamma]}^{N_{21}}(y) \eea All  superscripts $N_{{\bf r}}$ (${\bf
r}=1,7,21,27,35$) have infinite range, since they should provide a
basis for arbitrary fields on the 7-sphere. The index ${\bf r}$
specifies the $SO(7)$ representation of the corresponding
spherical harmonic. For example, $Y_{\alpha\beta\gamma}^{N_{35}}$
is in the third rank totally antisymmetric representation of
$SO(7)$ with dimension 35, while $Y_{(\alpha\beta)}^{N_{27}}$
is in the symmetric traceless 27-dimensional representation.
Derivatives of $Y$'s appear in the expansions since any tensor can
be decomposed into its transverse and longitudinal parts. After
fixing all local symmetries which do not correspond to gauge
invariances of the final $d=4$ theory and by choosing de Donder
type, $D^\alpha h_{(\alpha \beta)}(x,y)=0$, and Lorentz type,
$D^\alpha h_{\alpha \mu}(x,y)=0$, conditions the last term in
$h_{\mu \alpha}$ and the last two terms in $h_{(\alpha \beta)}$
drop out. To fix the local symmetries of the antisymmetric tensor
fields we choose the Lorentz conditions $D^\alpha
A_{\alpha\beta\gamma}(x,y)=D^\alpha
A_{\alpha\beta\mu}(x,y)=D^\alpha A_{\alpha\mu\nu}(x,y)=0$. As a
consequence, also these fields have only transverse harmonics
$a_{\mu\nu}^{N_1} (x)=a_\mu^{N_7}(x)=a^{N_{21}}(x)=0$.
Substituting the resulting expansions into the $d=11$ field equations,
the coefficients of each independent spherical harmonic yield the
$d=4$ field equations.

In the Einstein equation for $R_{\mu\nu}$ only $Y^{N_1}$ spherical
harmonics appear without derivatives. Thus there is only one field
equation, \ie one KK tower, for traceless symmetric tensors in
$AdS_4$.

Examining the Einstein equation for $R_{\alpha\beta}$ one can see
that the vector fields $B_\mu ^{N_7}$ are massive and transversal,
except for the lowest lying state corresponding to the Killing
vectors on $S^7$. The spin-0 fields $\phi ^{N_{27}}$ have a mass
matrix $\Delta _y+12$ ($\Delta _y$ is the Hodge-de Rham operator).
By a judicious gauge choice one can
eliminate $H_{\mu}^{N_1 \, \mu}$ in favour of $\pi ^{N_1}$ namely
$H_{\mu}^{N_1 \, \mu}=\frac{9}{7}\pi ^{N_1}$.

Collecting the coefficients of the spherical harmonics $Y_\alpha
^{N_7}$ and $D_\alpha Y^{N_1}$ in the Einstein equation for
$R_{\mu\alpha}$, one finds that the spin-1 spectrum consists of
linear combinations of $B_\mu ^{N_7}$ and $C_\mu ^{N_7}$ (from
$a_{\rho\sigma}^{N_7}$) and that one can eliminate the divergence
$D^\mu H^{N_1}_{\mu\nu}$ in favour of $\pi ^{N_1}$,
$a_{\rho\sigma\tau}^{N_1}$ except when $Y^{N_1}$ is a constant.

Similarly, inspecting the equations for $p$-form field strengths
($p=1,2,3,4$), one concludes that field expansions in
spherical harmonics can be chosen such that only the first terms
in the expansions survive with $Y$s being transversal and
traceless.

In particular, from the three-form field strength equation one
finds that $a_{\mu\nu\rho}^{N_1}=\varepsilon
_{\mu\nu\rho\lambda}D^\lambda \sigma^{N_1}$. This implies that the
divergence of $H_{\mu\nu}^{N_1}$ is proportional to a gradient.

From the four-form field strength equation one gets an equation
for $\Box _x \sigma^{N_1}$. Taking the trace of the equations for
$R_{\mu\nu}$ and $R_{\alpha\beta}$, an equation involving $\Box _x
\sigma^{N_1}$ and $\Box _x H_\mu ^{N_1\mu}$ arises. Resolving the
mixing between $a_{\mu\nu\rho}^{N_1}$ and $H_\mu ^{N_1\mu}$
produces to independent combinations and as many KK towers of
scalars.

From the two-form field strength equation one finds $D^\mu
a_{\mu\nu}^{N_7}=0$, which implies
$a_{\mu\nu}^{N_7}=\varepsilon_{\mu\nu}^{\,\,\,\,\,\,\,\rho\sigma}D_\rho
C_\sigma ^{N_7}$. Using one of the three-form field strength
equations one finds that $C^{N_7}_\mu$ and $B^{N_7}_\mu$ mix.
Resolving the mixing one finds two KK towers, one of which starts
with a massless vector corresponding to the internal Killing
vectors of $S^7$.

After diagonalizing the bosonic field equations one obtains the
mass spectrum summarized in Table \ref{boson}.
\begin{table}
\begin{center}
  \begin{tabular}{ | l | l | l | l | l | l | l |}\hline
Spin & Field & $SO(7)$ & $SO(8)$ &  $4(ML)^2$ & $\Delta$ & $\ell$
\\ \hline
$2^+$ & $h_{(\mu \nu)}^\prime$ & $N_1$ & $(\ell,0,0,0)$ & $\ell (\ell +6)$ & $\Delta=\frac{\ell}{2}+3$  & $ 
\ell\geq 0$ \\
\hline
$1_1^-$ & $h_{\mu \alpha}$ & $N_7$ & $(\ell,1,0,0)$ & $\ell(\ell +2)$ & $\Delta=\frac{\ell}{2}+2$ & 
$\ell\geq 0$ \\
$1_2^-$ & $A_{\mu \nu \alpha}$ & $N_7$ & $(\ell-2,1,0,0)$ & $(\ell +6)(\ell +4)$ & $\Delta=\frac{\ell}{2}+4$ 
& $\ell\geq
2$\\
\hline
$1^+$ & $A_{\mu \alpha \beta}$ & $N_{21}$ &
$(\ell-1,0,1,1)$ & $(\ell +2)(\ell +4)$ &
$\Delta=\frac{\ell}{2}+3$ & $\ell\geq 1$\\ \hline
$0_1^+$ & $A_{\mu \nu \rho}$ & $N_1$ & $(\ell+2,0,0,0)^*$ & $(\ell +2)(\ell
-4)$ & $\Delta=\frac{\ell}{2}+1$ & $\ell\geq 0$\\
$0_2^+$ & $h_{\alpha \alpha},\,h_{\lambda \lambda}^\prime$ & $N_1$ & $(\ell-2,0,0,0)$ & $(\ell +10)(\ell 
+4)$ &
$\Delta=\frac{\ell}{2}+5$ & $\ell\geq 2$\\ \hline
$0_3^+$ & $h_{(\alpha \beta)}$ & $N_{27}$ & $(\ell-2,2,0,0)$ & $\ell (\ell
+6)$ & $\Delta=\frac{\ell}{2}+3$ & $\ell\geq 2$\\ \hline
$0_1^-$ &$A_{\alpha \beta \gamma}$ & $N_{35}$ & $(\ell,0,2,0)$ & $(\ell
-2)(\ell +4)$ & $\Delta=\frac{\ell}{2}+2$ & $\ell\geq 0$\\ \hline
$0_2^-$ & $A_{\alpha \beta \gamma}$ & $N_{35}$ & $(\ell-2,0,0,2)$
& $(\ell +8)(\ell +2)$ & $\Delta=\frac{\ell}{2}+4$ & $\ell\geq
2$\\ \hline
    \end{tabular}
\caption{Bosonic KK towers after compactification on $S^7$}
\end{center}
\label{boson}
\end{table}
The resulting bosonic spectrum includes the massless graviton,
${\bf 28}$ massless vectors of $SO(8)$, corresponding to a
combination of $B_\mu$ (in $h_{\mu\a}$) and $C_\mu$ (in
$A_{\mu\nu\a}$), ${\bf 35}_v$ scalars ($\Delta =1$) and ${\bf
35}_s$ ($\Delta =2$) pseudoscalars with $(ML_{AdS})^2 = -2$. In
the supergravity literature \cite{Sezgin:1983ik,
Castellani:1984vv, VanNieuwenhuizen:1985be} masses of scalars are
often shifted by $-R/6$ so that $(ML_{AdS})^2 \rightarrow
(\tilde{M}L_{AdS})^2 = (ML_{AdS})^2 + 2$. The 70 (pseudo)scalars
in the $\cN = 8$ supergravity multiplet are `massless' in the
sense that $(\tilde{M}L_{AdS})^2=0$. Moreover, there are three
families of scalars and two families of pseudoscalar excitations.
Three of them ($0_2^+$, $0_3^+$ and $0_2^-$) contain only states
with positive mass square and correspond to irrelevant operators
in the dual CFT. The remaining families $0_1^+$ and $0_1^-$
contain states with positive, zero and negative mass squared
corresponding to irrelevant, marginal and relevant operators,
respectively.

A similar analysis can be performed for fermionic fluctuations. In
Table \ref{fermion} we summarize the fermionic mass spectrum.
\begin{table}
\begin{center}
 \begin{tabular}{ | l | l | l | l | l |}
    \hline
Spin &  $SO(8)$ &  $ 4(ML)^2$ & $\Delta$ & $\ell$ \\ \hline
$(\frac{3}{2})_1$ & $(\ell,0,0,1)$ & $ (\ell +2)^2$ & $\Delta=\frac{\ell}{2}+\frac{5}{2}$  & $ \ell\geq 0$ 
\\
\hline
$(\frac{3}{2})_2$ & $(\ell -1,0,1,0)$ & $ (\ell +4)^2$ & $\Delta=\frac{\ell}{2}+\frac{7}{2}$  & $ \ell\geq 
1$ \\
\hline
$(\frac{1}{2})_1$ & $(\ell +1,0,1,0)^*$ & $ \ell ^2$ & $\Delta=\frac{\ell}{2}+\frac{3}{2}$  & $ \ell\geq 0$ 
\\
\hline
$(\frac{1}{2})_2$ & $(\ell -1,1,1,0)$ & $ (\ell +2) ^2$ & $\Delta=\frac{\ell}{2}+\frac{5}{2}$  & $ \ell\geq 
1$ \\
\hline
$(\frac{1}{2})_3$ & $(\ell -2,1,0,1)$ & $ (\ell +4)^2$ & $\Delta=\frac{\ell}{2}+\frac{7}{2}$  & $ \ell\geq 
2$ \\
\hline
$(\frac{1}{2})_4$ & $(\ell -2,0,0,1)$ & $ (\ell +6) ^2$ & $\Delta=\frac{\ell}{2}+\frac{9}{2}$  & $ \ell\geq 
2$ \\
\hline
 \end{tabular}
 \caption{Fermionic KK towers after compactification on $S^7$}
\end{center}
\label{fermion}
\end{table}

The KK spectrum does not include the states with $*$ for $\ell
=-1$, since they do not propagate in the bulk but live on the
conformal boundary of $AdS_4$. They correspond to the singleton
representation of $Osp(8|4)$ that consists of $8_v$ bosons $X^i$
with $\Delta =\frac{1}{2}$, $(ML)^2=-\frac{5}{4}$ and $8_c$
fermions $\psi ^{\dot{a}}$ with $\Delta =1$, $ML=\frac{1}{2}$,
both at the unitary bound.

The KK excitations on $S^7$ can be put in one-to-one
correspondence with `gauge-invariant' composite operators on the
boundary. The dictionary for bosonic operators schematically
reads: \bea && s=2^+ \quad  \quad T^{i_1...i_\ell}_{\mu \nu,
\Delta = {\ell \over 2} + 3}= (\partial _\mu X_i
\partial _\nu X^i+\bar{\psi}\gamma _\mu \partial _\nu
\psi)X^{i_1}...X^{i_\ell}\\
&&  s=1^-_1 \quad \quad J_{\mu, \Delta = {\ell \over 2} + 2
}^{[ij]i_1...i_\ell}= (X^{[i}{\partial} _\mu X^{j]}+ \bar{\psi}
\Gamma^{ij}\gamma_\mu
\psi)X^{i_1}...X^{i_\ell}\\
&& s=1^-_2 \quad  \quad J^{[ij]i_1...i_{\ell-2}}_{\mu, \Delta =
{\ell \over 2} + 4} =\partial _\mu X_i\partial _\nu X^i \bar{\psi}
\gamma ^\nu \Gamma ^{ij}\psi X^{i_1}...X^{i_{\ell-2}}\\
&& s=1^+ \quad  \quad J^{a\dot{b}i_1...i_{\ell-1}}_{\mu,\Delta =
{\ell \over 2} + 3}= \bar{\psi}\Gamma _{jk}\partial
_\mu \psi (X_i \Gamma ^{ijk})^{a\dot{b}}X^{i_1}...X^{i_{\ell-1}} \\
&& s=0_1^+ \quad  \quad \Phi ^{ij i_1...i_\ell}_{\Delta = {\ell
\over 2} + 1}= X^{i} X^j
X^{i_1}...X^{i_\ell}\\
&& s=0_2^+ \quad  \quad \Phi ^{i_1...i_{\ell-2}} _{\Delta = {\ell
\over 2} + 5}=
\partial _\mu X^i \partial _\nu X_i
\bar\psi\gamma ^\mu \partial ^\nu \psi X^{i_1}...X^{i_{\ell-2}} \\
&& s=0_3^+ \quad  \quad \Phi ^{[ij][kl]i_1...i_{\ell-2}}_{\Delta =
{\ell \over 2} + 3}= (\bar{\psi} \Gamma ^{ij}
\gamma _\mu \psi X^{[k} \partial ^\mu X^{l]})X^{i_1}...X^{i_{\ell-2}}\\
&& s=0_1^- \quad  \quad
\Phi^{(\dot{a}\dot{b})i_1...i_\ell}_{\Delta = {\ell \over 2} + 2}=
\bar\psi ^{\dot{a}}
\psi ^{\dot{b}} X^{i_1}...X^{i_\ell}\\
&& s=0_2^- \quad \quad \Phi^{(a b)i_1...i_{\ell-2}}_{\Delta =
{\ell \over 2} + 4 }= (\Gamma ^{ijkl})^{ab} X_i\partial ^\mu
X_j\bar{\psi}\Gamma _{kl}\partial _\mu \psi
X^{i_1}...X^{i_{\ell-2}}\nn\\ \eea A similar dictionary can be compiled
for fermions.

\section{Polynomial representations for $SO(8)$ and $U(4)$}
\label{PolyRep}
In order to decompose KK harmonics on $S^7=SO(8)/SO(7)$ into KK
harmonics on $\mC\mP^3=U(4)/U(3)\times U(1)$, we will present the
construction of arbitrary representations of $SO(8)$ in the space
of polynomials of $12$ variables. The latter are the coordinates
of the subgroup $Z_+^{SO(8)}$ generated by the raising operators
of $SO(8)$. We will then describe a technique which allows to
identify which of the above polynomials correspond to highest
weight states of representations of $U(4)\subset SO(8)$. The method we use
is quite standard in representation theory of Lie groups (see \eg Chapter 16
of \cite{Zhelobenko}).

It is convenient to start with $SO(8,\mC)$ defined as the group of
$8\times 8$ complex matrices which leave invariant the quadratic
form $X^TC^{(8)}X$, where $X$ is a complex (column) vector whose
components will be enumerated as $X^1,X^2,X^3,X^4,X^{\tilde
4},X^{\tilde 3},X^{\tilde 2},X^{\tilde 1}$ and $C^{(8)}$ is an
$8\times 8$ matrix with $1$'s on SW-NE (anti)diagonal: \be
C^{(8)}_{ij}=C^{(8)}_{{\tilde i}{\tilde j}}=0, \hspace{1cm}
C^{(8)}_{i{\tilde j}}=C^{(8)}_{{\tilde j}i} = \delta_{ij},
\hspace{0.5cm} i,j=1,2,3,4 \ee By definition all matrices  $g\in
SO(8)$ satisfy the condition $g^TC^{(8)}g=C^{(8)}$. Eventually, in
order to select the compact real form $SO(8)$ of our interest, one
should identify the coordinates $X^{{\tilde i}}$ with ${\bar X}^i$
(bar means complex conjugate). A generic $SO(8)$ matrix $g$  can
be (uniquely) decomposed as (Gauss decomposition): \be g=\zeta
\lambda z, \ee where $\zeta \in Z_ -$, $z \in Z_+ $, $\lambda \in
\Lambda $ with $Z_+$ ($Z_-$) being the subgroup of lower (upper)
triangular matrices with $1$'s on the diagonal and $\Lambda$ is
the subgroup of diagonal matrices (Cartan subgroup). Let's set
$\lambda =Diag(\lambda_1,
\lambda_2,\lambda_3,\lambda_4,\lambda_4^{-1},\lambda_3^{-1},
\lambda_2^{-1},\lambda_1^{-1})$. We will realize the irreducible
representations of the group $SO(8)$ on some spaces of functions
defined on it. In particular, the role of the highest weight
vector will be played by the function : \be \alpha
(g)=\lambda_1^{m_1} \lambda_2^{m_2} \lambda_3^{m_3}
\lambda_4^{m_4} \label{charm} \ee where $m_1\ge m_2\ge m_3\ge
|m_4| $ ($m_i$ are either all integers or all half-integers)
uniquely characterize the irrep. The eigenvalues $\lambda_i$ can
be expressed  in terms of the matrix elements of $g$ explicitly:
 \be
 \lambda_p=\frac{\Delta_p}{\Delta_{p-1}},\hspace{0.5cm} p=1,2,3,4
 \ee
where $\Delta_0=1$ and $\Delta_p$, $p=1,2,3,4$ are the diagonal
minors \bea
 \Delta_p=\left|
\begin{array}{ccc}
g_{11}&\cdots &g_{1p}\\
\vdots&\cdots &\vdots\\
g_{p1}&\cdots &g_{pp}\\
\end{array}
 \right|.
 \eea
Introducing the notation $S_-=\frac{\Delta_3}{\sqrt{\Delta_4}}$,
$S_+=\sqrt{\Delta_4}$ (it is easy to see that $S_{+,-}$
polynomially depend on the matrix elements of $g$) we can rewrite
eq. (\ref{charm}) as \be \alpha (g)=\Delta_1^{\ell_1} \Delta_2^{\ell_2}
S_-^{\ell_3} S_+^{\ell_4} \label{charm1} \ee where $\ell_1=m_1-m_2$,
$\ell_2=m_2-m_3$, $\ell_3=m_3-m_4$ and $\ell_4=m_3+m_4$ are non-negative
integers commonly referred as the Dynkin labels of the irrep.
Consider the space ${\cal R}_\alpha$ of all linear combinations of
the functions $\alpha(gg_0)$,  $g_0\in SO(8)$. $SO(8)$ is
represented in ${\cal R}_\alpha$ simply by the right multiplication of
the argument. As already mentioned the function $\alpha(g)$ plays the role of the
highest weight state. For any function $f(g)\in {\cal R}_\alpha $
we have $f(\zeta\lambda z)=\alpha (\lambda) f(z)$ which shows that
to restore its full $g$-dependence it is sufficient to only know
the values the function assumes on the subgroup $Z_+$. This is why
actually we get representation on a space of functions of $z$, in fact
polynomials due to the polynomial dependence on $g$ of $\alpha(g)$
mentioned earlier.

There is an elegant way to characterize this space of polynomials.
Consider the four raising generators corresponding to the simple
roots\bea
e_1=E_{12}-E_{{\tilde 2}{\tilde 1}}; \hspace{0.5cm}
e_2=E_{23}-E_{{\tilde 3}{\tilde 2}}\nonumber \\
e_-=E_{34}-E_{{\tilde 3}{\tilde 4}}; \hspace{0.5cm}
e_+=E_{3{\tilde 4}}-E_{{\tilde 4}3} \label{risinggen} \eea where
$E_{pq}$ denotes  the $8\times 8$ matrix whose only non-zero entry
$1$ is at the position $(p,q)$. Denote their  {\it left} action on
${\cal R}_\alpha $ by ${\cal D}_1$, ${\cal D}_2$, $ {\cal D}_-$,
${\cal D}_+$. It is not difficult to prove that \bea
{\cal D}_1^{\ell_1 +1}\alpha(g)=0\nonumber\\
{\cal D}_2^{\ell_2 +1}\alpha(g)=0\nonumber\\
{\cal D}_-^{\ell_3 +1}\alpha(g)=0\nonumber\\
{\cal D}_+^{\ell_4 +1}\alpha(g)=0. \eea The key observation is
that the same equations are valid also for arbitrary functions
$f\in {\cal R}_\alpha $, since they are all generated by
$\alpha(g)$ through right multiplications which commute with left
multiplications. Below we will use a convenient explicit
parametrization of $Z_+\subset SO(8)$ in terms of two $4\times 4$
matrices $\eta $ and $a$ \bea \eta= \left(
\begin{array}{cccc}
1&\eta_{12} &\eta_{13}&\eta_{14}\\
0&1 &\eta_{23}&\eta_{24}\\
0&0 &1&\eta_{34}\\
0&0 &0&1\\
\end{array}
 \right);\hspace{0.5cm}
a= \left(
\begin{array}{cccc}
a_{14}&a_{13} &a_{12}&0\\
a_{24}&a_{23}&0&-a_{12}\\
a_{34}&0&-a_{23}&-a_{13}\\
0&-a_{34}&-a_{24}&-a_{14}
\end{array}
 \right).
\eea Let us further introduce the $8\times 8$ matrices which in
$2\times 2$ block form read \bea
 z_0=
\left(
\begin{array}{cc}
\eta &0\\
0&{\tilde \eta}\\
\end{array}
\right)
 ;\hspace{0.5cm}
z^\prime= \left(
\begin{array}{cc}
1&a\\
0&1
\end{array}
 \right),
\eea
where
\bea
{\tilde \eta}=
\left(
\begin{array}{cccc}
1& -\eta _{34} & -\eta _{24}+\eta _{23} \eta _{34} & -\eta _{14}+\eta _{12} \eta _{24}+\eta _{13} \eta _{34}-\eta _{12}
\eta _{23} \eta _{34} \\
0& 1 & -\eta _{23} & -\eta _{13}+\eta _{12} \eta _{23} \\
0& 0 & 1 & -\eta _{12}\\
0&0&0&1
\end{array}
 \right).
 \label{etaa}
\eea An arbitrary $z\in Z_+$ can be (uniquely) represented as
\be z=z^\prime z_0. \label{etaadecomposition} \ee Left
multiplication by raising generators (\ref{risinggen}) induces
infinitesimal motion on the parameters $a$, $\eta$. A
straightforward algebra shows that \eg \bea (1+\epsilon
e_1)z(a,\eta)=z(a+\delta  a,\eta+\delta\eta)+O(\epsilon^2), \eea
where the non-trivial variations are \bea \delta\eta _{12}=
\epsilon ,\,\,\delta \eta _{13}=\epsilon \eta _{23},\,\,\delta\eta
_{14}=\epsilon \eta _{24},\,\, \delta a_{13}=\epsilon a_{23},\,\,
\delta a_{14}=\epsilon a_{24}.\nonumber \eea Similarly examining the
remaining three generators we find \bea {\cal
D}_1&=&\partial_{\eta_{12}}+\eta_{23}
\partial_{\eta_{13}}+a_{23}\partial_{a_{13}}+
a_{24}\partial_{a_{14}} \nonumber\\
{\cal D}_2&=&\partial_{\eta_{23}}+\eta_{34} \partial_{\eta_{24}}+a_{13}\partial_{a_{12}}+
a_{34}\partial_{a_{24}} \nonumber\\
{\cal D}_-&=&\partial_{\eta_{34}}+a_{14}\partial_{a_{13}}+
a_{24}\partial_{a_{23}} \nonumber\\
{\cal D}_+&=&\partial_{a_{34}}. \eea
Thus any irreducible
representation of $SO(8)$ is realized on the space of polynomials
 of $12$ variables $a$, $\eta$ subject to the constraints
\bea
&&\left(\partial_{\eta_{12}}+\eta_{23} \partial_{\eta_{13}}+a_{23}\partial_{a_{13}}+
a_{24}\partial_{a_{14}}\right)^{\ell_1 +1}f(a,\eta)=0 \nonumber\\
&&\left(\partial_{\eta_{23}}+\eta_{34} \partial_{\eta_{24}}+a_{13}\partial_{a_{12}}+
a_{34}\partial_{a_{24}}\right)^{\ell_2 +1}f(a,\eta)=0  \nonumber\\
&&\left(\partial_{\eta_{34}}+a_{14}\partial_{a_{13}}+
a_{24}\partial_{a_{23}}\right)^{\ell_3 +1}f(a,\eta)=0  \nonumber\\
&&\left(\partial_{a_{34}}\right)^{\ell_4 +1}f(a,\eta)=0 .
\label{indicatorsystem}
\eea

Note that the constant polynomial always satisfies
(\ref{indicatorsystem}) and corresponds to the highest weight
state. Considering right multiplication it is not difficult to
find explicit expressions for the generators of $SO(8)$ as
operators acting on the space of polynomials. For our later
proposes let us specify how the diagonal part $\Lambda\subset
SO(8)$ is represented. Since \be z(a,\eta)\lambda=\lambda
\lambda^{-1}z(a,\eta)\lambda =\lambda z(a^\prime ,\eta^\prime),
\ee where \bea a^\prime_{ij}=\lambda_j^{-1}\lambda_i^{-1}a_{ij};
\hspace{0.5cm} \eta^\prime_{ij}=\lambda_j\lambda_i^{-1}\eta_{ij}
\eea we simply get \be \lambda \circ
f(a,\eta)=\lambda_1^{m_1}\lambda_2^{m_2}\lambda_3^{m_3}\lambda_4^{m_4}f(a^\prime
,\eta^\prime) \ee Notice that the variable $a_{ij}$ shifts the
weights as $m_i\rightarrow m_i-1$, $m_j\rightarrow m_j-1$
while the variable $\eta_{ij}$ shifts them as $m_i\rightarrow
m_i-1$, $m_j\rightarrow m_j+1$.

Consider now the $GL(4,C)\subset SO(8,C)$ subgroup whose
off-diagonal blocks in $2\times 2$ block notation are zero. This
subgroup does not mix the coordinates $X^i$ with $X^{{\tilde i}}$
and after restriction to the real sector it becomes the subgroup
$U(4)\subset SO(8)$.

In other words, for the reduction from $S^7$ to $S^7/\mZ_k$ or
$\mC\mP^3 \ltimes S^1$ we are interested in, the decomposition
$SO(8)\rightarrow SO(6)\times SO(2)$ is given by the embedding \be
{\bf 8_v}(1,0,0,0) \rightarrow {\bf 4}_{+1}[0,1,0] + {\bf
4^*}_{-1}[0,0,1] \label{8v} \ee where $(\ell _1,\ell _2,\ell
_3,\ell _4)$ and $[k,l,m]$ denote $SO(8)$ and $SO(6)$ Dynkin
labels respectively. As a result, for the Adjoint representation
one has \be {\bf 28}(0,1,0,0) \rightarrow {\bf 15}_{0}[0,1,1] +
{\bf 1}_{0}[0,0,0] + {\bf 6}_{+2}[1,0,0] + {\bf 6}_{-2}[1,0,0] \ee
while \be {\bf 8_s}(0,0,0,1) \rightarrow {\bf 6}_0[1,0,0] + {\bf
1}_{+2}[0,0,0] + {\bf 1}_{-2}[0,0,0] \label{8s} \ee \be {\bf
8_c}(0,0,1,0)\rightarrow {\bf 4}_{-1}[0,1,0] + {\bf
4^*}_{+1}[0,0,1] \label{8c} \ee for the spinorial representations.

Our goal is to identify the highest weight states of this subgroup
inside the space of polynomials of a given representation of
$SO(8)$. It is evident from the decomposition
(\ref{etaadecomposition},\ref{etaa})  that the right action by the
raising operators of $GL(4)$ subgroup $e_1$, $e_2$, $e_-$ (see eq.
(\ref{risinggen})) shifts the parameters $\eta$ and leave the
parameters $a$ untouched. Thus, in order to be a highest weight
state, a polynomial, besides satisfying the equations
(\ref{indicatorsystem}) should be independent of $\eta$. The {\it
indicator} system for the highest weight states becomes \bea
&&\left(a_{23}\partial_{a_{13}}+
a_{24}\partial_{a_{14}}\right)^{\ell_1 +1}f(a)=0 \nonumber\\
&&\left(a_{13}\partial_{a_{12}}+
a_{34}\partial_{a_{24}}\right)^{\ell_2 +1}f(a)=0  \nonumber\\
&&\left(a_{14}\partial_{a_{13}}+
a_{24}\partial_{a_{23}}\right)^{\ell_3 +1}f(a)=0  \nonumber\\
&&\left(\partial_{a_{34}}\right)^{\ell_4 +1}f(a)=0 .
\label{hwindicatorsystem}
\eea
Solving these equations one can fully decompose KK
harmonics on $S^7$ into KK harmonics of $\mC\mP^3\times S^1$ which is our
next task.

\section{From $S^7$ to $\mC\mP^3\ltimes S^1$}

$S^7$ is a $U(1)$ bundle over $\mC\mP^3$. The $\mC\mP^3$ solution of the
$d=10$ theory can be obtained from the $S^7$ solution of the
$d=11$ theory by Hopf fibration, \ie keeping only $U(1)$ invariant
states \cite{Nilsson:1984bj}. The compactification on $\mC\mP^3$ of the
$d=10$ theory yields a four dimensional theory with $\cN=6$
supersymmetry and with gauge group $SO(6)\times SO(2)$.

The truncation from $S^7$ to $\mC\mP^3 \ltimes S^1$ cannot be thought
of as spontaneous (super)symmetry breaking and one has to really
discard the states that are projected out by $\mZ_k$ or $SO(2)$ for
$k\rightarrow \infty$ even if it acts freely. In particular we
will later check that no Higgsing can account for the breaking of
$SO(8)$ to $SO(6)\times SO(2)$ but rather the coset vectors are
dressed with monopole operators and become massive for $k\neq 1,2$
\cite{Aharony:2008gk, Aharony:2008ug, Benna:2008zy, Benna:2009xd,
Klebanov:2009sg}.

Let us start with the KK towers of bosons. Using the procedure
described in the previous section or otherwise, for scalar
spherical harmonics with Dynkin labels $(\ell,0,0,0)$ one finds as
independent polynomials $\{a_{14}^m \,| \,m=0,...,\ell\}$. Thus
the following decomposition holds: \bea N_1: \quad (\ell,0,0,0)\rightarrow
\oplus [0,\ell -m,m]_{\ell -2m} \eea where the subscript is the
$SO(2)$ charge $Q$ of the appropriate representation.

For vector spherical harmonics with $SO(8)$ Dynkin labels
$(\ell-2,1,0,0)$ one gets $\{a_{12}a_{14}^m , \, a_{24}a_{14}^m
,\, (a_{13}a_{24}-a_{14}a_{23})a_{14}^m , \, a_{14}^m \, | \,
m=0,...,\ell\}$ as independent polynomials. The $SO(8)$
representation decomposes into $SO(6)$ representations as: \bea N_7: \quad
(\ell,1,0,0)\rightarrow && \oplus[0,\ell -m,m]_{\ell -2m}
\oplus [0,\ell -m+1,m+1]_{\ell -2m }\nonumber\\
&& \oplus [1,\ell -m,m]_{\ell -2m-2} \oplus [1,\ell -m,m]_{\ell
-2m+2} \eea One obtains the decomposition of the representation
$(\ell -2, 1, 0, 0)$ from the previous one by shifting $\ell$ to
$\ell-2$. In what follows we will simply omit the decompositions
which differ by shifts of the parameter $\ell$.

For two-form spherical harmonics with $SO(8)$ Dynkin labels $(\ell
-1,0,1,1)$ one finds $\{a_{14}^m , \, a_{23}a_{14}^m , \,
a_{34}a_{14}^m , \, a_{23}a_{34}a_{14}^m ,\, (a_{34}a_{12} -
a_{13}a_{24})a_{14}^m ,\, a_{23}(a_{23}a_{14} + a_{34}a_{12} -
a_{13}a_{24})a_{14}^m),\, a_{13}a_{14}^n , \, a_{34}a_{13}a_{14}^n
, \, (a_{34}a_{12}-a_{13}a_{24})a_{13}a_{14}^n \, | \,
m=0,...,\ell -1,\, n=0,..., \ell -2\}$ as independent polynomials.
One then finds the following decomposition: \bea N_{21}: \quad (\ell
-1,0,1,1)\rightarrow &&\oplus [0,\ell -m,m]_{\ell -2m-4}
\oplus [0,\ell -m-1,m+1]_{\ell -2m+2}\nonumber\\
&&\oplus [1,\ell -m,m]_{\ell -2m-2}
\oplus [1,\ell -m-1,m+1]_{\ell -2m}\nonumber\\
&&\oplus [0,\ell -m,m]_{\ell -2m}
\oplus [0,\ell -m-1,m+1]_{\ell -2m-2}\nonumber\\
&&\oplus [1,\ell -n-2,n]_{\ell -2n-4}
\oplus [2,\ell -n-2,n]_{\ell -2n-2}\nonumber\\
&&\oplus [1,\ell -n-2,n]_{\ell -2n}
\eea
The decomposition of the KK towers corresponding to $0_1^+$ and $0_2^+$
can be found from the decomposition of $2^+$ via
appropriate shifts.

For second rank symmetric traceless harmonics with Dynkin labels
$(\ell -2,2,0,0)$ the polynomials are: $\{a_{14}^m $, $
a_{12}a_{14}^m $, $a_{12}(a_{23}a_{14}-a_{13}a_{24})a_{14}^m$, $
a_{12}^2a_{14}^m$, $a_{12}a_{24} a_{14}^m $,$a_{24}a_{14}^m$,
$a_{24}(a_{23}a_{14}-a_{13}a_{24})a_{14}^m $, $(a_{13}a_{24}-
a_{14}a_{23})a_{14}^m$, $(a_{14}a_{23}-a_{13}a_{24})^2a_{14}^m$,
$a_{24}^2 a_{14}^m $, $|\, m=0,...,\ell -2\}$. The $SO(6)$
representations decomposed from $SO(8)$'s are: \bea N_{27}: \quad
(\ell -2,2,0,0)\rightarrow && \oplus [2,\ell -m-2,m]_{\ell -2m+2}
\oplus [1,\ell -m-2,m]_{\ell -2m}\nonumber\\
&& \oplus [1,\ell -m-2,m]_{\ell -2m-4}
\oplus [0,\ell -m-1,m+1]_{\ell -2m-2}\nonumber\\
&& \oplus [0,\ell -m-2,m]_{\ell -2m-2}
\oplus [1,\ell -m-1,m+1]_{\ell -2m}\nonumber\\
&& \oplus [1,\ell -m-1,m+1]_{\ell -2m-4}
\oplus [2,\ell -m-2,m]_{\ell -2m-2}\nonumber\\
&& \oplus [2,\ell -m-2,m]_{\ell -2m-6} \oplus [0,\ell
-m,m+2]_{q=\ell -2m-2}\nn \\ \eea

For the three-form spherical harmonic with $SO(8)$ Dynkin labels $(\ell ,0,2,0)$ one finds
$\{(a_{14}^m + a_{23} a_{14}^m + a_{23}^2 a_{14}^m) ,\,
a_{13}(a_{14}^n + a_{23} a_{14}^n) ,\, a_{13}^2 a_{14}^p\, | \, m=0,...,\ell ,\, n=0,...,\ell -1 , \, 
p=0,...,\ell -2\}$
polynomials. The representation $(\ell ,0,2,0)$ decomposes as:
\bea
N_{35}: \quad (\ell ,0,2,0)\rightarrow && \oplus [0,\ell -m,m+2]_{\ell -2m+2}
\oplus [0,\ell -m+1,m+1]_{\ell -2m}\nonumber\\
&& \oplus [0,\ell -m+2,m]_{\ell -2m-2}
\oplus [1,\ell -n-1,n+1]_{\ell -2n}\nonumber\\
&& \oplus [1,\ell -n,n]_{\ell -2n-2}
\oplus [2,\ell -p-2,p]_{\ell -2p-2}
\eea

For the three-form spherical harmonic with $SO(8)$ Dynkin labels
$(\ell -2,0,0,2)$ one has $\{(a_{14}^m ,\, (a_{14} a_{23} + a_{12}
a_{34} - a_{13} a_{24})a_{14}^m ,\, (a_{12} a_{34} - a_{13} a_{24}
+ a_{14} a_{23})^2 a_{14}^m ,\, a_{34} a_{14}^m ,\, $ $a_{34}
(a_{24} a_{13} - a_{34} a_{12} - a_{14} a_{23}) a_{14}^m ,\,
a_{34}^2 a_{14}^m \, | \, m=0,...,\ell -2 )$ and the following
decomposition: \bea N_{35}^\prime : \quad (\ell -2,0,0,2)\rightarrow && \oplus [0,\ell
-m-2,m]_{\ell -2m-2}
\oplus [0,\ell -m-2,m]_{\ell -2m+2}\nonumber\\
&& \oplus [1,\ell -m-2,m]_{\ell -2m-4}
\oplus [1,\ell -m-2,m]_{\ell -2m}\nonumber\\
&& \oplus [2,\ell -m-2,m]_{\ell -2m-2} \oplus [0,\ell
-m-2,m]_{\ell -2m-6}\nn\\ \eea

Let us now consider the fermionic KK towers. There are two
gravitini in the $SO(8)$ representations $(\ell, 0,0,1)$ and
$(\ell-1, 0,1,0)$.

For the $SO(8)$ representation $(\ell, 0,0,1)$ one finds
$\{a_{14}^m , \, (a_{14}a_{23}+a_{12}a_{34}-a_{13}a_{24})a_{14}^m
, \,a_{34}a_{14}^m \, | \, m=0,..,\ell\}$ as polynomials and the
following decomposition holds  \bea (\ell ,0,0,1)\rightarrow \oplus
[0,\ell -m,m]_{\ell -2m+2} \oplus [0,\ell -m,m]_{\ell -2m-2}
\oplus [1,\ell -m,m]_{\ell -2m}\nn\\ \eea

For the $SO(8)$ representation $(\ell-1, 0,1,0)$ the independent
polynomials are $\{ a_{14}^m ,\, a_{23}a_{14}^m ,\,
a_{13}a_{14}^{n} \, | \, m=0,...,\ell -1, \, {n}=0,...,\ell
-2\}$ and is decomposed as: \bea
(\ell -1,0,1,0)\rightarrow &&\oplus [0,\ell -m-1,m+1]_{\ell -2m} \oplus [0,\ell -m,m]_{\ell 
-2m-2}\nonumber\\
&& \oplus [1,\ell -{n}-2,{n}]_{\ell -2{n}-2} \eea There are
other fermions in the representations $(\ell+1,0,1,0)$,
$(\ell-2,0,0,1)$, $(\ell-1,1,1,0)$ and $(\ell-2,1,0,1)$.

For the $SO(8)$ representation $(\ell-1,1,1,0)$ the polynomials
have the form $\{a_{14}^m ,\, a_{23}a_{14}^m ,\, a_{23}(a_{13}a_{24}-a_{14}a_{23})a_{14}^m ,\, 
a_{24}a_{14}^m ,\,
a_{13}a_{24}a_{14}^m ,\, a_{23}a_{24}a_{14}^m ,\, a_{12}a_{14}^m,\,$ $a_{12}a_{23}a_{14}^m, a_{13}a_{14}^n ,
\,
a_{13}(a_{13}a_{24}-a_{23}a_{14})a_{14}^n ,\,
a_{12}a_{13}a_{14}^n\, |\, m=0,...,\ell -1, n=0,...,\ell-2\}$ and
one has the following decomposition: \bea (\ell
-1,1,1,0)\rightarrow && \oplus [1,\ell -m-1,m+1]_{\ell -2m+2}
\oplus [1,\ell -m,m]_{\ell -2m}\nonumber\\
&& \oplus [1,\ell -m,m]_{\ell -2m-4}
\oplus [0,\ell -m,m+2]_{\ell -2m}\nonumber\\
&& \oplus [1,\ell -m-1,m+1]_{\ell -2m-2}
\oplus [0,\ell -m+1,m+1]_{\ell -2m-2}\nonumber\\
&& \oplus [0,\ell -m-1,m+1]_{\ell -2m}
\oplus [0,\ell -m,m]_{\ell -2m-2}\nonumber\\
&& \oplus [2,\ell -n-2,n]_{\ell -2n}
\oplus [2,\ell -n-2,n]_{\ell -2n-4}\nonumber\\
&& \oplus [1,\ell -n-2,n]_{\ell -2n-2}
\eea

Finally for the $SO(8)$ representation $(\ell-2,1,0,1)$ the
polynomials have the form $\{ a_{14}^m ,\,
(a_{14}a_{23}-a_{13}a_{24})a_{14}^m ,\,
(a_{13}a_{24}-a_{12}a_{34}-a_{14}a_{23})(a_{14}a_{23}-
a_{13}a_{24})a_{14}^m ,\, a_{12}a_{14}^m ,\,
a_{12}(a_{12}a_{34}-a_{13}a_{24}+a_{14}a_{23})a_{14}^m ,\,
a_{24}a_{14}^m ,\,
a_{24}(a_{12}a_{34}-a_{13}a_{24}+a_{14}a_{23})a_{14}^m ,\,
a_{34}a_{14}^m ,\, a_{34}(a_{13}a_{24}- a_{14}a_{23})a_{14}^m ,\,
a_{34}a_{24}a_{14}^m ,\, a_{34}a_{12}a_{14}^m \,|\, m=0,...,\ell
-2\}$ and the decomposition reads \bea (\ell -2,1,0,1)\rightarrow
&& \oplus [1,\ell -m-2,m]_{\ell -2m+2}
\oplus [1,\ell -m-2,m]_{\ell -2m-2}\nonumber\\
&& \oplus [1,\ell -m-2,m]_{\ell -2m-6}
\oplus [0,\ell -m-2,m]_{\ell -2m}\nonumber\\
&& \oplus [0,\ell -m-2,m]_{\ell -2m-4}
\oplus [0,\ell -m-1,m+1]_{\ell -2m}\nonumber\\
&&\oplus [0,\ell -m-1,m+1]_{\ell -2m-4}
\oplus [2,\ell -m-2,m]_{\ell -2m}\nonumber\\
&&\oplus [2,\ell -m-2,m]_{\ell -2m-4}
\oplus [1,\ell -m-1,m+1]_{\ell -2m-2}\nonumber\\
&& \oplus [1,\ell -m-2,m]_{\ell -2m-2}
\eea

The relevant $SO(8)\rightarrow SO(6)\times SO(2)$ decomposition is
given by the embedding (\ref{8v}), (\ref{8s}), (\ref{8c}).  In
particular this implies \bea
{\bf 35_v}(2,0,0,0) \rightarrow && {\bf 15}_0[0,1,1] + {\bf 10}_{+2}[0,2,0] + {\bf 10^*}_{-2}
[0,0,2]\nonumber\\
{\bf 35_c}(0,0,2,0) \rightarrow && {\bf 15}_0[0,1,1] + {\bf
10^*}_{+2}[0,0,2] + {\bf 10}_{-2}[0,2,0]\nonumber\\
{\bf 35_s}(0,0,0,2)\rightarrow && {\bf 20'}_0[2,0,0] + {\bf 6}_{+2}[1,0,0] + {\bf 6}_{-2}[1,0,0]+\nonumber\\
&& {\bf 1}_{0}[0,0,0] + {\bf 1}_{+4}[0,0,0]+ {\bf 1}_{-4}[0,0,0]
\eea that are necessary to analyze the spectrum of scalars.

The zero charge spectrum \ie the states which constitute the KK
spectrum of Type IIA supergravity on $\mC\mP^3$ can be easily
identified in the above decompositions. For completeness and
comparison with the original literature \cite{Nilsson:1984bj}, we
collect the relevant formulae in an Appendix.

\section{A closer look at the KK spectrum}

As already observed, the $\mZ_k$ orbifold projection from $S^7$ to
$S^7/\mZ_k \approx \mC\mP^3 \ltimes S^1$ cannot be thought of as
spontaneous (super)symmetry breaking. `Untwisted' states that are
projected out do not simply become `massive' but are rather
eliminated from the spectrum. In particular in the large $k$ limit
only $SO(2)$ singlets survive. It is amusing to observe that only
states with $\ell$ even on $S^7$ give rise to neutral states. This
suggests that the parent theory could be either a compactification
on $S^7$ or on $\mR\mP^7=S^7/\mZ_2$. Indeed both lead to $SO(8)$ gauged
supergravity corresponding to the `massless' multiplet \be
\{g_{\mu\nu}, 8 \psi_\mu, 28 A_\mu, 56 \lambda, 35^+ + 35^-
\varphi \} \ee

Massless scalars, corresponding to marginal operators with $\Delta
=3$ on the boundary, only appear in higher KK multiplets, \ie in
the ${\bf 840'}=(2,0,0,2)$ and ${\bf 1386}=(6,0,0,0)$. None of
these can play the role of St\"uckelberg field for the 12 coset
vectors in the ${\bf 6}_{+2} + {\bf 6}_{-2}$ of $SO(8) /
SO(6)\times SO(2)$.

Indeed, using the group theory techniques described in Section
\ref{PolyRep} or otherwise, the decomposition of ${\bf
840'}=(2,0,2,0)$ under $SO(8) \rightarrow SO(6)\times SO(2)$ reads
\bea {\bf 840_{vc}}(2,0,2,0) \rightarrow && {\bf 84}_{+4}[0,2,2] +
{\bf 70}_{+2}[0,3,1]+{\bf 70}_{+2}[0,1,3]+ {\bf 64}_{+2}
[1,1,1]\nonumber\\
&& +{\bf 84}_{0}[0,2,2]+{\bf 45}_{0}[1,2,0]+{\bf 45}_{0}[1,0,2]\\
&& +{\bf 35}_{0}[0,4,0]+{\bf 35}_{0}[0,0,4]+{\bf 20}'_{0}[2,0,0]\nonumber\\
&& +{\bf 84}_{-4}[0,2,2] + {\bf 70}_{-2}[0,3,1]+{\bf
70}_{-2}[0,1,3]+ {\bf 64}_{-2} [1,1,1]\nonumber \eea This means
that the massless scalars in the ${\bf 840_{vc}}(2,0,2,0)$ cannot
account for the `needed' St\"uckelberg fields in the ${\bf
6}_{+2}+ {\bf 6}_{-2}$. Yet one can recognize massless scalars
neutral under $SO(2)$ that survive in $k\rightarrow \infty$ limit
and transform non-trivially under $SO(6)$. Turning them on in the
bulk, \eg in domain-wall solutions, should trigger RG flows to
theories with lower supersymmetry on the boundary.

The same applies to the other massless scalars in the ${\bf
1386}(6,0,0,0)$, the totally symmetric product of 6 ${\bf 8_v}
\rightarrow {\bf 4}_{+1}+{\bf 4^*}_{-1}$. The relevant
decomposition reads \bea {\bf 1386}(6,0,0,0) \rightarrow
&& {\bf 84}_{+6}[0,6,0] + {\bf 189}_{+4}[0,5,1]+ {\bf 270}_{+2}[0,4,2]
\nonumber\\
&& +{\bf 300}_{0}[0,3,3]\nonumber\\
&& +{\bf 84}_{-6}[0,0,6]+ {\bf 189}_{-4}[0,1,5]+{\bf 270}_{-2}[0,2,4]\nonumber\\
\eea
Once again there are no ${\bf 6}_{+2}+
{\bf 6}_{-2}$. In this case, `neutral' fields appear in the $ {\bf
300}$ representation of $SO(6)$.

In the KK spectrum, neutral (wrt to $SO(2)$) singlets (of $SO(6)$)
appear in the decomposition of ${\bf 35}_{s}$ parity odd scalars
$0_2^-$ with $M^2L_{AdS}^2 = 10$ that reads \bea {\bf
35_s}(0,0,0,2)\rightarrow && {\bf 20'}_0[2,0,0] + {\bf
6}_{+2}[1,0,0]
+ {\bf 6}_{-2}[1,0,0]\nonumber\\
&& +{\bf 1}_{0}[0,0,0] + {\bf 1}_{+4}[0,0,0]+ {\bf 1}_{-4}[0,0,0]
\eea They correspond to boundary operators with dimension $\Delta
= 5$. The only other neutral singlets arise from the $SO(8)$
singlet parity even scalar with $M^2L_{AdS}^2 = 18$, \ie $\Delta
=6$. Neither ones belongs in the supergravity
multiplet\footnote{After gauging $SO(8)$, the 70 scalars give rise
to ${\bf 35_v}(2,0,0,0)$ and ${\bf 35_c}(0,0,2,0)$ which in turn
decompose into ${\bf 35_v}(2,0,0,0) \rightarrow {\bf 15}_0[0,1,1]
+ {\bf 10}_{+2}[0,2,0] + {\bf 10^*}_{-2}[0,0,2]$ and $ {\bf
35_c}(0,0,2,0) \rightarrow {\bf 15}_0[0,1,1] + {\bf
10^*}_{+2}[0,0,2] + {\bf 10}_{-2}[0,2,0]$.}. They correspond to
the `stabilized' complexified K\"ahler deformation $\cJ + i B$ and
as such couple to the Type IIA world-sheet instanton recently
identified in \cite{Cagnazzo:2009zh}. Indeed the bosonic action
schematically reads $S_{wsi} = \int \cJ + i B = L^2/\ap$ since
$B=0$ in the ABJM model, while $B=l/k$ with $l=1,...,k-1$ for the
ABJ model involving fractional M2-branes. Effects induced by
world-sheet instantons in Type IIA on $\mC\mP^3$ should be dual to
the non-perturbative corrections discussed in
\cite{Hosomichi:2008ip}. It may be worth to observe that in
`ungauged' $\cN=6$ supergravity, arising from freely acting
asymmetric orbifolds of Type II superstrings on tori, world-sheet
and other asymmetric brane instantons \cite{Bianchi:2009mu,
Bianchi:2008cj} should correct $\cR^4$ terms very much as in their
parents with $\cN=8$ local supersymmetry.

Other non-perturbative effects are induced by E5-brane instantons
that should mediate the process of annihilation of $k$ D0-branes
into $N$ D4-branes wrapping $\mC\mP^2$
\cite{Aharony:2008gk,Aharony:2008ug}. In order to determine the
action of such an instanton it is worth recalling that the
pseudo-scalar mode $B_2 = \beta(x)J_2(y)$ is eaten by the vector
field $A_\mu^H = kA_\mu^{D4} - N A_\mu^{D0}$ that becomes massive.
The complete E5-brane instanton action should be $S_{E5} =
L^6/g_s^2(\ap)^3 + i \beta$ that indeed shifts under $U(1)_{H}$
gauge transformations and as such can compensate for the `charge'
violation in the above process as in similar cases with unoriented
D-brane instantons \cite{Bianchi:2009ij}.

\chapter{Singleton, partition functions and Higher Spins}
\label{chapter6}

In this section, we would like to discuss the higher spin (HS)
extension of $\cN = 6$ gauged supergravity.
Higher spin extensions of various supergravity theories in $AdS_4$
have been studied in \cite{Sezgin:1998gg, Sezgin:2002rt,
Engquist:2002vr} but to the best of our knowledge the case of $\cN
= 6$  has been overlooked.

Let us start by briefly recalling some basic features of higher
spin theories in $AdS_4$\footnote{See \eg \cite{Bianchi:2004ww,
Bianchi:2004xi, Bekaert:2005vh, Francia:2006hp} for recent
reviewes of both Vasiliev's and geometric approaches.}.
 In the non supersymmetric case the
HS algebra represents an extension of the conformal group
$SO(3,2)$ that admits two {\it singleton} representations
$\cD(1/2,0)$ (free boson) and $\cD(1,1/2)$ (free fermion). The two
labels denote conformal dimension $\Delta$ and spin $s$. Indeed
the maximal compact subgroup of $SO(3,2)$ is $SO(3)\times SO(2)
\approx SU(2)\times U(1)$ while `Lorentz' transformations and
dilatations commute and generate $SO(2,1)\times SO(1,1) \subset
SO(3,2)$. We will continue and call $\Delta$ the dimension and $s$
or $j$ spin. In `radial' quantization the `Hamiltonian' $\cH$ has
eigenvalues $\Delta$.

For later use let us collect here the partition functions of the
two singletons that take into account their conformal descendants
\ie non vanishing derivatives. For free bosons such that $\de^2
X=0$ one has \be \cZ_B(q) = Tr q^{2\cH} = {q - q^5 \over
(1-q^2)^3} ={q + q^3 \over (1-q^2)^2}\ee For free fermions
$\slash{\,\de} \Psi=0$ one has \be \cZ_F(q) = Tr q^{2\cH}= 2{q^2 - q^4 \over
(1-q^2)^3} = 2{q^2\over (1-q^2)^2} \ee Combining $n_b=8_v$ free
bosons and $n_f=8_c$ free fermions one finds the singleton
representation of $Osp(8|4) \supset SO(8)\times SO(3,2)$, whose
Witten index reads \bea \cZ_{\tiny\yng(1)}(q)= Tr(-)^F q^{2\cH} =
8_v \cZ _B(q) - 8_c \cZ_F(q)\eea One can also keep track of the
spin of the states in the spectrum by including a chemical
potential $y=e^{i\alpha}$ ($y^{J_3} =e^{i\alpha J_3}$) and find
\bea \cZ _B(q, \alpha)=\frac{q(1-q^4)}{(1-q^2) (1- e ^{i\alpha}
q^2) (1- e ^{-i\alpha} q^2)}= \frac{q(1+q^2)}{(1-2 q^2 \cos\alpha
+q^4)} \eea \bea \cZ_F(q, \alpha)=\frac{q^2 (1 - q^2)\chi _{1\over
2}(\alpha)}{(1-q^2) (1-2 q^2 \cos\alpha + q^4)} \eea where \bea
\chi _{1\over 2}(\alpha) = 2 \cos \frac{\alpha}{2} = tr_{1/2}
e^{i\alpha J_3} \eea is the character of the fundamental
representation of the `Lorentz' group $SU(2)$.

Before switching to higher spins, notice that $\mZ_k$ acts on the
singleton simply as \be 8_v\rightarrow 4\omega + 4^* \bar{\omega}
\qquad 8_c\rightarrow 4 \bar{\omega} + 4^* \omega \qquad
8_s\rightarrow 6 +  {\omega}^2 + \bar\omega^2 \ee with
$\omega=e^{2\pi i/k}$ playing the role of chemical potential or
rather fugacity for the $SO(2)\approx U(1)_B$ charge $Q$ commuting
with $SO(6)$ R-symmetry. One can introduce another three chemical
potentials $\beta_i$ or fugacities $x_i = e^{i\beta_i}$ in order
to keep track of the three Cartan's of $SO(6)\approx SU(4)$. We
refrain from doing so here.

\section{Doubleton and higher spin gauge fields}

{\it Doubleton} representations can be obtained as tensor products
of two singletons \cite{Ferrara:1997dh, Sundborg:1999ue,
HaggiMani:2000ru}. \be \cD(1/2,0)\otimes \cD(1/2,0) =
\oplus_{s=0}^\infty \cD(\Delta=s+1,s) \ee or \be \cD(1,
1/2)\otimes \cD(1,1/2) = \cD(\Delta=2,s=0) + \oplus_{s\neq
0}^\infty \cD(\Delta=s+1,s) \ee A consistent truncation, giving
rise to minimal HS theories with even spins only, stems from
restricting to symmetric tensors for bosons \be [\cD(1/2,0)\otimes
\cD(1/2,0)]_S = \oplus_{k=0}^\infty \cD(\Delta=2k+1,s=2k) \ee or
anti-symmetric for fermions \be [\cD(1, 1/2)\otimes \cD(1,1/2)]_A
= \cD(\Delta=2,s=0) + \oplus_{k\neq 0}^\infty
\cD(\Delta=2k+1,s=2k) \ee Odd spin states appear in the product
with opposite symmetry \be [\cD(1/2,0)\otimes \cD(1/2,0)]_A =
\oplus_{k=0}^\infty \cD(\Delta=2k+2,s=2k+1) \ee for bosons and \be
[\cD(1, 1/2)\otimes \cD(1,1/2)]_S = \oplus_{k=0}^\infty
\cD(\Delta=2k+2,s=2k+1) \ee for fermions. Generators of the HS
symmetry algebra can be realized as polynomials of bosonic
oscillators $y_\a, y_{\dot\a}=(y_\a)^\dagger$ satisfying $[y_\a,
y_\b] = i\varepsilon_{\a\b}$ and $[y_{\dot\a}, y_{\dot\b}] =
i\varepsilon_{\dot\a\dot\b}$.

The supersymmetric extensions require the introduction of
fermionic oscillators $\xi^i$ with $i=1, ..., \cN$, satisfying
$\{\xi^i, \xi^j\} = \delta^{ij}$. The resulting HS superalgebra
denoted by $shs^E(\cN|4)$ contains $Osp(\cN|4)$ whose bosonic
generators span $SO(3,2)\cong Sp(4,R)$ (conformal group) and
$SO(\cN)$ R-symmetry \cite{Sezgin:1998gg, Sezgin:2002rt,
Engquist:2002vr}.

In particular for $\cN =8$, with $SO(8)$ R-symmetry, $Osp(8|4)$ is
the maximal finite dimensional subalgebra of the HS gauge algebra
$shs^E(8|4)$, which is a Lie superalgebra. The relevant
super-singleton consists in\footnote{Different conventions for the
$SO(8)$ representations of bosons and fermions appear in the
literature which are related to the present one, chosen for
compatibility with our previous analysis, by $SO(8)$ triality.}
\be \widehat\cD_{\cN=8} = \cD(1/2,0;{\bf 8}_v) \oplus
\cD(1,1/2;{\bf 8}_c) \ee The (graded) symmetric product of two
singletons $[\widehat\cD_{\cN=8} \otimes
\widehat\cD_{\cN=8}]_{\hat{S}}$ yields \bea &&\{[\cD(1/2,0;{\bf
8}_v) \oplus \cD(1,1/2;{\bf 8}_c)] \otimes [\cD(1/2,0;{\bf 8}_v)
\oplus
\cD(1,1/2;{\bf 8}_c)]\}_{\hat{S}} = \nn \\
&&\cD(1,0;{\bf 1} + {\bf 35}_v) \oplus \cD(2,0;{\bf 1} + {\bf
35}_c) \oplus_k \cD(k+{3\over 2},k+{1\over 2}; {\bf 8}_s + {\bf
56}_s)\nn \\&&\oplus_{k\neq 0} \cD(2k+1,2k; {\bf 1} + {\bf 35}_v +
{\bf 1} + {\bf 35}_c) \oplus_k \cD(2k+2,2k+1; {\bf 28} + {\bf 28})\nn\\
\eea

It is reassuring to recognize above the `massless' states of $\cN
= 8$ gauged supergravity on $AdS_4$. The remaining states with
spin $s\le 2$ belong to the `short' Konishi multiplet and a
`semishort' multiplet with spin ranging from 2 to 6
\cite{Bianchi:1999ge, Bianchi:2000hn, Bianchi:2001cm }. Holography
allows to relate AdS compactifications of supergravity and
superstring theories to singleton field theories on the 3-d
boundary. As a first step, these field theories can be constructed
on the boundary of AdS as free superconformal theories. A
remarkable property of singletons is that the symmetric product of
two super-singletons gives an infinite tower of massless higher
spin states. In the limit $\lambda \rightarrow 0$, all higher spin
states become massless. After turning on interactions, a
pantagruelic Higgs mechanism, named {\it Grande Bouffe} in
\cite{Bianchi:2003wx, Beisert:2003te, Beisert:2004di,
Bianchi:2005ze}, takes place. All but a handful of HS gauge fields
become massive after `eating' lowest spin states. The boundary
counterpart of this phenomenon is the appearance of anomalous
dimensions for HS currents and their superpartners. One should
keep in mind that genuinely massive states are already present in
the spectrum at $\lambda \rightarrow 0$ and arise in the product
of three and more singletons.

Interacting theories for massless HS gauge fields, thus only
describing the doubleton, have been proposed by Vasiliev
\cite{Bekaert:2005vh} that capture some aspects of the holographic
correspondence in the extremely stringy (high AdS curvature)
regime. Only vague glimpses of an interacting theory incorporating
the {\it Grande Bouffe} have been offered so far
\cite{Bianchi:2003wx, Beisert:2003te, Beisert:2004di,
Bianchi:2005ze}.

Barring these subtle issues, let us discuss how to perform a
$\mZ_k$ projection of the spectrum giving rise to an $\cN =6$ HS
supergravity in $AdS_4$. In the limit $k\rightarrow \infty$ only
$SO(2)$ singlets survive \bea &&\{[\cD(1/2,0;{\bf 8}_v) \oplus
\cD(1,1/2;{\bf 8}_c)]
\}^{\otimes 2_{\hat{S}}}_{SO(2) singlets} = \nn \\
&&\cD(1,0;{\bf 1} + {\bf 15}) \oplus \cD(2,0; {\bf 1} + {\bf 15})
\oplus_k \cD(k+{3\over 2},k+{1\over 2}; {\bf 6} + {\bf 6} + {\bf
10} + {\bf 10}^*)\nn
\\&&\oplus_{s\neq 0} \cD(s+1,s; {\bf 1} + {\bf 15} + {\bf 1} +
{\bf 15})  \eea where indicated in bold-face are the surviving
representations of the $SO(6)$ R-symmetry. Candidate bosonic HS
operators on the boundary in the ${\bf 1} + {\bf 15}$ of $SO(6)$
are \be \cJ_{\mu_1...\mu_s}{}^i{}_j = X^i \de_{\mu_1}\de_{\mu_2}
... \de_{\mu_s} \bar{X}_j +  \bar{\Psi}^i \gamma_{\mu_1} \de_{\mu_2} ...
\de_{\mu_s} \Psi_j + ... \ee where dots stand for
symmetrization and subtraction of the traces and the coefficients
of the linear combination are to be chosen appropriately.

At finite $k$ and $\lambda$, states with $SO(2)$ charges $Q = k n$
survive. One can exploit orbifold technique to deduce the `free'
spectrum\footnote{Although $k$ is finite, one can take $k>>N$, so
that $\lambda<< 1$, in order to identify states that eventually
become massive.}.

The partition function or rather Witten index for the
super-singleton of $OSp(8|4)$ reads: \bea \cZ_{\tiny\yng(1)}
=\frac{8q}{(1+q)^2} \eea the $\mZ_k$ projection reads \be
\cZ^{\mZ_k}_{\tiny\yng(1)} = {1\over k} \sum_{r=0}^{k-1}
\cZ^{(r)}_{\tiny\yng(1)} \ee where

\be \cZ^{(r)}_{\tiny\yng(1)} =\frac{(4 \omega^r + 4\bar\omega^r)
q}{(1+q)^2} \ee with $\omega = e^{2\pi i/k}$. Clearly
$\cZ^{\mZ_k}_{\tiny\yng(1)} = 0$ since $\Sigma_{r=0}^{k-1}
\omega^r = 0$.

A non-trivial spectrum arises from the doubleton partition
function. Prior to the $\mZ_k$ projection one has \bea
\cZ_{\tiny\yng(2)}=\frac{1}{2}(\cZ_{\tiny\yng(1)} ^2(q) +
\cZ_{\tiny\yng(1)}(q^2))=4q^2 (8 (1+q)^{-4}+(1+q^2)^{-2}) \eea for
the (graded) symmetric doubleton, giving rise to precisely the
spectrum of $hs(8|4)$ discussed above.

Performing the $\mZ_k$ projection on the symmetric doubleton one
finds \bea &&\cZ^{\mZ_k}_{\tiny\yng(2)}=\frac{1}{2k}\sum_r
(\cZ^{(r)}_{\tiny\yng(1)}(q,\omega)^2 +
\cZ^{(r)}_{\tiny\yng(1)}(q^2,\omega^2)) \nn \\ && =
4q^2\left[4\left( 1 + \sum_r{\omega^{2r} + \bar\omega^{2r} \over
2k}\right)(1+q)^{-4}+\sum_r{\omega^{2r} + \bar\omega^{2r} \over
2k}(1+q^2)^{-2}\right]\nn\\ \eea for the (graded) symmetric doubleton,
giving rise to precisely the `massless' HS gauge fields of
$hs(6|4)$ for $k\neq 2$ and $hs(8|4)$ for $k = 1, 2$, as expected
$Z_{HS}=Z_{\tiny\yng(2)}$! Indeed \bea && \cZ_{HS} = \frac{36
(q^2 +q^4) + 72 \sum _{s=2k\neq 0}F_s(q) + 56 \sum _{s=2k+1}F_s(q)
- 64 \sum _{s=k+ \frac{1} {2}}F_s(q)}{(1-q^2)^3}\nonumber\\ \eea
with $F_s(q) = (2s+1) q^{2(s+1)} - (2s-1) q^{2(s+1)+2}$ taking
into account the presence of null descendants for conserved spin
$s$ currents of dimension $\Delta = s +1$. The relevant characters
read \bea \cX _s ^{\Delta =s+1} = \frac{q ^{2 \Delta}(2s+1) -
q^{2(\Delta +1)} (2s-1) }{(1 - q^2)^3} =
\frac{q ^{2 \Delta} [\chi _s(\alpha) - q^2 \chi _{s-1}(\alpha)
]}{(1-q^2) (1-2 q^2 \cos\alpha + q^4)}\nn\\ \eea up to some $SO(8)$
multiplicity $d_{(\ell ,...)}^{SO(8)}$.

The situation is summarized in the following Tables, where $s$
denotes spin and $h$ the `string' level.
\begin{table}
\begin{center}
 \begin{tabular}{ | l | l | l | l | l |}
  \hline
    $s \backslash h$ & 0 & 1 & 2 & 3 \\ \hline
    0 & 70 & 1+1 &  & \\ \hline
    $1\over 2$ & 56 & 8 &  &\\ \hline
    1 & 28 & 28 &  &\\ \hline
    $3\over 2$ & 8 & 56 &  &\\ \hline
    2 & 1 & 70 & 1 &\\ \hline
    $5\over 2$ &  & 56 & 8 &\\ \hline
    3 &  & 28 & 28 &\\ \hline
    $7\over 2$ &  & 8 & 56 &\\ \hline
    4 &  & 1 & 70 & 1 \\ \hline
    ... &  &  & ... & ...\\ \hline
    \end{tabular}
    \caption{$\cN=8$ $hs(8|4)\supset Osp(8|4)$}
\end{center}
\end{table}

The decomposition into charged sectors reads \bea
\cZ_{\tiny\yng(2)}=&&\frac{1}{(1-q^2)(1-2q^2\cos \alpha +q^4)}\, \{ \left[10 \left(\omega
^2+\omega_c^2\right)+16\right]\left(q^2+q^4\right)\chi_0(y)\nonumber\\
&&+\sum_{j\in 1,3,...}\left[12\left(\omega
^2+\omega_c^2\right)+32\right]
\left[\chi_j(y) q^{2 (j+1)}-\chi_{j-1}(y) q^{2 (j+1)+2}\right] \nonumber\\
&&+\sum_{j\in 2,4,...}\left[20\left(\omega
^2+\omega_c^2\right)+32\right] \left(\chi_{j}(y) q^{2
(j+1)}-\chi_{j-1}(y) q^{2
(j+1)+2}\right)\nonumber\\
&&-\sum_{j\in 1/2,3/2,...}16(\omega +\omega_c)^2\left(\chi_{j}(y)
q^{2 (j+1)}-\chi_{j-1}(y) q^{2 (j+1)+2}\right)\}. \eea
\begin{table}
\begin{center}
 \begin{tabular}{ | l | l | l | l | l | }
    \hline
    $s\backslash h$ & 0 & 1 & 2 & 3 \\ \hline
    0 & 15+15 & 1+1 &  &\\ \hline
    $1\over 2$ & $10+10^*+6$ & 6 &  &\\ \hline
    1 & 15+1 & 15+1 & & \\ \hline
    $3\over 2$ & 6 & $10+10^*+6$ &  &\\ \hline
    2 & 1 & 15+15 & 1 &\\ \hline
    $5\over 2$ &  & $10+10^*+6$ & 6 & \\ \hline
    3 &  & 15+1 & 15+1 & \\ \hline
    $7\over 2$ &  & 6 & $10+10^*+6$ & \\ \hline
    4 &  & 1 & 15+15 & 1  \\ \hline
    $9\over 2$ &  &  & $10+10^*+6$ & 6  \\ \hline
    5 &  &  & 15+1 & 15+1 \\ \hline
    $11\over 2$ &  &  & 6 & $10+10^*+6$ \\ \hline
    6 &  &  & 1+1 & 15+15  \\ \hline
    ... &  &  & ... & ...  \\ \hline
    \end{tabular}
\caption{$SO(2)$ neutral HS for $\cN=6$: $hs(6|4)\supset Osp(6|4)$}
\end{center}
\end{table}

\begin{table}
\begin{center}
 \begin{tabular}{ | l | l | l | l | l |}
    \hline
    $s\backslash h$ & 0 & 1 & 2 &\\ \hline
    0 & $(10+10^*)_{\pm 2}$ &  &  &\\ \hline
    $1\over 2$ & $15_{\pm 2}$ & $1_{\pm 2}$ &  &\\ \hline
    1 & $6_{\pm 2}$ & $6_{\pm 2}$ &  &\\ \hline
    $3\over 2$ & $1_{\pm 2}$ & $15_{\pm 2}$ &  &\\ \hline
    2 &  & $(10+10^*)_{\pm 2}$ & &\\ \hline
    $5\over 2$ &  & $15_{\pm 2}$ & $1_{\pm 2}$ &\\ \hline
    3 &  & $6_{\pm 2}$ & $6_{\pm 2}$ &\\ \hline
    $7\over 2$ &  & $1_{\pm 2}$ & $15_{\pm 2}$ &\\ \hline
    4 &  &  & $(10+10^*)_{\pm 2}$ &\\ \hline
    ... &  &  & ...& ... \\ \hline
    \end{tabular}
    \caption{Charged HS for $\cN=6$: $hs(8|4)/hs(6|4)\supset Osp(8|4)/Osp(6|4)$}
\end{center}
\end{table}

\section{Tripletons and higher $n$-pletons}

For higher multipletons one has to resort to Polya theory
\cite{Bianchi:2003wx, Beisert:2003te, Beisert:2004di,
Bianchi:2005ze}. Consider a set of `words' $A,B,...$ of $n$
`letters' chosen within the alphabet $\{a_i\}$ with $i = 1,... p$.
When $p\rightarrow \infty$, let us denote by $\cZ_1(q)$ the single
letter `partition function'. Let also $G$ be a group action
defining the equivalence relation $A\sim B$ for $A = gB$ with $g$
an element of $G\subset S_n$. Elements $g \in S_n$ can be divided
into conjugacy classes $[g] = (1)^{b_1}... (n)^{b_n}$, according
to the numbers $\{b_k(g)\}$ of cycles of length $k$. Polya theorem
states that the set of inequivalent words are generated by the
formula \be \cZ_n^G = {1\over |G|} \sum_{g\in G}
\prod_{k=1}^n\cZ_1(q^k)^{b_k(g)} \ee

In particular, for the cyclic group $G = Z_n$, conjugacy classes
are $[g] = (d)^{n/d}$ for each divisor $d$ of $n$. The number of
elements in a given conjugacy class labelled by $d$ is given by
Eulers totient function $\cE(d)$, equal to the number of integers
relatively prime to $d$. For $d = 1$ one defines $\cE(1)=1$. \be
\cZ_n^{Z_n} = {1\over n} \sum_{d | n} \cE(d)\cZ_1(q^d)^{n/d} \ee
For the full symmetric group one has \be \cZ_n^{S_n} = {1\over n!}
\sum_{n_r: \sum_r r n_r = n} {n! \over \prod_r r^{n_r} n_r!}
\prod_r \cZ_1(q^{r})^{n_r} \ee

Let us consider the product of three singletons. \bea
\cZ_{\tiny\yng(1)}^3 = \cZ_{\tiny\yng(1) \times \tiny\yng(1)
\times \tiny\yng(1)} \rightarrow
\cZ_{\tiny\yng(3)}+\cZ_{\tiny\yng(1,1,1)}+\cZ_{\tiny\yng(2,1)}
\eea There are thus three kinds of tri-pletons.

The totally symmetric tripleton is coded in the partition function
\bea && \cZ_{\tiny\yng(3)} = \frac{1}{6}(\cZ_{\tiny\yng(1)}^3(u) +
3 \cZ_{\tiny\yng(1)}(u)\cZ_{\tiny\yng(1)}(u^2) + 2
\cZ_{\tiny\yng(1)}(u^3))\eea where $u$ collectively denotes the
`fugacities' $q,y=e^{i\alpha},\omega\approx t, ...$.

For the cyclic tripleton one has \bea && \cZ_{cycl} =
\cZ_{\tiny\yng(3)} + \cZ_{\tiny\yng(1,1,1)} =
\frac{1}{3}(\cZ_{\tiny\yng(1)}^3(u) + 2\cZ_{\tiny\yng(1)}
(u^3))\eea

For totally anti-symmetric tripletons one finds \bea &&
\cZ_{\tiny\yng(1,1,1)} = \cZ _{cycl}- \cZ_{\tiny\yng(3)} =
\frac{1}{6}(\cZ_{\tiny\yng(1)}^3(u) + 2\cZ_{\tiny\yng(1)} (u^3) -
3\cZ_{\tiny\yng(1)}(u) \cZ_{\tiny\yng(1)}(u^2)\nn\\ \eea
while for mixed symmetry, incompatible with the cyclicity of the
trace, one eventually finds \bea &&\cZ_{\tiny\yng(2,1)} =
\cZ_{\tiny\yng(1)}^3 (u) - \frac{1}{3} \cZ_{\tiny\yng(1)}^3 (u) -
\frac{2}{3}\cZ_{\tiny\yng(1)} (u^3) = \frac{2}{3}
(\cZ_{\tiny\yng(1)}^3 (u) - \cZ_{\tiny\yng(1)} (u^3))\nn\\ \eea

Recalling the singleton partition function \bea
\cZ_{\tiny\yng(1)}(q, \alpha , \omega) = &&\frac{(4 \omega + 4^*
\bar{\omega}) q (1+q^2)}{(1-2 q^2 \cos\alpha + q^4)} -
\frac{(4 \bar{\omega} + 4^* \omega) q^2 \chi _{\frac{1}{2}} (\alpha)}{(1-2 q^2 \cos\alpha + q^4)}\nonumber\\
&& =\frac{4 (\omega + \bar{\omega})q}{(1-2 q^2 \cos\alpha + q^4)}
[1 + q^2 - \chi _{\frac{1}{2}} (\alpha) q] \eea where $\omega =
e^{2\pi i/k}$ and $\chi _{\frac{1}{2}} (\alpha)= tr_{1/2} \exp
(i\alpha J_3)$, one eventually finds \bea \cZ_{\tiny\yng(3)} =
&&\frac{1}{6}\left ( \frac{4^3 (\omega + \bar{\omega})^3 q^3 (1 +
q^2 -q \chi _{\frac{1}{2}}
(\alpha) )^3}{(1-2 q^2 \cos\alpha + q^4)^3} + \right .\nonumber\\
&& \left .3 \frac{4 (\omega + \bar{\omega}) q 4 (\omega ^2 +
\bar{\omega} ^2) q^2 (1 + q^2 -q\chi _{\frac{1}{2}} (\alpha))
(1 + q^4 - q^2\chi _{\frac{1}{2}} (2\alpha))}{(1-2 q^2 \cos\alpha + q^4) (1-2 q^4 \cos\alpha + q^8)} + 
\right .\nonumber\\
&& \left .2 \frac{4 (\omega ^3+ \bar{\omega}^3) q^3 (1 + q^6 -
q^3\chi _{\frac{1}{2}} (3 \alpha))}{(1-2 q^6 \cos\alpha + q^{12})}
\right ) \eea
for the totally symmetric tripleton.
Let us analyze the spectrum arising in this case. Except for the 1/2 BPS states, we will consider later
on, only `massive' representations above the unitary bound, whose
characters read \bea && \cX _s^{\Delta \neq s+1} = \frac{q ^{2
\Delta} \chi _s(\alpha)}{(1-q^2) (1-2 q^2 \cos\alpha + q^4)}
\rightarrow_{\alpha\rightarrow 0}\frac{q ^{2 \Delta}(2s+1)}{(1 -
q^2)^3} \eea appear in the decomposition \bea
\cZ_{\tiny\yng(3)}(q, \alpha , \omega) =\sum _{s, \Delta , Q}
c(s,\Delta, Q)\frac{q ^{2 \Delta} \chi _s(\alpha) \omega ^Q}
{(1-q^2) (1-2 q^2 \cos\alpha + q^4)} \eea Indeed it is easy to see
that no current like (twist $\tau =1$) fields appear beyond the
double-ton, since the twist \be \tau = \Delta -s = {n_X\over 2} +
n_{\de} + n_\Psi - n_{\de} - {n_\Psi\over 2} = {n_X\over 2} +
{n_\Psi\over 2} >1 \ee whenever $n_X + n_\Psi > 2$.

Using orthogonality of the $SU(2)$ characters \bea
\frac{1}{\pi}\int _0^{2\pi} \chi _s(\alpha)  \chi
_{s^\prime}(\alpha) \sin ^2\frac{\alpha}{2} d\alpha =\delta
_{2s+1,2s^\prime +1} \eea

one can decompose the partition function according to \bea
\sum_{Q,\Delta} \frac{c(s,\Delta, Q) \omega ^Q q ^{2
\Delta}}{(1-q^2)}=\frac{1}{\pi}\int _0^{2\pi}(1-2 q^2 \cos\alpha +
q^4) \sin ^2\frac{\alpha}{2} \chi _s(\alpha)\cZ_{\tiny\yng(3)}(q,
\alpha , \omega)  d\alpha \nn \\\eea

It is clear that only states with charge $Q=\pm 3, \pm 1$ are
present in the tri-pleton spectrum. Setting $y=e^{i\alpha}$, for
states with $Q=\pm1$ one finds \bea
\cZ_{\tiny\yng(3)}^{Q=\pm 1}&=&\sum_{k=0}^{\infty}\left[(40+256 k) q^{4 k+3}+(104+256 k) q^{4 k+5}\right] 
\chi_{2 k}
(y)\nonumber\\
&-&\left[(104+256 k) q^{4 k+4}+(152+256 k) q^{4 k+6}\right] \chi_{2 k+\frac{1}{2}}(y)\nonumber\\
&+&\left[(152+256 k) q^{4 k+5}+(216+256 k) q^{4 k+7}\right] \chi_{2 k+1}(y)\nonumber\\
&-&\left[(216+256 k) q^{4 k+6}+(296+256 k) q^{4 k+8}\right]
\chi_{2 k+\frac{3}{2}}(y) \eea these states are always projected
out by $\mZ_k$ since $\pm 1\neq n k$. For states with $Q=\pm3$ one
finds instead \bea
\cZ_{\tiny\yng(3)}^{Q=\pm 3}&=&\sum_{k=0}^{\infty}\left[\left[(20+256 k) q^{12 k+3}+(40+256 k) q^{12 
k+5}\right] \chi_{6
k}(y)\right.\nonumber\\
&-&\left[(40+256 k) q^{12 k+4}+(44+256 k) q^{12 k+6}\right] \chi_{6 k+\frac{1}{2}}(y)\nonumber\\
&+&\left[(44+256 k) q^{12 k+5}+(68+256 k) q^{12 k+7}\right] \chi_{6 k+1}(y)\nonumber\\
&-&\left[(68+256 k) q^{12 k+6}+(104+256 k) q^{12 k+8}\right] \chi_{6 k+\frac{3}{2}}(y)\nonumber\\
&+&\left[(104+256 k) q^{12 k+7}+(124 +256 k) q^{12 k+9}\right] \chi_{6 k+2}(y)\nonumber\\
&-&\left[(124+256 k) q^{12 k+8}+(132+256 k) q^{12 k+10}\right] \chi_{6 k+\frac{5}{2}}(y)\nonumber\\
&+&\left[(132+256 k) q^{12 k+9}+(152+256 k) q^{12 k+11}\right] \chi_{6 k+3}(y)\nonumber\\
&-&\left[(152+256 k) q^{12 k+10}+(188+256 k) q^{12 k+12}\right] \chi_{6 k+\frac{7}{2}}(y)\nonumber\\
&+&\left[(188+256 k) q^{12 k+11}+(212+256 k) q^{12 k+13}\right] \chi_{6 k+4}(y)\nonumber\\
&-&\left[(212+256 k) q^{12 k+12}+(216+256 k) q^{12 k+14}\right] \chi_{6 k+\frac{9}{2}}(y)\nonumber\\
&+&\left[(216+256 k) q^{12 k+13}+(236+256 k) q^{12 k+15}\right] \chi_{6 k+5}(y)\nonumber\\
&-&\left. \left[(236+256 k) q^{12 k+14}+(276+256 k) q^{12
k+16}\right] \chi_{6 k+\frac{11}{2}}(y)\right].\nn\\ \eea These states
survive only for $k=3$, \ie $\mZ_3$ projection. It is amusing to
observe how the number of representations of given spin $s=6
k+\frac{n}{2}$ grows with $k$ at the rate $256k$ for any $n$. This
is due to the possible distributions of derivatives among three
fields up to symmetries and total derivatives and to the structure
of higher spin supermultiplets \cite{Bianchi:2006ti}.

For higher multi-pletons the analysis is similar. It is clear that
only states with charge $Q=\pm n, \pm (n-2), ...$ are present in
the n-pleton spectrum. In particular $Q=0$ states are only present
when $n$ is even as already observed. We defer a detailed analysis to the future. For
the time being let us only display the partition functions for the
cyclic tetrapleton \bea
\cZ_{4,cycl}=\frac{1}{4}(\cZ_{\tiny\yng(1)}(q)^4 +
\cZ^2_{\tiny\yng(1)}(q^2)+ 2 \cZ_{\tiny\yng(1)}(q^4)) \eea and for the
totally symmetric tetrapleton \bea
\cZ_{\tiny\yng(4)}=\frac{1}{4!}(\cZ_{\tiny\yng(1)} ^4(q) + 6
\cZ^2_{\tiny\yng(1)}(q) \cZ_{\tiny\yng(1)}(q^2)+3 \cZ^2_{\tiny\yng(1)}
(q^2)+8 \cZ_{\tiny\yng(1)}(q^3)\cZ_{\tiny\yng(1)}(q)+6
\cZ_{\tiny\yng(1)}(q^4))\nn\\ \eea

The $\mZ_k$ projection on n-pletons reads \be \cZ_n^{\mZ_k} = {1\over
k} \sum_r \cZ_n^{(r)}(q,\omega^r) \ee and corresponds to keeping
only states with $Q=k n$ \ie integer multiples of $k$.

\section{KK excitations}

Let us now focus on the KK excitations, which deserve a separate
treatment. One can indeed write down the single-particle partition
function on $S^7$, decompose it into super-characters and identify
the $SO(2)$ charge sectors, relevant for the subsequent $\mZ_k$
projection \ie compactification on $\mC\mP^3$.

Introducing a chemical potential for the charge $Q$ ($t^Q$), the
super-character of an ultra-short 1/2 BPS representation of
$Osp(8|4)$ reads: \bea &&
\cX^{1/2BPS}_{\ell}(q,t)=\frac{t^{-2-\ell} q^{2+\ell}}{6
\left(1-t^2\right)^5 (1+q)^3}
\left[\ell^3 \left(-1+t^2\right)^2 (-1+q)^3 \right.\nonumber\\
&& \times \left(t^{6+2 \ell} \left(t^2-q\right)^2-\left(-1+t^2 q\right)^2\right)-6 \ell^2 \left(-1+t^2\right) (-1+q)^2
\nonumber\\
&& \times \left(t^{6+2 \ell} \left(t^2-q\right)^2 \left(-3+2 t^2+q\right)+\left(2+t^2 (-3+q)\right) \left(-1+t^2
q\right)^2\right)\nonumber\\
&& +6 t^{6+2 \ell} \left(t^2-q\right)^2 \left(-35+q (35+(-9+q) q)+2 t^4 \left(-5+q^2\right)\right.\nonumber\\
&& \left.+t^2 (35+q (-13+(-7+q) q))\right)
-\left(2 \left(-5+q^2\right)\right. \\
&&\left. +t^4 (-35+q (35+(-9+q) q))+t^2 (35+q (-13+(-7+q) q))\right)\nonumber\\
&& \times 6 \left(-1+t^2 q\right)^2 -\ell (-1+q)\left(t^{6+2 \ell} \left(t^2-q\right)^2 \left(-107+(70-11 q) q \right.
\right.\nonumber\\
&& \left. +t^4 (-47+(-2+q) q)-2 t^2 (-71+q (22+q))\right)
+\left(-1+t^2 q\right)^2 \nonumber\\
&&\left.\left.\times \left(47-(-2+q) q+2 t^2 (-71+q (22+q))+t^4
(107+q (-70+11 q))\right)\right)\right]\nonumber \eea

For $\ell =0$, corresponding to the gauged supergravity multiplet,
there is further shortening (null descendants) due to the presence
of conserved `currents' \ie stress-tensor, $SO(8)$ vector currents
and ${\bf 8}_s$ supercurrents. Taking this into account one finds
the following super-character \bea
\cX^{1/2BPS}_{\ell =0}(q)=\frac{1}{(1-q^2)^3}&& [(10t^2 +15 +10t^{-2})q^2-\nonumber\\
&& 2(15t^2 +10 +6 +10 +15t^{-2})q^3+\nonumber\\
&& (10t^2 +15 +10t^{-2} +3(6t^2+ 15+ 1+ 6t^{-2}))q^4-\nonumber\\
&& 4(t^2+ 6+ t^{-2})q^5 -(6t^2+ 15+ 1- 5+ 6t^{-2})q^6+\nonumber\\
&& 2(t^2+ 6+ t^{-2})q^7 -3q^8] \eea the denominator takes into
account derivatives (descendants). Quite remarkably this formula
coincides with the previous one when $\ell =0$.

After some algebra, putting $t=1$, one finds \be
\cX^{1/2BPS}_{\ell =0}(q) = {q^2 (3 q^3 - 7 q^2 - 7 q +35) \over
(1+q)^3}\ee a factor $(1-q)^2$
 cancels between numerator and denominator
meaning that not only $n_b=n_f$ and the sum with $\Delta^1$
vanishes but also the sum with $\Delta^2$ should vanish. This should
be related to the absence of quantum corrections to the negative vacuum energy,
\ie cosmological constant in the bulk.

The 1/2 BPS partition function is given by \be
\cZ^{\cN=8}_{1/2BPS} = \sum_\ell \cX^{1/2BPS}_\ell = {35 q^2 \over
(1-q^2)^2} \ee The simplicity of the result is due to `miraculous'
cancellations between bosonic and fermionic operators with the
same scaling dimensions in different KK multiplets \ie with
different $\ell$'s. This does not happen in $AdS_5/CFT_4$
holography, whereby (protected) bosonic operator have integer
dimensions and (protected) fermionic operators have half-integer
dimensions \cite{Bianchi:2003wx, Beisert:2003te, Beisert:2004di,
Bianchi:2004ww, Bianchi:2004xi, Bianchi:2005ze, Bianchi:2006ti}.

In order to perform the $\mZ_k$ projection it is useful to decompose
into $SO(2)$ charge sectors according to \be \cZ^{\cN=8
\rightarrow \cN=6}_{1/2BPS}(q,t) = {q^2 [(1+q^6) P_2(t) - (q +
q^5) P_3(t) + (q^2 + q^4) P_4(t) - q^3 P_5(t) ] \over (1-q t)^4
(1-q t^{-1})^4
(1+ q)^2} \ee where \bea && P_2(t) = 10 t^{+2} + 15 + 10 t^{-2} \nn\\
&& P_3(t) = 20 t^{+3} + 10 t^{+2} + 64 t^{+1} + 22 + 64 t^{-1} +
10 t^{-2} + 20 t^{-3}\nn\\&& P_4(t) = 15 t^{+4} + 8 t^{+3} + 104
t^{+2} + 48 t^{+1} + 175 + 48 t^{-1} + 104 t^{-2} + 8 t^{-3}+ 15 t^{-4}\nn\\
&& P_5(t) = 4 t^{+5} + 2 t^{+4} + 64 t^{+3} + 40 t^{+2} + 196
t^{+1} + 88 + \nn\\ &&\qquad + 196 t^{-1} + 40 t^{-2} + 64 t^{-3}+
2 t^{-4} + 4 t^{-5} \nn\\\eea

Depending on the choice of $k$ one can recognize the surviving 1/2
BPS states as those with $Q=kn$. In formulae one has to replace
$t$ with $\omega^r$ and sum over $r=0, ..., k-1$.

\chapter{Instantons in $\mC \mP ^3$}
\label{chapter7}
An interesting characterstic of $AdS_4\times \mC\mP ^3$, which $AdS_5\times S^5$ is lacking, is the 
existence of stringy instantons in $\mC\mP ^3$ which arise from the string worldsheet which wraps a 
topologically non-trivial two-cycle $\mC\mP ^1 \simeq S^2$ of $\mC \mP ^3$ in the Wick rotated theory. This 
is a stringy counterpart of the instantons of two dimensional $\mC\mP ^n$ sigma-models. The two-cycle 
corresponds to the closed K\"ahler two-form $J_2$ on $\mC \mP ^3$. The consistent 
gauge fixing of kappa-symmetry does not allow to reduce the string action to the supercoset sigma model, \ie 
to eliminate the eight fermionic modes corresponding to the broken supersymmetries. This stringy instanton 
has twelve fermionic zero modes all corresponding to unbroken supersymmetries of the background and there 
are no zero modes associated with broken supersymmetries. These twelve zero modes are divided into eight 
massive fermionic zero modes, which are four copies of the two component Killing spinor on $S^2$, and four 
other modes, which are two copies of massless chiral and anti-chiral fermion on $S^2$ electrically 
coupled to the electromagnetic potential created on $S^2$ by a monopole placed in the center of $S^2$. The 
presence of the stringy instanton and its fermionic zero modes may generate non-perturbative corrections to 
the string effective action, which may affect its properties and if so should be taken into account in 
studying, \eg the $AdS_4/CFT_3$ correspondence. The instantons may contribute to the 
worldsheet S-matrix.
Due to the presence of the fermionic zero modes, the non-perturbative amplitude should contain operator 
insertions up all fermionic zero modes.  

The Green-Schwarz superstring action in a generic Type IIA supergravity background has the following form:
\be 
S=-\frac{1}{4\pi \alpha ^\prime}\int d^2\xi \sqrt{-h}h^{IJ}\cE _I^{\,\,\, A}\cE _J^{\,\,\, B}\eta 
_{AB}-\frac{1}{2\pi\alpha ^\prime}\int B_2
\ee
where $\xi ^I$ ($I=0,1$) are the worldsheet coordinates, $h_{IJ}(\xi)$ is an intrinsic worldsheet metric, 
$\cE _I^{\,\,\, A}$ are worldsheet pullbacks of target superspace vector supervielbeins and $B_2$ is the 
pullback of the NS-NS two-form. One has to substitute the expressions for the vielbeins and the NS-NS two-
form up to second order in fermions in the above action and keep only terms up to quadratic order in 
fermions. After Wick rotation the action takes the following form:
\bea
S_E &=&\frac{e^{\frac{2}{3}\phi _0}}{4\pi\alpha ^\prime}\int d^2\xi \sqrt{h}h^{IJ}(e_I^{\,\,\,a}e_J^{\,\,
\,b}\delta _{ab} + e_I^{\,\,\,a^\prime}e_J^{\,\,\,b^\prime}\delta _{a^\prime b^\prime} )\label{SE}\\
&+& \frac{e^{\frac{2}{3}\phi _0}}{2\pi\alpha ^\prime}\int d^2\xi \Theta (\sqrt{h}h^{IJ} + i\varepsilon 
^{IJ}\Gamma _{11})[ie_I^{\,\,\,A}\Gamma _A \nabla _J \Theta -\frac{1}{R}e_I^{\,\,\,A}e_J^{\,\,
\,B}\Gamma _A \cP _6 \gamma ^5\Gamma _B\Theta] \nn
\eea
and the kappa-symmetry matrix is
\bea
\Gamma =-\frac{i}{2\sqrt{\det G_{IJ}}}\varepsilon ^{IJ}\cE _I^{\,\,\,A}\cE _J^{\,\,\,B}\Gamma _{AB}
\Gamma _{11}, \quad \Gamma ^2=1
\eea
where $\Gamma _{11}$ is given in terms of $D=4$ gamma matrices $\Gamma _{11}=\gamma ^5\otimes \gamma ^7$, 
$\Gamma ^A$ are $D=10$ gamma matrices,
$e^{\frac{2}{3}\phi _0}$ is the vacuum expectation value of the dilaton, where $R$ is the radius of the 
$S^7$ sphere whose base is $\mC\mP ^3$ and $l_p$ is the eleven dimensional Planck length related to the 
string tension as $l_p=e^{\frac{1}{3}\phi _0}\sqrt{\alpha ^\prime}$. The 32-component fermionic variable 
$\Theta ^\alpha$ is split by projectors $\cP _6$ and $\cP _2$ into the 24-component spinors $\vartheta ^
{\alpha a^\prime}$ ($\alpha =1,..,4; a^\prime =1,...,6$) corresponding to 
the 24 supersymmetries of the $AdS_4\times\mC\mP ^3$ solution and the 8-component spinors $v^{\alpha q}$ 
($q=1,2$) corresponding to the broken supersymmetries. $e^a(x)$ and $e^{a\prime}(y)$ are the vielbein
for $AdS_4$ of radius $R/ 2$ and for $\mC\mP ^3$, respectively. The induced metric on the worldsheet is 
$G_{IJ}=\cE_I^{\,\,\,A}\cE_J^{\,\,\,B}\eta _{AB}$. To identify the stringy instanton in $\mC\mP ^3$ it is 
convenient to consider the Fubini-Study metric on $\mC\mP ^3$ \cite{Pope:1984bd}
\bea
ds^2=R^2\left(\frac{1}{4}(d\theta ^2 + \sin ^2 \theta (d\varphi +\frac{1}{2}\sin ^2\alpha \sigma _3)^2)+\sin 
^2 \frac{\theta}{2}d\alpha ^2 \right. \nn\\
\left. +\frac{1}{4}\sin ^2 \frac{\theta}{2}\sin ^2\alpha (\sigma _1^2+\sigma _2^2+ 
\cos ^2\alpha \sigma _3^2) \right)
\label{Fubini-Study}
\eea
where $0\leq \theta \leq \pi$, $0\leq \varphi \leq 2\pi$ and $0\leq \alpha \leq \frac{\pi}{2}$, and $\sigma 
_1$, $\sigma _2$, $\sigma _3$ are three left-invariant one forms on $SU(2)$ and obey $d\sigma _1=-\sigma 
_2\sigma_3$ etc. Notice that now $\theta$ and $\varphi$ parametrize a two-sphere of radius $R/2$, which is 
topologically non-trivial and associated to the K\"ahler form on $\mC\mP ^3$. If a stringy instanton wraps 
this sphere once then $\theta$ and $\varphi$ can be identified with the string worldsheet coordinates, while 
all the other $\mC\mP ^3$ and $AdS_4$ coordinates are worldsheet constants for the instanton solution. Thus 
the pullback on the strngy instanton of the metric (\ref{Fubini-Study}) of $\mC\mP ^3$ of radius $R$ is the 
metric of radius $R/2$:
\be
ds^2=\frac{R^2}{4}(d\theta ^2 + \sin ^2\theta d\varphi ^2)
\ee
In this coordinate system the $S^2$ vielbein $e^i$ and the spin connection $w^{ij}_{S^2}$ ($i,j=1,2$) can be 
chosen to have the following form:
\be
e^1=\frac{R}{2}d\theta , \quad e^2=\frac{R}{2}\sin\theta d\varphi , \quad w^{12}_{S^2}=\cos \theta d\varphi
\ee
and the $S^2$ curvature 2-form is 
\be
R^{ij}=dw ^{ij}_{S^2}=\frac{4}{R^2}e^ie^j.
\ee
Let us consider the bosonic part of the Wick rotated action [\ref{SE}], which is:
\be
S_E=\frac{T}{2}\int d^2\xi \sqrt{h}h^{IJ}e_I^{\,\,\,i}e_J^{\,\,\,j}\delta _{ij}
\ee
where $T=\frac{e^{\frac{2}{3}\phi _0}}{2\pi\alpha ^\prime}$ and $e^i$ are the vielbeins on $S^2$. To 
discuss the instanton it is convenient to introduce complex coordinates both on the worldsheet and in the 
target space. In the ($z,\bar{z}$) coordinate system on the worldsheet the action takes the form
\be
S_E=\frac{T}{2}\int d^2z e_z^{\,\,\,i}e_{\bar{z}}^{\,\,\,j}\delta _{ij}
\ee
where it is taken into account that in the conformal gauge $\sqrt{h}h^{IJ}=\delta ^{IJ}$. The Fubini-Study 
metric on $\mC\mP^1$ is
\be
ds^2_{\mC\mP^1}=\frac{d\zeta d\bar{\zeta}}{(1+|\zeta |^2)^2}
\label{cp1metric}
\ee
If we choose 
\be
\zeta =\tan \frac{\theta}{2}e^{i\varphi}
\ee
then the Fubini-Study 
metric on $\mC\mP^1$ gets the form of the metric on $S^2$ of radius $\frac{1}{2}$:
\be
ds^2=\frac{1}{4}(d\theta ^2+\sin ^2\theta d\varphi ^2)
\ee
In the $\zeta, \bar{\zeta}$ coordinate system the string action takes the following form:
\be
S_E=\frac{TR^2}{4}\int d^2 z \frac{|\partial \zeta |^2 + |\bar{\partial} \zeta |^2}{(1+|\zeta |^2)^2}
\ee
The local minimum is at $\bar{\partial} \zeta =0$ or at $\partial \zeta =0$, which means that the embedding 
is given by a holomorphic function $\zeta =\zeta (z)$ for the instanton or by an anti-holomorphic function 
$\zeta =\zeta (\bar{z})$ for the anti-instanton. This is the classical instanton solution of the 
two-dimensional $O(3)$ sigma-model \cite{Polyakov:1975yp} or rather its extension to $\mC\mP^3$ 
\cite{Golo:1978de, Golo:1978dd, D'Adda:1978uc}. The remaining 
part of the action can be shown to be a topological invariant. The Virasoro constraints should also be 
satisfied by the classical string solution. The Virasoro constraints in the conformal gauge have the form
\be
\frac{\partial \zeta \partial \bar{\zeta}}{(1+|\zeta |^2)^2}=0
\ee
which obviously are satisfied by the (anti)instanton solution. 

To discuss fermionic sector it will be convenient to choose the $\mC\mP^3$ gamma matrices as:
\bea
\gamma ^{a^\prime}=(\rho^i\otimes {\bf 1}, \rho^3\otimes\gamma ^{\tilde{a}}), \quad \gamma _7=\rho ^3\otimes 
\gamma _{\tilde{5}},\quad \gamma _{\tilde{5}}=\frac{1}{4!}\varepsilon 
_{\tilde{a}\tilde{b}\tilde{c}\tilde{d}}\gamma ^{\tilde{a}\tilde{b}\tilde{c}\tilde{d}}
\eea
$\gamma _{\tilde{a}}$, $\tilde{a}=3,4,5,6$, are $4\times 4$ Dirac gamma matrices corresponding to 
the four-dimensional subspace of $\mC\mP^3$ orthogonal to $\mC\mP^1$ and $\gamma _{\tilde{5}}^2=1$.
In the fermionic sector it is natural to impose on the fermionic fields the conventional kappa-symmetry 
gauge-fixing condition, which is the following:
\be
\frac{1}{2}(1+\Gamma)\Theta =\frac{1}{2}(1-\gamma _5\gamma _{\tilde{5}})\Theta =0
\label{fermionickappa}
\ee
This means that the fermions split into two sectors according to their chiralities in $AdS_4$ and in the 
four dimensional subspace of $\mC\mP^3$ orthogonal to $\mC\mP^1$.
\bea 
&&\Theta _+: \quad \gamma _5\Theta _+=\gamma _{\tilde{5}}\Theta _+=\Theta _+\\
&&\Theta _-: \quad \gamma _5\Theta _-=\gamma _{\tilde{5}}\Theta _-=-\Theta _-
\eea
The supersymmetry projection matrices $\cP_2$ and $\cP_6$ act on these two sets as:
\bea
&&\cP_6\Theta _+ =\Theta _+ =\vartheta _+\\ 
&&\cP_2\Theta _+ = v_+=0\\
&&\cP_6\Theta _-=\frac{1}{2}(1-\rho ^3\tilde{J})\Theta _-=\vartheta _-\\
&&\cP_2\Theta _-=\frac{1}{2}(1+\rho ^3\tilde{J})\Theta _-=v
\eea
It follows that all the eight $\vartheta _+$ are fermions corresponding to unbroken supersymmetries of the 
$AdS_4\times \mC\mP^3$ background. In the $\Theta_-$ sector four fermions $\vartheta _-$ correspond to 
unbroken supersymmetries and the other four $v$ to the broken ones. Since the kappa-symmetry projector 
commutes with the supersymmetry projectors, it is not possible to choose the kappa-symmetry gauge-fixing 
condition in such a way to put to zero all the eight broken supersymmetry fermions.

Examining fermionic equations for instanton configuration one concludes 
that in the $\Theta_+$ sector the stringy instanton has eight fermionic zero modes. In the spherical 
coordinates they have the following form:
\be
\vartheta _+=e^{-\frac{i}{2}\theta \rho^1}e^{\frac{i}{2}\varphi \rho^3}\epsilon _+=
\left( \cos \frac{\theta}{2}-i\rho^1 \sin \frac{\theta}{2}\right)\left( \cos \frac{\varphi}{2}+i\rho^3 \sin 
\frac{\varphi}{2}\right)\epsilon _+
\ee
where $\rho^1=\sigma ^1$, $\rho^2=\sigma ^3$ and $\rho^3=-\sigma ^2$ are Pauli matrices and $\epsilon _+$ is 
an arbitrary constant spinor satisfying the chirality conditions 
$\gamma_5\epsilon_+=\gamma_{\tilde{5}}\epsilon _+=\epsilon _+$.
The other equation of motion requires to put $v=0$. In the $\vartheta_-$ sector the stringy instanton has 
four zero modes which have the form:
\be
\vartheta _- =\frac{1}{2}e^{-\frac{i}{2}\rho _3\varphi}\left[(1+\rho ^3)\lambda _-(\zeta)
+(1-\rho ^3)\mu _-(\bar{\zeta})\right]
\ee
where $\lambda _-(\zeta)$ and $\mu _-(\bar{\zeta})$ are holomorphic and anti-holomorphic spinors in the 
projective coordinates $\zeta$ and $\bar{\zeta}$ of $S^2\simeq \mC\mP ^1$. They are anti-chiral in the 
directions transverse to the instanton. For the anti-instanton the solution takes the same form but with 
$\lambda_-(\bar{\zeta})$ and $\mu_-(\zeta)$. These fermionic zero modes do not contribute to the bosonic 
equations. 

Thus, the stringy instanton wrapping the non-trivial two-cycle inside $\mC\mP^3$ has twelve fermionic zero 
modes. The eight fermionic fields $\vartheta_+$ and four $\vartheta_-$ correspond to twelve (of the twenty 
four) supersymmetries of the $AdS_4\times \mC\mP^3$ background which are linearly realized on the string 
worldsheet. Thus, the fermionic zero modes play the role similar to goldstinos, which break supersymmetry.

\chapter*{Conclusions}

In this thesis we have discussed non-perturbative phenomena emerging in gauge 
and in string/supergravity theories. We discussed instantons in supersymmetric Yang-Mills theories. 
We computed the partition function of 5D minimal supersymmetric $U(1)$ gauge theory with extra adjoint 
matter in general $\Omega$-background and have shown in particular that unlike the case with no extra 
matter, the partition function with extra adjoint at some special values of the parameters directly 
reproduces the generating function for the Poincar\`e polynomial of the moduli space of instantons.
We discussed instantons and their effects in string theories. In particular
`gauge' instantons may generate a VY-ADS-like superpotential of the form
$$W \approx {\Lambda^{\beta} \over \phi^{\beta -3}}$$
where $\beta$ is the one-loop coefficient in the expansion of the $\beta$ function and
$\Lambda^{\beta} = M_s^\beta e^{-T(C)}$.
`Exotic' instantons may generate a non-perturbative
superpotential of the form $$W \approx M_s^{3-n} e^{-S_{EDp'}(C')} \phi^n
\quad (n=0,1,...)$$
Combining the two kinds of superpotentials one can achieve (partial) moduli stabilization and SUSY breaking! 
The same may happen when only one kind of superpotential is generated in the presence of fluxes, which we 
discussed in details, and in another dynamical effects, such as FI terms \cite{MBJFMFF}.
When extra zero-modes are present, threshold corrections to (higher-derivative) terms may arise. We 
illustrated this possibility for a compactification to $D=6$ on $T^6/\mZ_2$, where a fully non-perturbative 
four hyperini amplitude (Fermi interaction) can be computed exploiting Heterotic - Type I duality.
Threshold corrections to gauge couplings in freely acting orbifolds $T^6/\mZ_2\times \mZ_2$
were computing by similar means.
A  by-product of the analysis in $D=6$, an economical mechanism of moduli
stabilization can be exploited whereby non-anomalous $U(1)$'s in $D=4$ eat would-be hypers due to anomalies 
in $D=6$.
The behaviour of D-brane instanton effects in the presence of fluxes or under wall crossing and the 
reformulation of (unoriented) D-brane instanton calculus in terms of localization are extremely active 
subjects. The part of the vast literature on the subject are \cite{Lerda_etal}, \cite{FMPT}, \cite{Blum}, 
\cite{BWnieist}, \cite{U(4)}, \cite{MBJFMFF}, \cite{BCKWreview}.

Then we have discussed $AdS/CFT$ correspondence and the role of instantons particularly in $AdS_5/CFT_4$.
We discussed another interesting application of the correspondence, $AdS_4/CFT_3$.
This allows to investigate the worldvolume theory of M2-branes, 
the basic objects of M-theory. In this context we considered $\cN=8$ supergravtiy on 
$AdS_4\times S^7$, which is the low energy limit of M-theory compactified on $S^7$.
We have re-analyzed the KK spectrum of $d=11$ supergravity on
$S^7$ and $S^7/\mZ_k$. The latter includes monopole operators dual
to charged states in Type IIA on $\mC\mP^3$. To this end we have
presented some group theoretic methods for the decomposition of
the $SO(8)$ into $SO(6)\times SO(2)$ valid also for other
cosets\cite{Caviezel:2008ik, Haack:2009jg, Kounnas:2007dd} where resolution of the mixing among
various fluctuations should be possible on the basis of symmetry
arguments. In particular, massless vectors associated to Killing
vectors in generic flux vacua with isometries have been recently
discussed in \cite{Bianchi:2010cy}.
We have then considered higher spin symmetry enhancement. We have
displayed the partition functions for singletons, doubletons and
tripletons and discussed in details higher spin fields and 1/2 BPS
states corresponding to KK excitations of $\cN =6$ gauged
supergravity. It would be worth pursuing the analysis to higher
n-pletons and to cases with lower supersymmetry, yet based on
internal coset manifolds. It would be also nice to try to get semi-realistic $AdS_4$ 
compactifications in the both cases when the Chern-Simons levels sum up to zero or not using orientifold 
projections \cite{Aharony:2008gk}, \cite{Forcella:2009jj}.
We discussed also instantons in $\mC\mP^3$.

 \vskip 1cm

\chapter*{Appendix A}
\addcontentsline{toc}{chapter}{Appendix A}
\setcounter{equation}{0}
\renewcommand{\theequation}{A.\arabic{equation}}
\quad

General formula
 \bea
d_{(\ell _1 , \ell _2, \ell _3, \ell
_4)}^{SO(8)}=&&\frac{1}{4320}\times
(1+\ell _1) (1+\ell _2) (1+\ell _3) (1+\ell _4)\nn\\
&& (2+\ell _1+\ell _2) (2+\ell _2+\ell _3) (2+\ell _2+\ell _4) \nn\\
&& (3+\ell _1+\ell _2+\ell _3) (3+\ell _1+\ell _2+\ell _4) (3+\ell_2+\ell _3+\ell _4)\nn\\
&& (4+\ell _1+\ell _2+\ell _3+\ell _4) (5+\ell_1+2 \ell _2+\ell _3+\ell _4) \eea

Specific cases (KK harmonics)

\bea && d_{(\ell , 0, 0, 0)}^{SO(8)}= \frac{1}{360} (1+\ell)
(2+\ell) (3+\ell)^2 (4+\ell) (5+\ell)
\quad \leftrightarrow \quad Y_{N_1}\nn\\
&& d_{(\ell, 1, 0, 0)}^{SO(8)}= \frac{1}{60} (1+\ell) (3+\ell)
(4+\ell)^2 (5+\ell) (7+\ell)
\quad \leftrightarrow \quad Y_{N_7}\nn\\ %
&& d_{(\ell, 0, 1, 1)}^{SO(8)}= \frac{1}{24} (1+\ell) (2+\ell)
(4+\ell)^2
(6+\ell) (7+\ell) \quad \leftrightarrow \quad Y_{N_{21}}\nn\\
&& d_{(\ell, 2, 0, 0)}^{SO(8)}= \frac{1}{18} (1+\ell) (4+\ell)
(5+\ell)^2 (6+\ell) (9+\ell) \quad \leftrightarrow \quad
Y_{N_{27}}\\
&& d_{(\ell, 0, 2, 0)}^{SO(8)}= d_{(\ell, 0, 0, 2)}^{SO(8)}=
\frac{1}{36} (1+\ell) (2+\ell) (3+\ell)
(5+\ell) (6+\ell) (7+\ell) \quad \leftrightarrow \quad Y_{N_{35}}\nn\\
&& d_{(\ell, 0, 1, 0)}^{SO(8)}= d_{(\ell, 0, 0,
1)}^{SO(8)}=\frac{1}{90} (1+\ell) (2+\ell) (3+\ell) (4+\ell)
(5+\ell) (6+\ell)\nn\\
&& d_{(\ell, 1, 1, 0)}^{SO(8)}= d_{(\ell, 1, 0, 1)}^{SO(8)}=
\frac{1}{18} (1+\ell) (3+\ell) (4+\ell) (5+\ell) (6+\ell) (8+\ell)\nn\eea

\chapter*{Appendix B}
\addcontentsline{toc}{chapter}{Appendix B}
\setcounter{equation}{0}
\renewcommand{\theequation}{B.\arabic{equation}}
\quad
\section*{Zero Charge states}

In this Appendix we list states with $Q=0$ in the KK towers of $S^7$ after the
decomposition of $SO(8)$ into $SO(6)\times SO(2)$.

Bosons:
\bea (\ell,0,0,0)_{\ell\geq 0}\rightarrow
\left[0,\frac{\ell}{2},\frac{\ell}{2}\right]\eea 
\bea
(\ell,1,0,0)_{\ell\geq 0}\rightarrow
\left[0,\frac{\ell}{2},\frac{\ell}{2}\right]+\left[0,\frac{\ell}{2}+1,\frac{\ell}{2}+1\right]\nonumber\\
+\left[1,\frac{\ell}{2}+1,\frac{\ell}{2}-1\right
]+\left[1,\frac{\ell}{2}-1,\frac{\ell}{2}+1\right ] 
\eea 
\bea
(\ell-1,0,1,1)_{\ell\geq 1}\rightarrow
\left[0,\frac{\ell}{2}+2,\frac{\ell}{2}-2\right
]+\left[0,\frac{\ell}{2}-2,
\frac{\ell}{2}+2\right]+\left[1,\frac{\ell}{2}+1,\frac{\ell}{2}-1\right
]\nonumber\\+\left[1,\frac{\ell}{2}-1,\frac{\ell} {2}+1\right
]+\left[0,\frac{\ell}{2},\frac{\ell}
{2}\right]+\left[0,\frac{\ell}{2},\frac{\ell}{2}\right
]\nonumber\\+\left[1,\frac{\ell}{2},\frac{\ell} {2}-2\right
]+\left[1,\frac{\ell}{2}-2,\frac{\ell}{2}\right
]+\left[2,\frac{\ell}{2}-1,\frac{\ell}{2}-1\right] \nn\\
\eea 
\bea
(\ell-2,2,0,0)_{\ell\geq 2}\rightarrow
\left[2,\frac{\ell}{2}-3,\frac{\ell}{2}+1\right
]+\left[1,\frac{\ell}{2}-2, \frac{\ell}{2}\right
]+\left[1,\frac{\ell}{2},\frac{\ell}{2}-2\right
]\nonumber\\+\left[0,\frac{\ell}{2},\frac{\ell} {2}\right]
+\left[0,\frac{\ell}{2}-1,\frac{\ell}{2}-1\right]+\left[1,\frac{\ell}{2}-1,\frac{\ell}
{2}+1\right]\nonumber\\+\left[1,\frac{\ell}{2}+1,
\frac{\ell}{2}-1\right]+\left[2,\frac{\ell}{2}-1,\frac{\ell}{2}-1\right]+\left[2,\frac{\ell}{2}+1,
\frac{\ell}{2}-3\right]+\left[0,\frac{\ell}{2}+1,\frac{\ell}{2}+1\right]\nn\\
 \eea
\bea (\ell,0,2,0)_{\ell\geq 0}\rightarrow
\left[0,\frac{\ell}{2}-1,\frac{\ell}{2}+3\right ]+\left[0,
\frac{\ell}{2}+1,\frac{\ell}{2}+1\right ]+\left[0,\frac{\ell}{2}+3,\frac{\ell}{2}-1\right]\nonumber\\
+\left[1,\frac{\ell}{2}-1,\frac{\ell}{2}+1\right ]
+\left[1,\frac{\ell}{2}+1,\frac{\ell}{2}-1\right
]+\left[2,\frac{\ell}{2}-1,\frac{\ell} {2}-1\right] \nn\\
\eea 
\bea
(\ell-2,0,0,2)_{\ell\geq 2}\rightarrow
\left[0,\frac{\ell}{2}-1,\frac{\ell}{2}-1\right]+\left[0,
\frac{\ell}{2}-3,\frac{\ell}{2}+1\right]+\left[1,\frac{\ell}{2},\frac{\ell}{2}-2\right]\nonumber\\
+\left[1,\frac{\ell}{2}-2,\frac{\ell}{2}\right]+\left[2,\frac{\ell}{2}-1,\frac{\ell}{2}-1
\right]+\left[0,\frac{\ell}{2}+1,\frac{\ell}{2}-3\right]\nn\\ 
\eea

Fermions:
\bea (\ell,0,0,1)_{\ell\geq 0}\rightarrow
\left[0,\frac{\ell}{2}-1,\frac{\ell}{2}+1\right]+
\left[0,\frac{\ell}{2}+1,\frac{\ell}{2}-1\right]+\left[1,\frac{\ell}{2},\frac{\ell}{2}\right]
 \eea
\bea (\ell-1,0,1,0)_{\ell\geq 1}\rightarrow
\left[0,\frac{\ell}{2}-1,\frac{\ell}{2}+1\right]+
\left[0,\frac{\ell}{2}+1,\frac{\ell}{2}-1\right]+\left[1,\frac{\ell}{2}-1,\frac{\ell}{2}-1\right]\nn\\
\eea
\bea (\ell-1,1,1,0)_{\ell\geq 1}\rightarrow
\left[1,\frac{\ell}{2}-2,\frac{\ell}{2}+2\right]+2\left[1,\frac{\ell}{2},\frac{\ell}{2}\right]
+\left[1,\frac{\ell}{2}+2,\frac{\ell}{2}-2\right]\nonumber\\
+\left[0,\frac{\ell}{2},\frac{\ell}{2}+2\right]+
\left[0,\frac{\ell}{2}+2,\frac{\ell}{2}\right]+
\left[0,\frac{\ell}{2}-1,\frac{\ell}{2}+1\right]\nonumber\\
+\left[0,\frac{\ell}{2}+1,\frac{\ell}{2}-1\right]+
\left[2,\frac{\ell}{2}-2,\frac{\ell}{2}\right]+
\left[2,\frac{\ell}{2},\frac{\ell}{2}-2\right]
+\left[1,\frac{\ell}{2}-1,\frac{\ell}{2}-1\right]\nn\\ 
\eea 
\bea
(\ell-2,1,0,1)_{\ell\geq 2}\rightarrow
\left[1,\frac{\ell}{2}-3,\frac{\ell}{2}+1\right]
+2\left[1,\frac{\ell}{2}-1,\frac{\ell}{2}-1\right]+
\left[1,\frac{\ell}{2}+1,\frac{\ell}{2}-3\right]\nonumber\\
+\left[0,\frac{\ell}{2}-2,\frac{\ell}{2}\right]+
\left[0,\frac{\ell}{2},\frac{\ell}{2}-2\right]+
\left[0,\frac{\ell}{2}-1,\frac{\ell}{2}+1\right]\nonumber\\
+\left[0,\frac{\ell}{2}+1,\frac{\ell}{2}-1\right]+
\left[2,\frac{\ell}{2}-2,\frac{\ell}{2}\right]+
\left[2,\frac{\ell}{2},\frac{\ell}{2}-2\right]
+\left[1,\frac{\ell}{2},\frac{\ell}{2}\right]\nn\\
 \eea

\chapter*{Appendix C}
\addcontentsline{toc}{chapter}{Appendix C}
\setcounter{equation}{0}
\renewcommand{\theequation}{C.\arabic{equation}}
\quad
\section*{Generating functions for $SO(8)$ representations}

The generating function for multiplicities of the scalar
spherical harmonics on $S^7$ is given by
\bea
\cF_{N_1}(q)=\frac{1+q}{(1-q)^7}
\eea
The coefficient of $q^\ell$ gives the dimension of the $SO(8)$
representation with Dynkin
label $(\ell,0,0,0)$.

The generating function for vector spherical harmonics with
$SO(8)$ Dynkin label $(\ell-1,1,0,0)$ reads:
\bea
\cF_{N_7}(q)=\frac{(28-36q+35q^2-21q^3+7q^4-q^5) q}{(1-q)^7} 
\eea

For two-form spherical harmonics with $SO(8)$ Dynkin label
$(\ell-1,0,1,1)$ the generating function is:
\bea
\cF_{N_{21}}(q)=\frac{(56-42q+22q^2-7q^3+q^4)q^2}{(1-q)^7} 
\eea

For second rank symmetric traceless harmonics the $SO(8)$
Dynkin index is $(\ell,2,0,0)$ and the generating function is
given by the following formula:
\bea 
\cF_{N_{27}}(q)=\frac{4(75-175q+203q^2-133q^3+47q^4-7q^5)q^2}{(1-q)^7}
\eea

Finally, for three-form spherical harmonics with $SO(8)$ Dynkin
label $(\ell-1,0,2,0)$ (or $(\ell-1,0,0,2)$) one has
\bea
\cF_{N_{35}}(q)=\frac{(35-21q+7q^2-q^3)q^2}{(1-q)^7}
\eea
Let us complete the description with the spectrum of spinor
spherical harmonics.

For gravitini with Dynkin labels $(\ell,0,0,1)_{\ell\geq0}$  and
$(\ell-1,0,1,0)_{\ell\geq1}$, the generating function is:
\bea
\cF_{gravitini}(q)=\frac{8q}{(1-q)^7} 
\eea

For spinors with Dynkin labels $(\ell-1,1,1,0)_{\ell\geq1}$ and
$(\ell-2,1,0,1)_{\ell\geq2}$ one has
\bea
\cF_{spinor}(q)=\frac{8q^2(20-35q+35q^2-21q^3+7q^4-q^5)}{(1-q)^7}.
\eea

\chapter*{Appendix D}
\addcontentsline{toc}{chapter}{Appendix D}
\setcounter{equation}{0}
\renewcommand{\theequation}{D.\arabic{equation}}
\quad
\section*{Generating functions for $SO(6)$ representations}

In this Appendix we present the decomposition of the $SO(8)$
generating functions under $SO(6)\times SO(2)$. Below a factor of
${(1-q t^{-1})^{-4} (1-q t)^{-4}}$ is always understood.

For $(\ell,0,0,0)$ one has: \be
\hat{\cF}_{graviton}(q)=1-q^2 \ee For $(\ell,1,0,0)$ one
has: \bea
&&\hat{\cF}_{gb1}(q,t)=6 t^2-4 t q-4 t^3 q+q^2+t^4 q^2\nonumber \\
&&\hat{\cF}_{gb2}(q,t)=1-q^2\\
 &&\hat{\cF}_{gb3}(q,t)=15+36 q^2- 4 q^3 t^{-3}-4 t^3 q^3+16 q^4+q^6 +(16 q^2+6 q^4)t^{-2}+\nonumber\\
&& t^2 (16 q^2+6 q^4)-(24 q+24 q^3+4 q^5)t^{-1}-t (24 q+24 q^3+4 q^5)\nonumber\\
&& \hat{\cF}_{gb4}(q,t)=6t^{-2}- 4 q t^{-3}- 4 qt^{-1}+q^2+
q^2t^{-4}\nonumber \eea 
For $(\ell-1,0,1,1)$ one has: \bea
&& \hat{\cF}_{gb}^1(q,t)=4 t^3 q-6 t^2 q^2-t^4 q^2+4 t q^3-q^4\nonumber\\
&& \hat{\cF}_{gb}^2(q,t)=4 t q-q^2-6 t^2 q^2+4 t^3 q^3-t^4 q^4\nonumber\\
&& \hat{\cF}_{gb}^3(q,t)=-35 q^2+4 t^3 q^3-16 q^4 - 6 q^4t^{-2}-q^6 - t^2(16 q^2+6 q^4)+\nonumber\\
&& (24 q^3+4 q^5)t^{-1}+t (20 q+24 q^3+4 q^5)\nonumber
\eea
\bea
&& \hat{\cF}_{gb}^4(q,t)=-35 q^2 +4 q^3t^{-3}-16 q^4-6 t^2 q^4-q^6 - (16 q^2+6 q^4)t^{-2}+\nonumber\\
&& t (24 q^3+4 q^5) +(20 q+24 q^3+4 q^5)t^{-1}\\
&& \hat{\cF}_{gb}^5(q,t)=4 qt^{-1}-q^2- 6 q^2 t^{-2}+ 4 q^3t^{-3} -q^4t^{-4}\nonumber\\
&& \hat{\cF}_{gb}^6(q,t)= 4 q t^{-3}- q^2t^{-4}- 6 q^2 t^{-2}+ 4 q^3 t^{-1}-q^4\nonumber\\
&& \hat{\cF}_{gb}^7(q,t)=6 t^2 q^2-4 t q^3-4 t^3 q^3+q^4+t^4 q^4\nonumber\\
&& \hat{\cF}_{gb}^8(q,t)= 6 q^4 t^{-2}+6 t^2 q^4- q^2 (20 q+ 4
q^3) t^{-1} - t q^2 (20 q+ 4 q^3) +\nn\\
&&q^2(20+15
q^2+q^4)\nonumber\\
&& \hat{\cF}_{gb}^9(q,t)= 6 q^2 t^{-2}- 4 q^3 t^{-3}- 4 q^3
t^{-1}+q^4+ q^4 t^{-4}\nonumber \eea 
For $(\ell-2,2,0,0)$' one
has: \bea
&& \hat{\cF}_{sc1}^1(q,t)=6 t^2 q^4+6 t^6 q^4 -t^3 q^2 (20 q+4 q^3) -t^5 q^2 (20 q+4 q^3)+\nn\\
&&t^4 q^2 (20+15 q^2+q^4)\nonumber\\
&& \hat{\cF}_{sc1}^2(q,t)=6 t^2 q^2-4 t q^3-4 t^3 q^3+q^4+t^4 q^4\nonumber\\
&& \hat{\cF}_{sc1}^3(q,t)= 6 q^2 t^{-2}- 4 q^3 t^{-3}- 4 q^3 t^{-1} +q^4 + q^4 t^{-4}\nonumber\\
&& \hat{\cF}_{sc1}^4(q,t)=-1+ 4 q t^{-1}+4 t q-q^2- 6 q^2 t^{-2}-6
t^2 q^2+ 4 q^3 t^{-3}+4 t^3 q^3-\nn\\
&& q^4 t^{-4}-t^4
q^4\nonumber\\
&& \hat{\cF}_{sc1}^5(q,t)=q^2-q^4\nonumber\\
&& \hat{\cF}_{sc1}^6(q,t)=-36 q^2-16 q^4- 6 q^4 t^{-2}-6 t^6 q^4-q^6-t^2 (6+32 q^2+12 q^4)+\nn\\
&& t (24 q+ 28 q^3+ 4 q^5)+t^3 (24 q+ 28 q^3+ 4 q^5)+ (24 q^3 +4 q^5) t^{-1}+\nn\\
&& t^5 (24 q^3+ 4 q^5)-t^4 (36 q^2+16 q^4+q^6)\nonumber\\
&& \hat{\cF}_{sc1}^7(q,t)=-36 q^2-16 q^4- 6 q^4 t^{-6}-6 t^2 q^4-q^6- (6+32 q^2+12 q^4)t^{-2}+\nn\\
&& (24 q^3+4 q^5)t^{-5}+t (24 q^3+4 q^5)+ (24 q+28 q^3+4 q^5) t^{-3}+\nn\\
&& (24 q+28 q^3+4 q^5)t^{-1}- (36 q^2+16 q^4+q^6)t^{-4}\nonumber\\
&& \hat{\cF}_{sc1}^8(q,t)= 6 q^4 t^{-2}+6 t^2 q^4- q^2 (20 q+ 4 q^3) t^{-1}- t q^2 (20 q+ 4 q^3)+\nn\\
&&q^2(20+15 q^2+q^4)\nonumber\\
&& \hat{\cF}_{sc1}^9(q,t)=6 q^4 t^{-6}+ 6 q^4 t^{-2}- q^2(20 q+ 4
q^3)t^{-5}- q^2 (20 q+4 q^3)t^{-3}+ \nn\\
&&q^2 (20+15
q^2+q^4)t^{-4}\nonumber\\
&& \hat{\cF}_{sc1}^{10}(q,t)=-15-156 q^2-176 q^4- 15 q^4 t^{-4}-15 t^4 q^4-31 q^6+\nn\\
&& (60 q^3+24 q^5) t^{-3}+t^3(60 q^3+24 q^5)- (90 q^2+ 106 q^4+ 16 q^6) t^{-2}- \nn\\
&&t^2 (90 q^2+ 106 q^4+ 16 q^6)+(60 q+184 q^3+84 q^5+4 q^7)t^{-1}+\nn\\
&& t (60 q+184 q^3+84 q^5+4 q^7)
\eea
For $(\ell,0,2,0)$ one has:
\bea
&& \hat{\cF}_{sc2}^1(q,t)=-4 t^3 q+15 q^2- 4 q^3 t^{-1}+q^4+ t^2 (10+6 q^2)-t (20 q+4 q^3)\nonumber\\
&& \hat{\cF}_{sc2}^2(q,t)=15+36 q^2- 4 q^3 t^{-3}-4 t^3 q^3+16 q^4+q^6+ (16 q^2+ 6 q^4) t^{-2}+\nonumber\\
&& t^2 (16 q^2+ 6 q^4)- (24 q+24 q^3+4 q^5) t^{-1}-t (24 q+24 q^3+4 q^5)\nonumber\\
&& \hat{\cF}_{sc2}^3(q,t)=- 4 q t^{-3}+15 q^2-4 t q^3+q^4+ (10+ 6 q^2)t^{-2}- (20 q+4 q^3)t^{-1}\nonumber\\
&& \hat{\cF}_{sc2}^4(q,t)=4 t^3 q^3- 6 q^4 t^{-2}- t^2 q (16 q+ 6 q^3)+ q (24 q^2+4 q^4)t^{-1}+\nonumber\\
&& t q (20+24 q^2+4 q^4)- q (35 q+ 16 q^3+ q^5)\\
&& \hat{\cF}_{sc2}^5(q,t)= 4 q^3 t^{-3}-6 t^2 q^4- q (16 q+6 q^3) t^{-2}+ q (20+24 q^2+4 q^4) 
t^{-1}+\nonumber\\
&&  t q (24 q^2+ 4 q^4)-q (35 q+16 q^3+q^5)\nonumber\\
&& \hat{\cF}_{sc2}^6(q,t)= 6 q^4 t^{-2}+6 t^2 q^4- q^2(20 q+4 q^3)
t^{-1}- t q^2 (20 q+4 q^3) +\nn\\
&&q^2(20 +15 q^2 +q^4)\nonumber \eea 
For $(\ell-2,0,0,2)$ one has: \bea
&& \hat{\cF}_{sc3}^1(q,t)=t^4(q^2 -q^4)\nonumber\\
&& \hat{\cF}_{sc3}^2(q,t)=q^2-q^4\nonumber\\
&& \hat{\cF}_{sc3}^3(q,t)=(q^2-q^4)t^{-4}\nonumber\\
&& \hat{\cF}_{sc3}^4(q,t)=6 t^2 q^2-4 t q^3-4 t^3 q^3+q^4+t^4 q^4\\
&& \hat{\cF}_{sc3}^5(q,t)=6 q^2t^{-2}-4 q^3t^{-3}-4 q^3t^{-1}+q^4+ q^4t^{-4}\nonumber\\
&& \hat{\cF}_{sc3}^6(q,t)=6 q^4t^{-2} +6 t^2 q^4 -q^2 (20 q+4
q^3)t^{-1} -t q^2 (20 q +4q^3) +\nn\\
&&q^2 (20+15 q^2+q^4)\nonumber \eea
For $(\ell,0,0,1)$ one has: \bea
&& \hat{\cF}_{gr1}^1(q,t)=t^2 (1-q^2)\nonumber\\
&& \hat{\cF}_{gr1}^2(q,t)=t^{-2}(1-q^2)\\
&& \hat{\cF}_{gr1}^3(q,t)=6- 4 q t^{-1}-4 t q+ q^2 t^{-2}+t^2
q^2\nonumber \eea 
For $(\ell-1,0,1,0)$ one has: \bea
&& \hat{\cF}_{gr2}^1(q,t)=4 t q-6 q^2-t^2 q^2+ 4 q^3 t^{-1}- q^4 t^{-2}\nonumber\\
&& \hat{\cF}_{gr2}^2(q,t)= 4 q t^{-1}-6 q^2- q^2 t^{-2}+4 t q^3-t^2 q^4\\
&& \hat{\cF}_{gr2}^3(q,t)=6 q^2- 4 q^3 t^{-1}-4 t q^3+ q^4
t^{-2}+t^2 q^4\nonumber \eea 
For $(\ell-1,1,1,0)$ one has: \bea
&& \hat{\cF}_{f1}^1(q,t)=4 t^5 q^3-6 q^4 -t^4 q (16 q +6 q^3)+ t q (24 q^2+4 q^4) +\nonumber\\
&& t^3 q(20+24 q^2 +4 q^4) -t^2 q (35 q+16 q^3+q^5)\nonumber\\
&& \hat{\cF}_{f1}^2(q,t)= 4 q^3 t^{-1}-6 t^4 q^4-q (16 q+6 q^3)+t q (20+24 q^2+4 q^4)+\nonumber\\
&& t^3 q (24 q^2+4 q^4)-t^2 q(35 q+16 q^3+q^5)\nonumber\\
&& \hat{\cF}_{f1}^3(q,t)= 4 q^3 t^{-5}-6 q^4- q (16 q+6 q^3)t^{-4}+ q(20+24 q^2+4 q^4)t^{-3}+ \nonumber\\
&& q(24 q^2+4 q^4)t^{-1}-q(35 q+16 q^3+q^5)t^{-2}\nonumber\\
&& \hat{\cF}_{f1}^4(q,t)=- 15 q^4 t^{-2}-10 t^4 q^4+ q(56 q^2+24 q^4) t^{-1}+t^3 q(40 q^2+20 q^4)-\nonumber\\
&& t^2 q(60 q+80 q^3+15 q^5)-q (74 q+90 q^3+16 q^5)+t q (36+120 q^2+60 q^4+4 q^6)\nonumber\\
&& \hat{\cF}_{f1}^5(q,t)=4 t q^3- 6 q^4 t^{-4} -q(16 q+6 q^3)+ q (24 q^2+4 q^4) t^{-3}+\nonumber\\
&& q (20+24 q^2+4 q^4) t^{-1}- q (35 q+16 q^3+q^5) t^{-2}\nonumber\\
&& \hat{\cF}_{f1}^6(q,t)=- 10 q^4 t^{-4}-15 t^2 q^4 +t q(56 q^2+24 q^4)+ q (40 q^2+20 q^4) t^{-3}-\nonumber\\
&&  q(60 q+80 q^3+15 q^5) t^{-2} - q(74 q+90 q^3+16 q^5)+\nn\\
&& q(36+120 q^2+60 q^4+4 q^6) t^{-1}\nonumber\\
&& \hat{\cF}_{f1}^7(q,t)=4 t q-6 q^2-t^2 q^2+ 4 q^3 t^{-1}- q^4 t^{-2}\nonumber\\
&& \hat{\cF}_{f1}^8(q,t)= 4 q t^{-1}-6 q^2- q^2 t^{-2}+4 t q^3-t^2 q^4\nonumber\\
&& \hat{\cF}_{f1}^9(q,t)=6 q^4+6 t^4 q^4 -t q^2 (20 q+4 q^3) -t^3 q^2 (20 q+4 q^3)+\nn\\
&&t^2 q^2 (20+15 q^2+q^4)\nonumber\\
&& \hat{\cF}_{f1}^{10}(q,t)=6 q^4+ 6 q^4 t^{-4}- q^2 (20 q+4 q^3)
t^{-3}- q^2 (20 q+4 q^3) t^{-1}+ \nn\\
&&q^2 (20+15
q^2+q^4) t^{-2}\nonumber\\
&& \hat{\cF}_{f1}^{11}(q,t)=6 q^2- 4 q^3 t^{-1}-4 t q^3+ q^4
t^{-2}+t^2 q^4 \eea 
Finally, for $(\ell-2,1,0,1)$ one
has: \bea
&& \hat{\cF}_{f2}^1(q,t)=6 t^4 q^2-4 t^3 q^3-4 t^5 q^3+t^2 q^4+t^6 q^4\nonumber\\
&& \hat{\cF}_{f2}^2(q,t)=6 q^2- 4 q^3 t^{-1}-4 t q^3+ q^4 t^{-2}+t^2 q^4\nonumber\\
&& \hat{\cF}_{f2}^3(q,t)=6 q^2 t^{-4}- 4 q^3 t^{-5}- 4 q^3 t^{-3}+ q^4 t^{-6}+ q^4 t^{-2}\nonumber\\
&& \hat{\cF}_{f2}^4(q,t)=t^2 (q^2-q^4)\nonumber\\
&& \hat{\cF}_{f2}^5(q,t)=(q^2-q^4)t^{-2}\nonumber\\
&& \hat{\cF}_{f2}^6(q,t)=4 t q+4 t^3 q-6 q^2-6 t^4 q^2+ 4 q^3 t^{-1}+4 t^5 q^3- \nn\\
&&q^4 t^{-2}-t^6 q^4-t^2 (1+q^2)\nonumber
\eea
\bea
&& \hat{\cF}_{f2}^7(q,t)=4 q t^{-3}+ 4 q t^{-1}-6 q^2- 6 q^2
t^{-4}+ 4 q^3 t^{-5}+ 4 t q^3-\nn\\
&& q^4 t^{-6}-t^2 q^4- (1+q^2)t^{-2}\nonumber\\
&& \hat{\cF}_{f2}^8(q,t)=6 q^4+6 t^4 q^4 -t q^2(20 q+4 q^3) -t^3
q^2 (20 q+4 q^3)+\nn\\
&& t^2 q^2 (20+15 q^2+q^4)\nonumber\\
&&\hat{\cF}_{f2}^9(q,t)=6 q^4+ 6 q^4 t^{-4}- q^2(20 q+4 q^3) t^{-3}-
q^2(20 q+4 q^3) t^{-1}+\nn\\
&& q^2(20+15 q^2+q^4)
t^{-2}\nonumber\\
&& \hat{\cF}_{f2}^{10}(q,t)=-6-32 q^2-12 q^4- 6 q^4 t^{-4}-6 t^4 q^4+ (24 q^3+4 q^5)t^{-3}+\nn\\
&& t^3 (24 q^3+4 q^5)+ (24 q+28 q^3+4 q^5)t^{-1} +t(24 q+28 q^3+4 q^5) -\nn\\
&& (36 q^2+16 q^4+q^6) t^{-2} -t^2(36 q^2+16 q^4+q^6)\nonumber\\
&& \hat{\cF}_{f2}^{11}(q,t)=6 q^2- 4 q^3 t^{-1} - 4 t q^3+ q^4
t^{-2}+t^2 q^4 \eea

\pagebreak


\begin{thebibliography}{99}
  
\bibitem{Bianchi:2009ij}
  M.~Bianchi and M.~Samsonyan,
  Int.\ J.\ Mod.\ Phys.\  A {\bf 24} (2009) 5737
  [arXiv:0909.2173 [hep-th]].
  
\bibitem{Poghossian:2008ge}
  R.~Poghossian and M.~Samsonyan,
  J.\ Phys.\ A  {\bf 42} (2009) 304024
  [arXiv:0804.3564 [hep-th]].
  
\bibitem{Bianchi:2010mg}
  M.~Bianchi, R.~Poghossian and M.~Samsonyan,
  JHEP {\bf 1010} (2010) 021
  [arXiv:1005.5307 [hep-th]].
  
\bibitem{BCKWreview}
  M.~Frau and A.~Lerda,
  Fortsch.\ Phys.\  {\bf 52}, 606 (2004)
  [arXiv:hep-th/0401062].
  A.~M.~Uranga,
  JHEP {\bf 0901}, 048 (2009)
  [arXiv:0808.2918 [hep-th]].
Annu.Rev.Nuc.Part.Sci (59) 2009, arXiv:0902.3251;
 M. Cvetic,
``D-brane Instantons'', Lectures delivered at the School ``New
Perspectives in String Theory'', GGI Arcetri (Florence), June
2009.

\bibitem{Kirbook}
  E.~Kiritsis,
{\it  Princeton, USA: Univ. Pr. (2007) 588 p}
  K.~Becker, M.~Becker and J.~H.~Schwarz,
{\it  Cambridge, UK: Cambridge Univ. Pr. (2007) 739 p}

\bibitem{ADHM}
M. Atiyah, V. Drinfeld, N. Hitchin and Yu. Manin, 
Phys. Lett. {\bf A 65} (185) 1978.

\bibitem{MKR}
M. Bianchi, S. Kovacs, G. Rossi, 
Lect.NotesPhys. {\bf 737} (303) 2008, arXiv:hep-th/0703142.

\bibitem{Douglas'95}
M. R. Douglas, ``Branes within Branes " arXiv:hep-th/9512077.
  E.~Witten,
  Nucl.\ Phys.\  B {\bf 460}, 335 (1996)
  [arXiv:hep-th/9510135].

\bibitem{Dine}
M. Dine, N. Seiberg, X. G. Wen and E. Witten 
Nucl. Phys. B {\bf 278}, (769) 1986;

M. Dine, N. Seiberg, X. G. Wen and E. Witten

Nucl. Phys. B {\bf 289}, (319) 1987.

\bibitem{DualitInst}
  A.~Sen,
  arXiv:hep-th/9802051.
  E.~Kiritsis,
  arXiv:hep-th/9906018.

\bibitem{25}
L. J Dixon, V. Kaplunovski and J. Louis, 
Nucl. Phys. B {\bf 355}, (649) 1991;

I. Antoniadis, E. Gava and K. S. Narain, 
Nucl. Phys. B {\bf 383} (93) 1992, arxiv:hep-th/9204030;

E. Kiritsis and C. Kounnas, 
Nucl. Phys. B {\bf 442} (472) 1995, arxiv:hep-th/9501020; Nucl. Phys. Proc. Suppl. {\bf 
41} (1995) 331, arxiv:hep-th/9410212.

\bibitem{Becker}
K. Becker, M. Becker and A. Strominger, 
Nucl. Phys. B {\bf 456}, (130) 1995, arxiv:hep-th/9507158.

\bibitem{Billo}
M. Billo, M. Frau, I. Pesando, F. Fucito, A. Lerda and A. Liccardo, ``Classical gauge instantons from open 
strings", JHEP {\bf 0302}, (045) 2003, arxiv:hep-th/0211250;

\bibitem{Billo.Inst.CalcinRR}
M.Billo, M. Frau, F. Fucito and A. Lerda, 
JHEP {\bf 0611}, (012) 2006, arxiv:hep-th/0606013.

\bibitem{Green:2000ke}
  M.~B.~Green and M.~Gutperle,
  JHEP {\bf 0002} (2000) 014
  [arXiv:hep-th/0002011].

\bibitem{Dorey}
N. Dorey, T.J. Hollowood, V.V. Khoze, M.P. Mattis and S. Vandoren, 
Nucl. Phys. B{\bf 552} (88) 1999, arXiv:hep-
th/9901128.

\bibitem{BKB}
C. Bachas, C. Fabre, E. Kiritsis, N. A. Obers and P. Vanhove,
Nucl. Phys. B {\bf 509} (33) 1998, arxiv:hep-th/9707126;

E. Kiritsis and N. A. Obers, 
JHEP {\bf 9710} (004) 1997, arxiv:hep-th/9709058;

C. Bachas, 
Nucl. Phys. Proc. Suppl. {\bf 68} (348) 1998, arxiv:hep-th/9710102.

\bibitem{WitPol}
J. Polchinski, E. Witten, 
Nucl. Phys. B {\bf 460} (525) 1996, arXiv:hep-th/9510169;

\bibitem{Kiritsis}
L. J. Dixon, V. Kaplunovsky, J. Louis, 
Nucl. Phys. B {\bf 329} (27) 1990;

C. Bachas, C. Fabre, E. Kiritsis, N. A. Obers, P. Vanhove, 
Nucl. Phys. B {\bf 509}, (33) 1998, arXiv:hep-th/9707126;

W. Lerche, S. Stieberger, 
Adv. Theor. Math. Phys. {\bf 3}, (1539) 1999, arXiv:hep-th/9907133.

\bibitem{MBEGJFMKN}
M. Bianchi, E. Gava, J. F. Morales, K. S. Narain, 
Nucl.Phys. B {\bf 547} (96) 1999, arXiv:hep-th/9811013.

\bibitem{Gutperetal}
M. Gutperle, 
JHEP {\bf 9905} (007) 
1999, arXiv:hep-th/9903010.

\bibitem{KFSS}
K. Foerger, S. Stieberger, 
Nucl.Phys. B{\bf 559}  (277) 1999, arXiv:hep-th/9901020.

\bibitem{Lerda_etal}
M. Billo', L. Ferro, M. Frau, L. Gallot, A. Lerda, I. Pesando, 
JHEP {\bf 0907} (092) 2009, arXiv:0905.4586;

\bibitem{Fucito:2009rs}
  F.~Fucito, J.~F.~Morales and R.~Poghossian,
  JHEP {\bf 0910} (2009) 041
  [arXiv:0906.3802 [hep-th]].

\bibitem{ABDFPT}
I. Antoniadis, C. Bachas, E. Dudas, 
Nucl.Phys. B{\bf 560} (93) 1999, arXiv:hep-th/9906039.

I. Antoniadis, C. Bachas, C. Fabre, H. Partouche, T.R. Taylor, 
Nucl.Phys. B{\bf 489} (160) 1997, arXiv:hep-th/9608012.

\bibitem{Blum}
R. Blumenhagen, M. Cvetic and T. Weigand, 
Nucl. Phys .B {\bf 771} (113) 2007, arxiv:hep-th/0609191;

L. E. Ibanez and A. M. Uranga, 
JHEP {\bf 0703} (052) 2007, arxiv:hep-th/0609213;

\bibitem{Blumik}
M. Haack, D. Krefl, D. Lust, A. Van Proeyen and M. Zagermann, 
JHEP {\bf 0701} (078) 2007, arxiv:hep-th/0609211;


B. Florea, S. Kachru, J. McGreevy and N. Saulina, 
JHEP {\bf 0705} (024) 2007, arxiv:hep-th/0610003.

\bibitem{Silverstein}
E. Silverstein and E. Witten, 
Nucl. Phys. B {\bf 444}, (161) 1995, arxiv:hep-th/9503212;

\bibitem{BWnieist}
C. Beasley and E, Witten, 
JHEP {\bf 0602} (060) 2006, arxiv:hep-th/0512039.

\bibitem{FW}
D. S. Freed and E. Witten, 
arxiv:hep-th/9907189.

\bibitem{KPT}
A. K. Kashani-Poor and A. Tomasiello, 
Nucl. Phys. B {\bf 728}, (135) 2005, arxiv:hep-th/0505208.

\bibitem{Billo':2008sp}
  M.~Billo', L.~Ferro, M.~Frau, F.~Fucito, A.~Lerda and J.~F.~Morales,
  JHEP {\bf 0810} (2008) 112
  [arXiv:0807.1666 [hep-th]].

\bibitem{Billo':2008pg}
  M.~Billo', L.~Ferro, M.~Frau, F.~Fucito, A.~Lerda and J.~F.~Morales,
  JHEP {\bf 0812} (2008) 102
  [arXiv:0807.4098 [hep-th]].

\bibitem{Billo:2004zq}
  M.~Billo, M.~Frau, I.~Pesando and A.~Lerda,
  JHEP {\bf 0405} (2004) 023
  [arXiv:hep-th/0402160].

\bibitem{Billo:2005jw}
  M.~Billo, M.~Frau, F.~Lonegro and A.~Lerda,
  JHEP {\bf 0505} (2005) 047
  [arXiv:hep-th/0502084].













\bibitem{U(4)}
M. Bianchi, F. Fucito, J. F. Morales, 
JHEP{\bf 0707} (038) 2007, arxiv:hep-th/0704.0784.

\bibitem{Anom&Tad}
M. Bianchi, J. F. Morales, 
JHEP {\bf 0003} (030) 2000, arXiv:hep-th/0002149.

\bibitem{AIMU}
G. Aldazabal, D. Badagnani, L.E. Ibanez, A.M. Uranga, 
JHEP {\bf 9906} (031) 1999, arXiv:hep-th/9904071.

\bibitem{Camara:2004jj}
  P.~G.~Camara, L.~E.~Ibanez and A.~M.~Uranga,
  Nucl.\ Phys.\  B {\bf 708} (2005) 268
  [arXiv:hep-th/0408036].


\bibitem{Lionettoetal}
P. Anastasopoulos, F. Fucito, A. Lionetto, G. Pradisi, A. Racioppi, Y. S. Stanev, 
Phys.Rev.{\bf D78}, (085014) 2008, arXiv:0804.1156 ,

F. Fucito, A. Lionetto, A. Mammarella, A. Racioppi, 
arXiv:0811.1953.

\bibitem{PWDH}
A. Dabholkar, 
Phys. Lett. B {\bf 357} (307) 1995, arXiv:hep-th/9506160;

C. M. Hull, 
 Phys. Lett. B {\bf 357} (545) 1995, arXiv:hep-
th/9506194.

\bibitem{EW}
E. Witten, 
Nucl. Phys. B {\bf 443} (85) 1995, arXiv:hep-th/9503124.

\bibitem{ABPSS}
C. Angelantonj, M. Bianchi, G. Pradisi, A. Sagnotti, Y. Stanev, 
Phys. Lett. B {\bf 385} (96) 1996, arXiv:hep-th/9606169.

\bibitem{ABFPT}
I. Antoniadis, C. Bachas, C. Fabre, H. Partouche, T.R. Taylor, 
Nucl. Phys. B {\bf 489} (160) 1997, arXiv:hep-th/9608012 .

\bibitem{X}
M. Berkooz, R. G. Leigh, J. Polchinski, J. H. Schwarz, N. Seiberg, E. Witten, 
Nucl. Phys. B {\bf 475} (115) 1996, arXiv:hep-th/9605184.

\bibitem{MSGP}
M. Bianchi, A. Sagnotti, 
Nucl. Phys. B {\bf 361} 519 (1991),
Phys. Lett. B {\bf 247} (517) 1990;

E. G. Gimon, J. Polchinski, 
Phys. Rev. D {\bf 54} (1667) 1996, arXiv:hep-th/9601038.

\bibitem{DPGJ}
A. Dabholkar, J. Park, 
Nucl. Phys. B {\bf 477} (701) 1996, arXiv:hep-th/9604178 and 
Nucl. Phys. B {\bf 472} 207 (1996), arXiv:hep-th/9602030.

E. G. Gimon, C. V. Johnson, 
Nucl. Phys. B {\bf 477} (715) 1996, arXiv:hep-th/9604129.

\bibitem{Duff}
M. J. Duff, 
Nucl. Phys. B {\bf 442} (47) 1995, arXiv:hep-th/9501030.

\bibitem{19}
M. Bianchi, G. Pradisi, A. Sagnotti, 
Nucl. Phys. B {\bf 376} (365) 1992;

Z. Kakushadze, 
Nucl. Phys. B {\bf 512} (221) 1998, arXiv:hep-th/9704059;

M. Bianchi,
Nucl. Phys. B {\bf 528} (73) 1998, arXiv:hep-th/9711201;

E. Witten, 
JHEP {\bf 9802} (006) 1998, arXiv:hep-th/9712028;

C. Angelantonj and R. Blumenhagen, 
Phys. Lett. B {\bf 473} (86) 2000, arXiv:hep-th/9911190.

\bibitem{Berkooz}
M. Berkooz, R. G. Leigh, J. Polchinski, J. H. Schwarz, N. Seiberg, E. Witten, 
Nucl. Phys. B {\bf 475} (115) 1996, arXiv:hep-th/9605184.

\bibitem{Anomalies&Tadpoles}
M. B. Green, J. H. Schwarz and P. C. West, 
Nucl. Phys. B {\bf 254} (327) 1985;

A. Sagnotti, 
Phys. Lett. B {\bf 294} (196) 1992, arXiv:hep-th/9210127.

\bibitem{Pascal}
P. Anastasopoulos, 
Phys. Lett. B {\bf 588} (119) 2004, arXiv:hep-th/0402105;

P. Anastasopoulos, M. Bianchi, E. Dudas, E. Kiritsis, 
JHEP {\bf 0611} (057) 2006, arXiv:hep-th/0605225.

\bibitem{DFMS}
L. J. Dixon, D. Friedan, E. J. Martinec, S. H. Shenker, 
Nucl. Phys. B {\bf 282} (13) 1987.

\bibitem{KM}
E. Kiritsis, N.A. Obers, B. Pioline, 
JHEP {\bf 0001} (029) 2000, arXiv:hep-th/0001083;

P. Mayr, S. Stieberger, 
Nucl. Phys.  B {\bf 407} (725) 1993, arXiv:hep-th/9303017.

\bibitem{CPKW}
M. Cvetic, I. Papadimitriou, 
Phys. Rev. D {\bf 68} (046001) 2003; Erratum-ibid. D70 (2004) 029903, arXiv:hep-th/0303083;

I. R. Klebanov and E. Witten, 
Nucl. Phys. B {\bf 664} (3) 2003, arXiv:hep-th/0304079;

\bibitem{GMB}
E.Gava, J.F.Morales, K.S.Narain, G.Thompson, 
Nucl.Phys. B {\bf 528} (95) 1998,arXiv:hep-th/9801128;

Massimo Bianchi, 
arXiv:hep-th/9702098;

C. Bachas, 
arXiv:hep-th/9701019.

\bibitem{Blumenhagen:2007zk}
  R.~Blumenhagen, M.~Cvetic, D.~Lust, R.~Richter and T.~Weigand,
  Phys.\ Rev.\ Lett.\  {\bf 100} (2008) 061602
  [arXiv:0707.1871 [hep-th]].

\bibitem{Argurio:2007vqa}
  R.~Argurio, M.~Bertolini, G.~Ferretti, A.~Lerda and C.~Petersson,
  JHEP {\bf 0706} (2007) 067
  [arXiv:0704.0262 [hep-th]].

\bibitem{Blumenhagen:2007bn}
  R.~Blumenhagen, M.~Cvetic, R.~Richter and T.~Weigand,
  JHEP {\bf 0710} (2007) 098
  [arXiv:0708.0403 [hep-th]].
  
\bibitem{MBJFMFF}
M. Bianchi, F. Fucito, J. F. Morales, 
JHEP {\bf 0908} (040) 2009, arXiv:0904.2156.


\bibitem{Bianchi:1998nk}
  M.~Bianchi, M.~B.~Green, S.~Kovacs and G.~Rossi,
  JHEP {\bf 9808} (1998) 013
  [arXiv:hep-th/9807033].

\bibitem{FPS} R.Flume, R.Poghossian, H.Storch 
Mod.Phys.Lett. A17 (2002) 327, arXiv:hep-th/0112211.

\bibitem{Nek}
N.~A.~Nekrasov 
Adv.\ Theor.\ Math.\ Phys.\  {\bf 7} (2004) 831,
arXiv:hep-th/0206161.

\bibitem{FP}
R.~Flume and R.~Poghossian
Int.\ J.\ Mod.\ Phys.\ A {\bf 18} (2003) 2541,
arXiv:hep-th/0208176.

\bibitem{Nek_Okoun}
N.~Nekrasov and A.~Okounkov
arXiv:hep-th/0306238.

\bibitem{Sei_Witt}
N.~Seiberg and E.~Witten
Nucl.\ Phys.\ B {\bf 426} (1994) 19, Erratum-ibid.\ B {\bf 430}
(1994) 485, arXiv:hep-th/9407087.

\bibitem{Iqbal}  A.~Iqbal, C.~Kozcaz, K.~Shabbir
arXiv:0803.2260

\bibitem{fucito}
U.~Bruzzo, F.~Fucito, J.~F.~Morales and A.~Tanzini,
JHEP {\bf 0305} (2003) 054, arXiv:hep-th/0211108.

\bibitem{Nak_lect} H.~Nakajima,
K.~Yoshioka, 
arXiv:math/0311058.

\bibitem{Nak_blowup}H.~Nakajima,
K.~Yoshioka,
arXiv:math/0306198.

\bibitem{FMPT}F.Fucito, J.F.Morales, R.Poghossian, A.Tanzini,
JHEP 0601 (2006) 031, arXiv:hep-th/0510173.

\bibitem{Nak_book}H.~Nakajima,
 American Mathematical Society, University Lectures
Series v.18 (1999).

\bibitem{Iqbal_Vafa} A.~Iqbal, C.~Kozcaz, C~.Vafa ,
arXiv:hep-th/0701156.


\bibitem{Schwarz:2004yj}
  J.~H.~Schwarz,
  JHEP {\bf 0411}, 078 (2004)
  [arXiv:hep-th/0411077].

\bibitem{Bagger:2006sk}
  J.~Bagger and N.~Lambert,
  Phys.\ Rev.\  D {\bf 75}, 045020 (2007)
  [arXiv:hep-th/0611108].
  
\bibitem{Bagger:2007jr}
  J.~Bagger and N.~Lambert,
  Phys.\ Rev.\  D {\bf 77}, 065008 (2008)
  [arXiv:0711.0955 [hep-th]].

\bibitem{Bagger:2007vi}
  J.~Bagger and N.~Lambert,
  JHEP {\bf 0802}, 105 (2008)
  [arXiv:0712.3738 [hep-th]].


\bibitem{Bagger:2008se}
  J.~Bagger and N.~Lambert,
  Phys.\ Rev.\  D {\bf 79}, 025002 (2009)
  [arXiv:0807.0163 [hep-th]].


\bibitem{Gustavsson:2007vu}
  A.~Gustavsson,
  Nucl.\ Phys.\  B {\bf 811}, 66 (2009)
  [arXiv:0709.1260 [hep-th]].

\bibitem{Alday:2009aq}
  L.~F.~Alday, D.~Gaiotto and Y.~Tachikawa,
  Lett.\ Math.\ Phys.\  {\bf 91} (2010) 167
  [arXiv:0906.3219 [hep-th]].
  
\bibitem{Aharony:2008ug}
  O.~Aharony, O.~Bergman, D.~L.~Jafferis and J.~Maldacena,
  JHEP {\bf 0810}, 091 (2008)
  [arXiv:0806.1218 [hep-th]].
  
\bibitem{Klebanov:2009sg}
  I.~R.~Klebanov and G.~Torri,
  arXiv:0909.1580 [hep-th].

\bibitem{Jafferis:2008qz}
  D.~L.~Jafferis and A.~Tomasiello,
  JHEP {\bf 0810}, 101 (2008)
  [arXiv:0808.0864 [hep-th]].


\bibitem{Jafferis:2008em}
  D.~L.~Jafferis and X.~Yin,
  arXiv:0810.1243 [hep-th].


\bibitem{Gaiotto:2009tk}
  D.~Gaiotto and D.~L.~Jafferis,
  arXiv:0903.2175 [hep-th].

\bibitem{Aganagic:2009zk}
  M.~Aganagic,
[arXiv:0905.3415 [hep-th]].
  
\bibitem{Forcella:2009jj}
  D.~Forcella and A.~Zaffaroni,
  JHEP {\bf 1005} (2010) 045
  [arXiv:0911.2595 [hep-th]].

\bibitem{Jafferis:2009th}
  D.~L.~Jafferis,
  arXiv:0911.4324 [hep-th].




\bibitem{Bhattacharya:2008zy}
  J.~Bhattacharya, S.~Bhattacharyya, S.~Minwalla and S.~Raju,
  JHEP {\bf 0802}, 064 (2008)
  [arXiv:0801.1435 [hep-th]].

\bibitem{Bhattacharya:2008bja}
  J.~Bhattacharya and S.~Minwalla,
  JHEP {\bf 0901}, 014 (2009)
  [arXiv:0806.3251 [hep-th]].

\bibitem{Kim:2009wb}
  S.~Kim,
  Nucl.\ Phys.\  B {\bf 821}, 241 (2009)
  [arXiv:0903.4172 [hep-th]].

\bibitem{Kim:2010vw}
  S.~Kim and J.~Park,
  arXiv:1003.4343 [hep-th].



\bibitem{Castellani:1984vv}
  L.~Castellani, R.~D'Auria, P.~Fre, K.~Pilch and P.~van Nieuwenhuizen,
  Class.\ Quant.\ Grav.\  {\bf 1}, 339 (1984).
  
\bibitem{Salam:1989fm}
  A.~Salam and E.~Sezgin,
  ``Supergravities in diverse dimensions.  {\bf V}. 1, 2,''
{\it  Amsterdam, Netherlands: North-Holland (1989) 1499 p.
Singapore, Singapore: World Scientific (1989) 1499 p}

\bibitem{Malda}
J.~M.~Maldacena
Adv.~Theor.~Math.~Phys.~{\bf 2} 231, (1998),
[arXiv:hep-th/9711200 [hep-th]].
  
\bibitem{Aharony:2008gk}
  O.~Aharony, O.~Bergman and D.~L.~Jafferis,
  JHEP {\bf 0811}, 043 (2008)
  [arXiv:0807.4924 [hep-th]].

\bibitem{Gaiotto:2009mv}
  D.~Gaiotto and A.~Tomasiello,
 [arXiv:0901.0969 [hep-th]].

\bibitem{Gaiotto:2009yz}
  D.~Gaiotto and A.~Tomasiello,
  J.\ Phys.\ A  {\bf 42}, 465205 (2009)
  [arXiv:0904.3959 [hep-th]].
  
\bibitem{Petrini:2009ur}
  M.~Petrini and A.~Zaffaroni,
  JHEP {\bf 0909} (2009) 107
  [arXiv:0904.4915 [hep-th]].



\bibitem{Nilsson:1984bj}
  B.~E.~W.~Nilsson and C.~N.~Pope,
  Class.\ Quant.\ Grav.\  {\bf 1}, 499 (1984).

\bibitem{Klebanov:2008vq}
  I.~Klebanov, T.~Klose and A.~Murugan,
  JHEP {\bf 0903}, 140 (2009)
  [arXiv:0809.3773 [hep-th]].

\bibitem{Benna:2009xd}
  M.~K.~Benna, I.~R.~Klebanov and T.~Klose,
  arXiv:0906.3008 [hep-th].
  
\bibitem{VanNieuwenhuizen:1985be}
  P.~Van Nieuwenhuizen,
{\it  In *Les Houches 1983, Proceedings, Relativity, Groups and
Topology, Ii*, 823-932}


\bibitem{Sezgin:1983ik}
  E.~Sezgin,
  Phys.\ Lett.\  B {\bf 138}, 57 (1984).
  
\bibitem{Zhelobenko}
D.~P.~Zhelobenko
``Compact Lie groups and their representations".
Translations of mathematical monographs, {\bf V}. 40, AMS (1973).  

\bibitem{Benna:2008zy}
  M.~Benna, I.~Klebanov, T.~Klose and M.~Smedback,
  JHEP {\bf 0809}, 072 (2008)
  [arXiv:0806.1519 [hep-th]].

\bibitem{Cagnazzo:2009zh}
  A.~Cagnazzo, D.~Sorokin and L.~Wulff,
  JHEP {\bf 1005}, 009 (2010)
  [arXiv:0911.5228 [hep-th]].

\bibitem{Hosomichi:2008ip}
  K.~Hosomichi, K.~M.~Lee, S.~Lee, S.~Lee, J.~Park and P.~Yi,
  JHEP {\bf 0811} (2008) 058
  [arXiv:0809.1771 [hep-th]].

\bibitem{Bianchi:2008cj}
  M.~Bianchi,
  Nucl.\ Phys.\  B {\bf 805}, 168 (2008)
  [arXiv:0805.3276 [hep-th]].

\bibitem{Bianchi:2009mu}
  M.~Bianchi,
  Fortsch.\ Phys.\  {\bf 57}, 356 (2009)
  [arXiv:0902.0650 [hep-th]].


\bibitem{Sezgin:1998gg}
  E.~Sezgin and P.~Sundell,
  JHEP {\bf 9811}, 016 (1998)
  [arXiv:hep-th/9805125].

\bibitem{Sezgin:2002rt}
  E.~Sezgin and P.~Sundell,
  Nucl.\ Phys.\  B {\bf 644}, 303 (2002)
  [Erratum-ibid.\  B {\bf 660}, 403 (2003)]
  [arXiv:hep-th/0205131].

\bibitem{Engquist:2002vr}
  J.~Engquist, E.~Sezgin and P.~Sundell,
  Class.\ Quant.\ Grav.\  {\bf 19}, 6175 (2002)
  [arXiv:hep-th/0207101].

\bibitem{Bianchi:2004ww}
  M.~Bianchi,
  Comptes Rendus Physique {\bf 5}, 1091 (2004)
  [arXiv:hep-th/0409292].

\bibitem{Bianchi:2004xi}
  M.~Bianchi,
  Fortsch.\ Phys.\  {\bf 53}, 665 (2005)
  [arXiv:hep-th/0409304].

\bibitem{Bekaert:2005vh}
  X.~Bekaert, S.~Cnockaert, C.~Iazeolla and M.~A.~Vasiliev,
  arXiv:hep-th/0503128.
  
\bibitem{Francia:2006hp}
  D.~Francia and A.~Sagnotti,
  J.\ Phys.\ Conf.\ Ser.\  {\bf 33}, 57 (2006)
  [arXiv:hep-th/0601199].

\bibitem{Ferrara:1997dh}
  S.~Ferrara and C.~Fronsdal,
  Class.\ Quant.\ Grav.\  {\bf 15}, 2153 (1998)
  [arXiv:hep-th/9712239].


\bibitem{Sundborg:1999ue}
  B.~Sundborg,
  Nucl.\ Phys.\  B {\bf 573}, 349 (2000)
  [arXiv:hep-th/9908001].
  
\bibitem{HaggiMani:2000ru}
  P.~Haggi-Mani and B.~Sundborg,
  JHEP {\bf 0004}, 031 (2000)
  [arXiv:hep-th/0002189].


\bibitem{Bianchi:1999ge}
  M.~Bianchi, S.~Kovacs, G.~Rossi and Y.~S.~Stanev,
  JHEP {\bf 9908}, 020 (1999)
  [arXiv:hep-th/9906188].

\bibitem{Bianchi:2000hn}
  M.~Bianchi, S.~Kovacs, G.~Rossi and Y.~S.~Stanev,
  Nucl.\ Phys.\  B {\bf 584}, 216 (2000)
  [arXiv:hep-th/0003203].

\bibitem{Bianchi:2001cm}
  M.~Bianchi, S.~Kovacs, G.~Rossi and Y.~S.~Stanev,
  JHEP {\bf 0105}, 042 (2001)
  [arXiv:hep-th/0104016].




\bibitem{Bianchi:2003wx}
M.~Bianchi, J.~F.~Morales, H.~Samtleben.
JHEP {\bf 0307} (2003) 062
[arXiv:0305052 [hep-th]].

\bibitem{Beisert:2003te}
  N.~Beisert, M.~Bianchi, J.~F.~Morales and H.~Samtleben,
  JHEP {\bf 0402}, 001 (2004)
  [arXiv:hep-th/0310292].


\bibitem{Beisert:2004di}
  N.~Beisert, M.~Bianchi, J.~F.~Morales and H.~Samtleben,
  JHEP {\bf 0407}, 058 (2004)
  [arXiv:hep-th/0405057].


\bibitem{Bianchi:2005ze}
  M.~Bianchi, P.~J.~Heslop and F.~Riccioni,
  JHEP {\bf 0508}, 088 (2005)
  [arXiv:hep-th/0504156].


\bibitem{Bianchi:2006ti}
  M.~Bianchi, F.~A.~Dolan, P.~J.~Heslop and H.~Osborn,
  Nucl.\ Phys.\  B {\bf 767}, 163 (2007)
  [arXiv:hep-th/0609179].


\bibitem{Kounnas:2007dd}
  C.~Kounnas, D.~Lust, P.~M.~Petropoulos and D.~Tsimpis,
  JHEP {\bf 0709}, 051 (2007)
  [arXiv:0707.4270 [hep-th]].
  
\bibitem{Caviezel:2008ik}
  C.~Caviezel, P.~Koerber, S.~Kors, D.~Lust, D.~Tsimpis and M.~Zagermann,
  Class.\ Quant.\ Grav.\  {\bf 26}, 025014 (2009)
  [arXiv:0806.3458 [hep-th]].


\bibitem{Haack:2009jg}
  M.~Haack, D.~Lust, L.~Martucci and A.~Tomasiello,
  JHEP {\bf 0910}, 089 (2009)
  [arXiv:0905.1582 [hep-th]].

\bibitem{Bianchi:2010cy}
  M.~Bianchi and W.~de Paula,
  JHEP {\bf 1004}, 113 (2010)
  [arXiv:1003.2536 [hep-th]].

\bibitem{Pope:1984bd}
  C.~N.~Pope and N.~P.~Warner,
  Phys.\ Lett.\  B {\bf 150} (1985) 352.

\bibitem{Polyakov:1975yp}
  A.~M.~Polyakov and A.~A.~Belavin,
  JETP Lett.\  {\bf 22} (1975) 245
  [Pisma Zh.\ Eksp.\ Teor.\ Fiz.\  {\bf 22} (1975) 503].

\bibitem{Golo:1978de}
  V.~L.~Golo and A.~M.~Perelomov,
  Phys.\ Lett.\  B {\bf 79}, 112 (1978).


\bibitem{Golo:1978dd}
  V.~L.~Golo and A.~M.~Perelomov,
  Lett.\ Math.\ Phys.\  {\bf 2}, 477 (1978).
  
\bibitem{D'Adda:1978uc}
  A.~D'Adda, M.~Luscher and P.~Di Vecchia,
  Nucl.\ Phys.\  B {\bf 146}, 63 (1978).
  

\end{thebibliography}
\end{document}